\def\doccolumn{2}

\if\doccolumn2
    \documentclass[lettersize,journal,twocolumn,twoside]{IEEEtran}
    \newcommand\subfigwidth{0.485\linewidth}
    
    \newcommand\subsubfigwidth{0.245\linewidth}
    \newcommand\halfsubfigwidth{0.475\linewidth}
    \newcommand\halfsubfigwidthA{0.85\linewidth}
    \newcommand\figwidth{\linewidth}
    \newcommand\figwidthA{0.9\linewidth}
    \newcommand\matsep{2pt}
    \newcommand\matstretch{1.0}
\else
    \documentclass[12pt,draftclsnofoot,onecolumn,twoside]{IEEEtran}
    \newcommand\subfigwidth{8.0cm}
    
    \newcommand\subsubfigwidth{4.0cm}
    \newcommand\halfsubfigwidth{4.8cm}
    \newcommand\halfsubfigwidthA{5.8cm}
    \newcommand\figwidth{7.5cm}
    \newcommand\figwidthA{7.5cm}
    \newcommand\matsep{3pt}
    \newcommand\matstretch{0.6}
\fi

\usepackage[utf8]{inputenc}
\usepackage[T1]{fontenc}
\usepackage[cmex10]{amsmath}
\usepackage{mleftright}       
\mleftright                   

\usepackage{amsfonts}
\usepackage{amssymb}
\usepackage{amsthm}
\usepackage{array}
\usepackage{bm}
\usepackage{bbm}
\usepackage{epsfig}
\usepackage{color}
\usepackage{xcolor}
\usepackage{cite}
\usepackage{graphicx}
\usepackage{multirow}
\usepackage{stfloats}
\usepackage{enumitem}
\usepackage{dsfont}
\usepackage{booktabs}
\usepackage{tabularx}
\usepackage{dsfont}
\usepackage{xpatch}
\usepackage{mathtools}
\usepackage{algorithm}
\usepackage{algpseudocode}
\usepackage{threeparttable}
\usepackage{verbatim}
\usepackage{textcomp}
\usepackage{multicol}
\usepackage{url}
\usepackage{ulem}
\usepackage{scalerel}
\usepackage[caption=false, font=footnotesize]{subfig}

\usepackage[ddmmyyyy]{datetime}
\PassOptionsToPackage{hyphens}{url}\usepackage{hyperref}

\def\rev{\textcolor{red}}

\if\doccolumn2
\usepackage[switch]{lineno}
\else
\usepackage{lineno}
\fi
\modulolinenumbers[2]
\setpagewiselinenumbers

\makeatletter
\DeclareRobustCommand\bfseriesitshape{%
  \not@math@alphabet\itshapebfseries\relax
  \fontseries\bfdefault
  \fontshape\itdefault
  \selectfont
}
\makeatother
\DeclareTextFontCommand{\textbfit}{\bfseriesitshape}

\if\doccolumn2
\newtheoremstyle{newthmstyle}{\topsep}{\topsep}{}{\parindent}{\itshape}{:}{ }{}
\fi
\theoremstyle{newthmstyle}
\newtheorem{thm}{Theorem}
\newtheorem{define}[thm]{Definition}

\newtheorem{prop}[thm]{Proposition}
\newtheorem{cor}[thm]{Corollary}
\newtheorem{exam}[thm]{Example}

\newtheorem{remark}[thm]{Remark}

\DeclareMathOperator*{\argmax}{arg\,max}

\DeclareMathOperator*{\bigboxdot}{\scalerel*{\boxdot}{\textstyle\sum}}
\DeclareMathOperator*{\bigboxplus}{\scalerel*{\boxplus}{\textstyle\sum}}
\newcommand\mmid{\,|\,}
\DeclareMathSymbol{\mdot}{\mathord}{symbols}{"01}

\makeatletter
\xpatchcmd\proof{\@addpunct{.}}{\@addpunct{:}}{}{}
\xpatchcmd\proof{\itshape}{\itshape}{}{}
\xpatchcmd\proof{\hskip}{\hskip3}{}{}
\makeatother

\begin{document}

\title{Successive Cancellation Decoding \\ With Future Constraints for Polar Codes \\ Over the Binary Erasure Channel}

\author{Min~Jang,~\IEEEmembership{Member,~IEEE},
Jong-Hwan Kim,
Seho Myung, \\
and Kyeongcheol Yang,~\IEEEmembership{Senior Member,~IEEE}
\\

\thanks{
    This work has been published to the IEEE Access \cite{Jang2023access} (Digital Object Identifier: 10.1109/ACCESS.2023.3312577).
    An earlier version of this paper will be presented in part at the 2023 IEEE Global Communications Conference (Globecom).
}
\thanks{M. Jang is with the Samsung Research, Samsung Electronics, Seoul 06765, South Korea (e-mail: \mbox{mn.jang@}samsung.com).}
\thanks{J.-H. Kim is with the Network Business, Samsung Electronics, Suwon, Gyeonggi 16677, South Korea (e-mail: \mbox{jhwan729.kim@}samsung.com).}
\thanks{S. Myung is with the Mobile eXperience Business, Samsung Electronics, Suwon, Gyeonggi 16677, South Korea (e-mail: \mbox{seho.myung@}samsung.com).}
\thanks{K. Yang is with the Deptartment of Electrical Engineering, Pohang University of Science and Technology (POSTECH), Pohang, Gyeongbuk 37673, South Korea (e-mail: \mbox{kcyang@}postech.ac.kr).}
}
\maketitle


\begin{abstract}
    In the conventional successive cancellation (SC) decoder for polar codes, all the future bits to be estimated later are treated as random variables.
    However, polar codes inevitably involve frozen bits,
    and their concatenated coding schemes also include parity bits (or dynamic frozen bits) causally generated from the past bits estimated earlier.
    We refer to the frozen and parity bits located behind a target decoding bit as its \textit{future constraints (FCs)}.
    Although the values of FCs are deterministic given the past estimates,
    they have not been exploited in the conventional SC-based decoders, not leading to optimality.
    In this paper,
    with a primary focus on the binary erasure channel (BEC), we propose SC-check (SCC) and belief propagation SCC (BP-SCC) decoding algorithms in order to leverage FCs in decoding.
    We further devise an improved tree search technique based on stack-based backjumping (SBJ) to solve dynamic constraint satisfaction problems (CSPs) formulated by FCs.
    Over the BEC, numerical results show that a combination of the BP-SCC algorithm and the SBJ tree search technique achieves the erasure recovery performance close to the dependence testing (DT) bound, a bound of achievable finite-length performance.
\end{abstract}



\IEEEpeerreviewmaketitle

\section{Introduction} \label{sec:intro}

\IEEEPARstart{P}{olar} codes are error-correcting codes first proved to achieve the symmetric capacity of an arbitrary binary-input discrete memoryless channel (B-DMC) under successive-cancellation (SC) decoding \cite{Arikan2009}.
By polar coding, physical parallel B-DMCs are transformed into virtual channels (called sub-channels)
whose symmetric capacities are either extremely high or low as the code length increases.
Based on this channel polarization,
the total capacity is readily achieved by assigning information bits to reliable sub-channels
while allocating frozen bits to unreliable ones.
Also, excellent performance is achieved in practical applications
by concatenating appropriate outer codes and employing improved decoding schemes
such as SC-list (SCL), SC-stack (SCS), SC-fip (SCF), SC-Fano, and SC inactivation (SCI) algorithms \cite{Tal2015,Niu2012a,Afisiadis2014,Jeong2019,Coskun2020}.
Due to the good performance especially for short code lengths,
polar codes are now being used to transmit control information in the 5th Generation (5G) communication standard, the 3rd Generation Partnership Project (3GPP) New Radio (NR) \cite{TS38212}.

In SC decoding,
encoder input bits are sequentially estimated, assuming that previous estimates are true.
Ar{\i}kan pointed out in his seminal work \cite{Arikan2009} that the SC decoder treats all future bits as random variables (RVs),
even though there are frozen bits whose values are known.
It was repeatedly noted in \cite{Arikan2016} that they are regarded as pure noise in sequential decoding.
This relaxation brings suboptimality with respect to the maximum likelihood (ML) decision,
but allows efficient computation using recursive formulas.
The recursive decoding operations are tractable for the performance analysis,
and the achievability of the channel capacity is proved despite its suboptimality.

The relaxation that assumes all future bits as RVs has also been generally taken in improved SC-based decoding schemes
including SCL, SCS, SCF, SC-Fano, and SCI algorithms \cite{Tal2015,Niu2012a,Afisiadis2014,Jeong2019,Coskun2020}.
These schemes estimate each information bit without using future bits with deterministic values,
but they compensate for the relaxation by directly processing them later.
For example, the SCI decoder \cite{Coskun2020} performs SC decoding without taking future bits into account
and replaces a bit decoded as an erasure by a dummy variable.
After estimating all bits,
the values of the dummy variables are identified by solving the linear equations formulated by gathering information from subsequently decoded frozen bits.

In this paper, we aim to settle the suboptimality of the SC decoder for finite-length polar codes.
Specifically, our goal is to immediately incorporate all future bits into decoding each bit, in contrast to the existing methods.
We define \textit{a future constraint (FC)} as a bit behind the target decoding bit, whose value is deterministic given the estimates for the past bits.
In general concatenated polar coding schemes,
there are two categories of FCs: frozen bits and parity bits.
Polar codes inevitably involve frozen bits whose values are generally fixed to zero.
Parity bits (also called dynamic frozen symbols \cite{Trifonov2016}) are typically generated by concatenating an outer code such as a cyclic redundancy check (CRC) code \cite{Niu2012}  and a parity-check (PC) code \cite{Wang2016}.
A recursive construction method for precoded polar codes was also proposed in \cite{Miloslavskaya2021} as a way of producing parity bits for SCL decoding.
They are exploited to increase the minimum distance and identify invalid candidate codewords during and after decoding.
Recently, polarization-adjusted-convolutional (PAC) codes were proposed by Ar{\i}kan \cite{Arikan2019} as a new concatenated coding scheme,
where the constrained bits after the first information bit are all to be generated as parity bits by convolutional coding.

Using the fact that the values of FCs are deterministic given the previous estimates,
we first show the advantages of exploiting them by presenting a bitwise maximum \textit{a posteriori} (MAP) SC decoding method.
Then, two elementary algorithms are proposed to immediately incorporate the FCs in sequential SC decoding for practical applications as follows:
\begin{itemize}
\item
In sequential decoding of a certain target bit,
we suspend its estimation until the next information bit.
Instead, two hypotheses for bit values 0 and 1 are established,
and the FCs in between are completed using the previous bits of each hypothesis.
The proposed decoder computes the likelihood and/or performs a validation check for the two hypotheses,
and then, chooses one of them depending on the decoding results.
In this way, the FCs ahead of the next information bit are to be directly captured.
We refer to this approach as \textit{an SC-check (SCC) algorithm}.
\item
To further cover the FCs behind the next information bit,
we derive conversion rules to find the constraints on the encoder output, equivalent to the FCs given on the encoder input.
The constraints appearing at the encoder output (i.e., decoder input) can now be incorporated into decoding of each bit.
In detail, a polar code graph is modified by connecting new vertices corresponding to the converted constraints,
and belief propagation (BP) decoding is performed over the graph.
Combining this idea with SCC decoding, we propose \textit{a BP-SCC algorithm}.
\end{itemize}

BP decoding is combined with SC decoding in existing hybrid BP-SC schemes \cite{Yuan2014,Zhou2018}.
Yuan and Parhi \cite{Yuan2014} proposed a hybrid architecture that uses SC and BP decoding individually,
with BP employed to enable early termination and potentially enhance the input for subsequent SC decoding.
Similarly, Zhou \textit{et. al.} \cite{Zhou2018} presented a single decoder architecture that mixes SC and BP algorithms to reduce the latency of serial operation.
Although the proposed BP-SCC algorithm may be regarded as a hybrid BP-SC scheme,
the aim of BP in this algorithm is to incorporate FCs into the decoding process, unlike the conventional hybrid BP-SC schemes.


The proposed SCC and BP-SCC decoding algorithms are closely related to the look-ahead algorithms in that decoding is performed by including future bits.
The most related work is the SC look-ahead (SC-LA) decoding algorithm with constant delay $D$ proposed in \cite{Vajha2019}.
In the SC-LA algorithm, $D$ bits are estimated at once after exploring all $2^D$ paths in the decision tree.
Some future frozen bits may naturally be involved in decoding a target information bit depending on the code structure due to the delay.
Also, fat-tree decoders (FTDs) were proposed in \cite{Schurch2017} to generalize SC decoding with look-ahead.
The FTDs perform message-passing over the cyclic-free factor graph derived from the generator matrix in order to compute posterior probabilities more accurately.
In simplified SC-based decoding schemes including \cite{Ardakani2019}, some future bits might be incorporated in simultaneous operation for fast processing.
Recently, a hypothesis-testing-based hard decision (HTHD) algorithm was proposed in \cite{Sun2020} to simplify the decoding operation of unstructured nodes, possibly including future bits.

\if\doccolumn2
\begin{figure}[t!]
    \centering
    \subfloat[Conventional SC decoding]
        {\includegraphics[height=6.2cm]{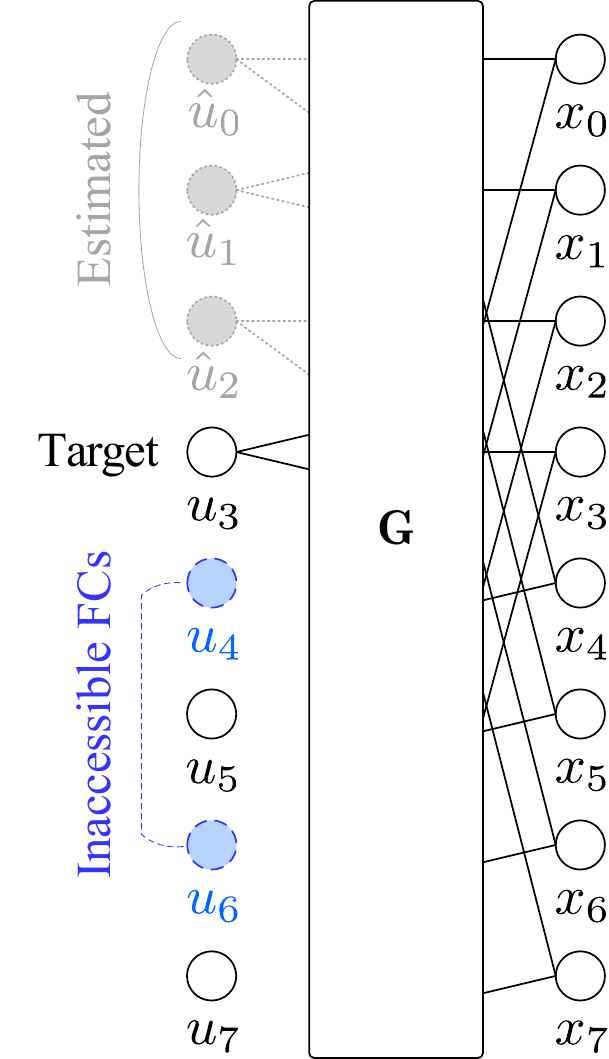}}
        \label{fig:concept1}\hfill
    \subfloat[Proposed FC-aided decoding]
        {\includegraphics[height=6.2cm]{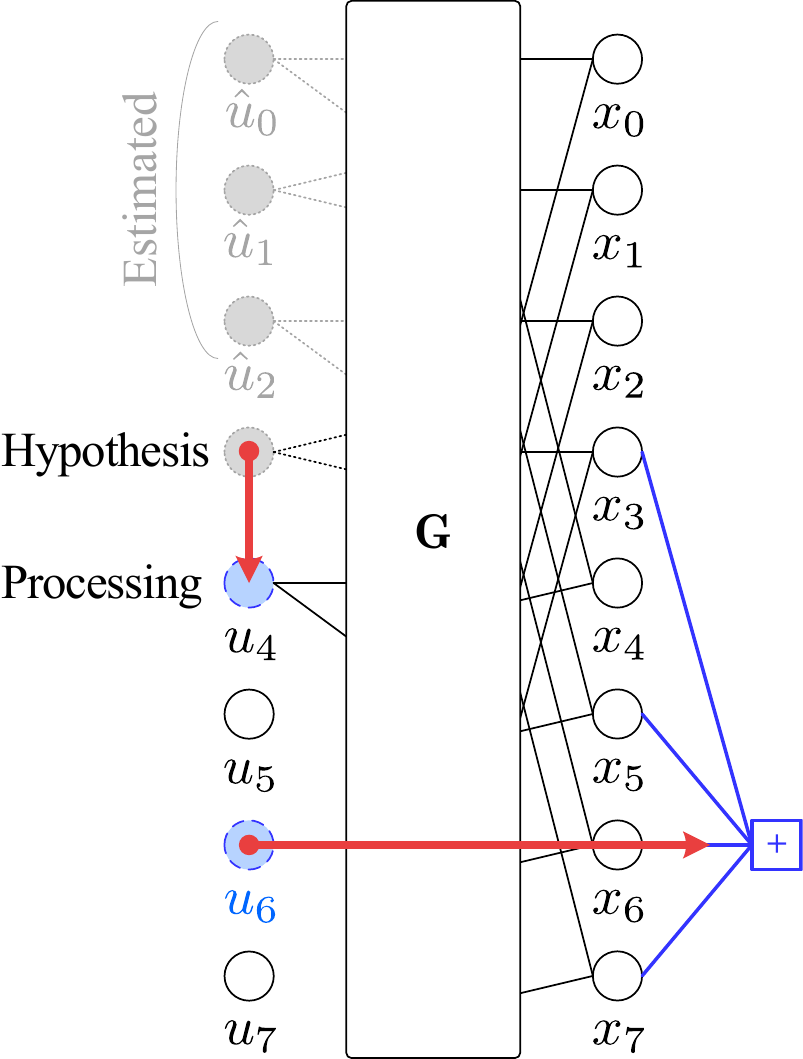}}
        \label{fig:concept2}\\
    \caption{Concept of the proposed FC-aided SC decoding algorithm.
      In conventional SC-decoding algorithms, FCs are not accessible, so they are just treated as RVs.
      In the proposed decoding algorithm for a target information bit,
      the FCs located before and after the next information bit are incorporated by SCC decoding and FC conversion, respectively.}
    \label{fig:concept}
\end{figure}
\else
\begin{figure}[t]
    \centering
    \subfloat[Conventional SC decoding]
        {\includegraphics[height=7cm]{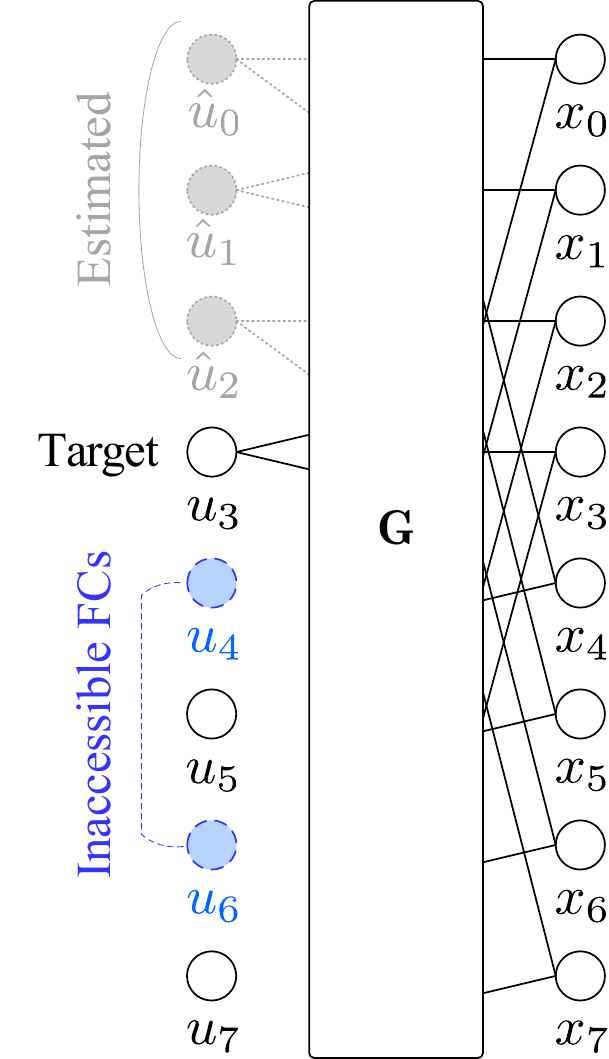}}
        \label{fig:concept1}\qquad\qquad
    \subfloat[Proposed FC-aided decoding]
        {\includegraphics[height=7cm]{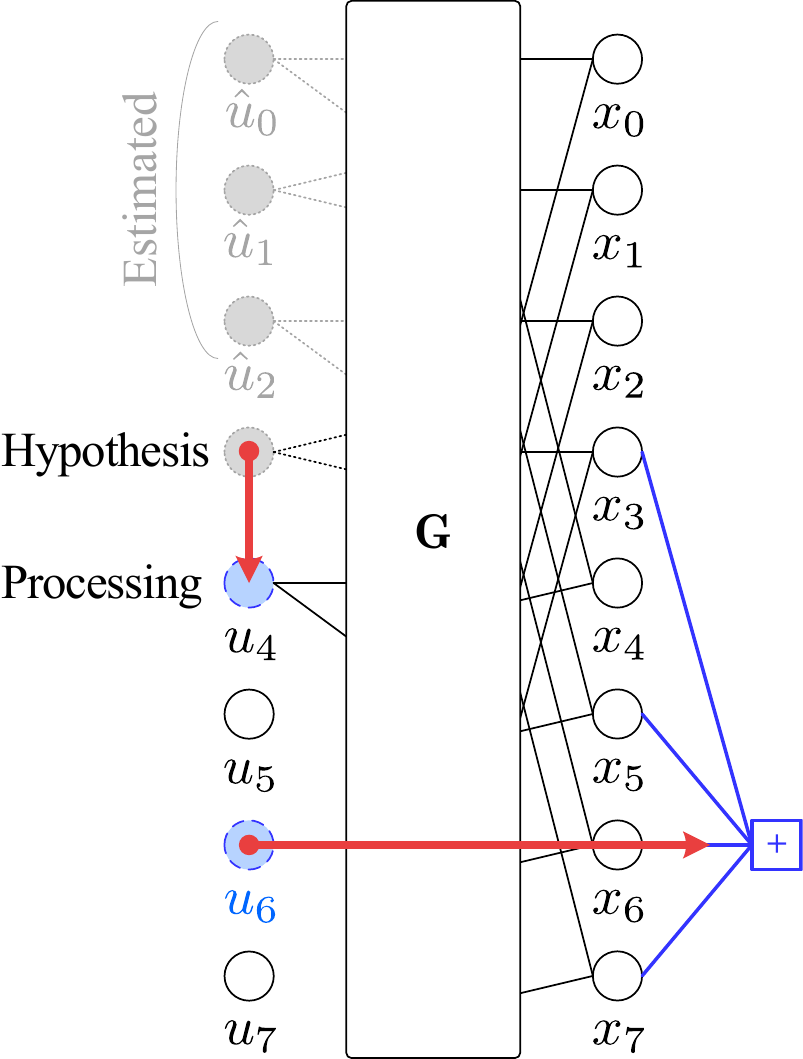}}
        \label{fig:concept2}\\
    \caption{Concept of the proposed FC-aided SC decoding algorithm.
      In conventional SC-decoding algorithms, FCs are not accessible, so they are just treated as RVs.
      In the proposed decoding algorithm for a target information bit,
      the FCs located before and after the next information bit are incorporated by SCC decoding and FC conversion, respectively.}
  \label{fig:concept}
\end{figure}
\fi

The most distinct feature of the proposed BP-SCC decoding algorithm is that it incorporates all future frozen bits and parity bits into the estimation of a target bit.
As depicted in Fig.~\ref{fig:concept},
the FCs close to the target bit are taken directly by the SCC algorithm,
while the others are incorporated by converting them into equivalent constraints on the decoder input so that they are leveraged by the BP operations.
For general concatenated coding schemes considered in this paper,
the FCs induced by parity bits play a significant role in improving their performance.

Based on the proposed BP-SCC algorithm,
an improved tree search technique is further developed.
Sequential estimation for polar codes is essentially accompanied by tree search,
and the SC and SCL decoding algorithms exploit a depth-first search (DFS) method and a breath-first search (BFS) method with limited branching (also called beam search), respectively \cite{Arikan2019}.
Parity-check equations involving FCs are adaptively changed in response to the decisions made during sequential estimation and the assignment of values is accordingly determined.
For this reason, sequential decoding with FCs may be viewed as constraint satisfaction problems (CSPs) \cite{Meseguer1989}
and the dynamic CPS framework \cite{Mittal1990} is considered in our work.
Inspired by a conflict-directed backjumping (CBJ) method \cite{Prosser1993},
a stack-based backjumping (SBJ) technique is proposed to efficiently explore the tree while avoiding invalid partial solutions in terms of the full space of FCs.

Targeting the binary erasure channel (BEC),
we present a specific decoding algorithm using the proposed FC-aided techniques.
The performance and computational complexity of the algorithm are then evaluated by considering the 3GPP NR polar coding system specified in \cite{TS38212},
where the polar codes are concatenated by the 11-bit CRC code.
In the proposed decoders, each of the frozen and parity bits in these concatenated polar codes is captured as an FC.
Numerical results show that FCs play a role in improving the performance of sequential decoding.
In particular, the BP-SCC algorithm accompanied by the SBJ tree search technique achieves performance close to the dependence testing (DT) bound for the BEC, an approximation of the best achievable performance by any code of given length and dimension \cite{Polyanskiy2010}.

\section{Preliminaries}\label{sec:pre}
\subsection{Notation}\label{sec:notation}
Throughout this paper, we use the following general rules to represent mathematical symbols.
\begin{itemize}
  \item[-] We use zero-based numbering, where index 0 is assigned to the initial element of sets, vectors, and matrices.
  \item[-] Calligraphic letters (e.g., $\mathcal{A}$) are used to denote sets.
  \item[-] The standard notation $\mathcal{A}^\mathsf{c}$ is used to denote the complementary set of $\mathcal{A}$.
  \item[-] Given $\mathcal{A}$ and $\mathcal{B}$, we write $\mathcal{A}\backslash \mathcal{B}$ to denote the relative complement of $\mathcal{A}$ with respect to $\mathcal{B}$.
  \item[-] Given a set $\mathcal{A}$ and any number $b$, we write $b+\mathcal{A}$ to denote $\{b+a\mmid a\in\mathcal{A}\}$.
  \item[-] We use the conventions $\mathbb{N}$, $\mathbb{Z}$, $\mathbb{R}$, and $\mathbb{C}$ to denote the sets of natural numbers, integers, real numbers, and complex numbers, respectively.
  \item[-] We write $\mathbb{F}_2$ to denote the binary field.
  \item[-] For $b\in\mathbb{F}_2$, let $\check{b}$ indicate its one's complement, that is, $\check{b}=b+1$ over $\mathbb{F}_2$.
  \item[-] For a non-negative integer $i$, denote $\mathbb{Z}_i$ as the set of consecutive integers from $0$ to $i-1$.
  \item[-] For two integers $i,j$ such that $i<j$, a set of consecutive integers from $i$ to $j$ is denoted by $\{i:j\}$.
      When this symbol is used in a subscript, parentheses may be omitted (i.e., $i:j$) for a concise expression.
  \item[-] Boldface lowercase letters (e.g., $\mathbf{a}$) and boldface uppercase letters (e.g., $\mathbf{A}$) are used to denote vectors and matrices, respectively.
  \item[-] Given a vector $\mathbf{a}$ and a set $\mathcal{A}$,
    let $\mathbf{a}_{\mathcal{A}}$ denote the subvector $(a_i:i\in\mathcal{A})$.
  \item[-] Given a vector $\mathbf{a}$ and two non-negative integers $i, j$ with $j>i$,
    let $\mathbf{a}_i^j\triangleq(a_i,a_{i+1},\ldots,a_j)$.
  \item[-] Given a matrix $\mathbf{A}$ and two non-negative integers $i, j$,
    let $\mathbf{A}_{i,j}$ denote the entry in row $i$ and column $j$ of $\mathbf{A}$.
  \item[-] Given a matrix $\mathbf{A}$ and two non-negative integer sets $\mathcal{R}, \mathcal{C}$,
    let $\mathbf{A}_{\mathcal{R}, \mathcal{C}}$ be the submatrix of $\mathbf{A}$ composed of the rows and columns whose indices are in $\mathcal{R}$ and $\mathcal{C}$, respectively.
  \item[-] For simple representation concerning matrix slicing, we use $\ast$ in shorthand to indicate all entries in a row or column dimension.
    To be specific, for an $n\times m$ matrix $\mathbf{A}$,
    let $\mathbf{A}_{\mathcal{R},\ast}\triangleq \mathbf{A}_{\mathcal{R},0:m-1}$,
    and $\mathbf{A}_{\ast,\mathcal{C}}\triangleq \mathbf{A}_{0:n-1,\mathcal{C}}$.
\end{itemize}

\subsection{Concatenated Polar Codes}

\if\doccolumn2
\begin{figure}[t]
    \centering
    \centering
    \subfloat[General concatenated coding scheme]
        {\includegraphics[width=\linewidth]{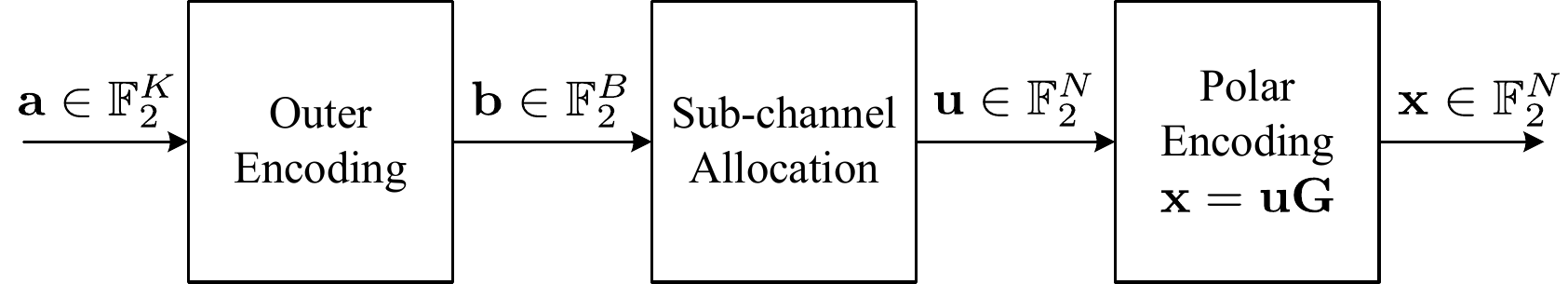}}
        \label{fig:blockA}\\ \bigskip
    \subfloat[Equivalent model with rate-profiling and precoding]
        {\includegraphics[width=\linewidth]{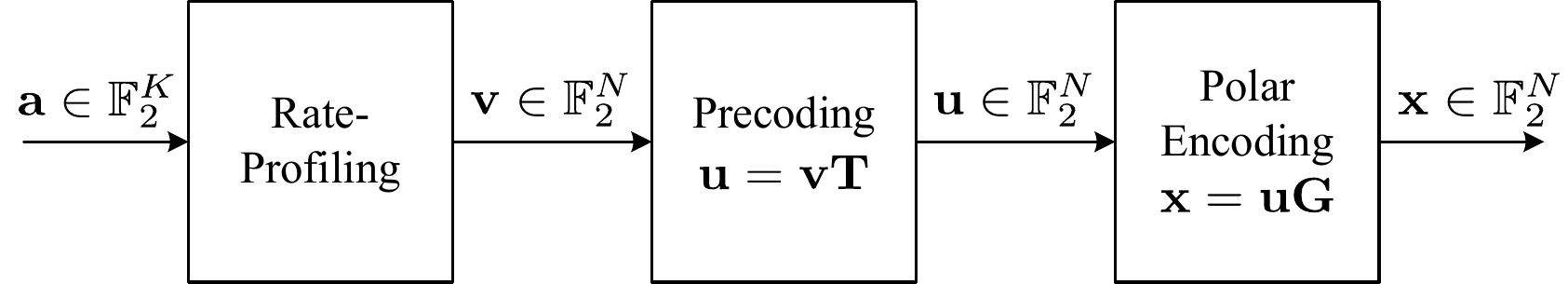}}
        \label{fig:blockB}\\
    \caption{Block diagrams for concatenated polar coding.}
  \label{fig:block}
\end{figure}
\else
\begin{figure}[t]
    \centering
    \subfloat[General concatenated coding scheme]
        {\includegraphics[width=\subfigwidth]{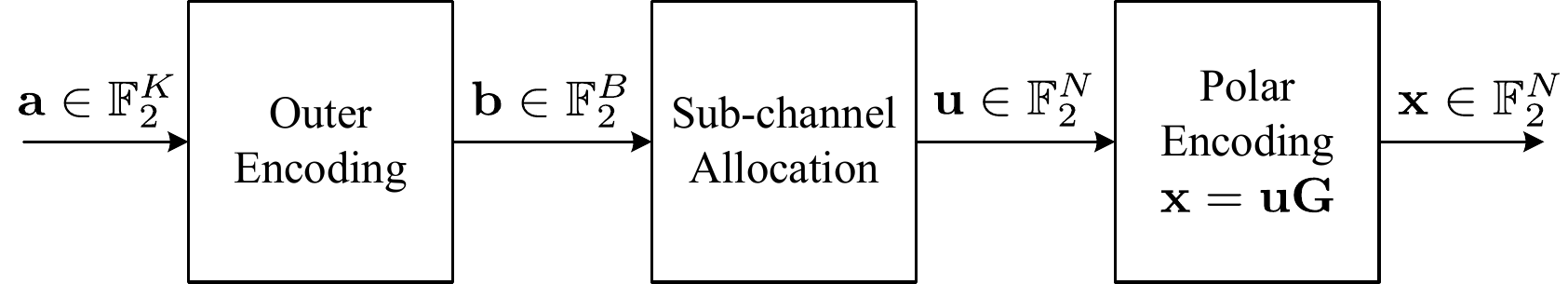}}
        \label{fig:blockA}\hfill
    \subfloat[Equivalent model with rate-profiling and precoding]
        {\includegraphics[width=\subfigwidth]{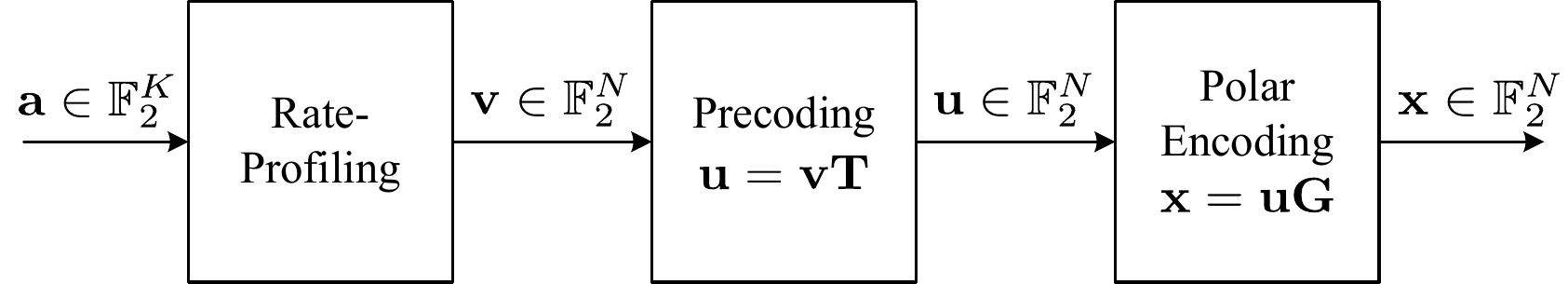}}
        \label{fig:blockB}\\
    \caption{Block diagrams for concatenated polar coding.}
  \label{fig:block}
\end{figure}
\fi

We briefly introduce an $(N,K)$ concatenated polar code,
where $N=2^n$ ($n\in\mathbb{N}$) is the code length and $K$ is the code dimension.
Fig.~\ref{fig:block} (a) shows the block diagram of the coding scheme.
Let $\mathbf{a}\in\mathbb{F}_2^{K}$ denote an input information vector.
It is encoded by an outer code to generate $\mathbf{b}\in\mathbb{F}_2^B$.
The outer code is systematic in the sense that $\mathbf{b}$ consists of $K$ information bits of $\mathbf{a}$ and $B-K$ parity bits generated by the outer code.
Then, the bits in $\mathbf{b}$ are mapped to a polar encoder input vector $\mathbf{u}\in\mathbb{F}_2^{N}$ under the operation of sub-channel allocation.
We write $\mathcal{A}, \mathcal{P}, \mathcal{F} \subset \mathbb{Z}_N$ to denote the index sets of information bits, parity bits, and frozen bits to be assigned to $\mathbf{u}$, respectively.
Finally, a polar codeword $\mathbf{x}\in\mathbb{F}_2^N$ is obtained by the linear transformation
\begin{equation}\label{eq:uG}
    \mathbf{x}=\mathbf{u}\mathbf{G},
\end{equation}
where $\mathbf{G}\in\mathbb{F}_2^{N\times N}$ is the generator matrix of a length-$N$ polar code.
The generator matrix is given by $\mathbf{G}=\mathbf{F}^{\otimes n}$,
where $\mathbf{F}^{\otimes n}$ is the $n$-th Kronecker power of the $2\times 2$ polarization kernel $\mathbf{F}=\left[\begin{smallmatrix} 1 & 0 \\ 1 & 1 \end{smallmatrix}\right]$.

In the concatenated polar coding scheme,
the outer encoding and sub-channel allocation are designed together
so that any parity bit $u_i$ ($i\in\mathcal{P}$) is generated from the information bits whose indices are smaller than $i$.
Specifically, $u_i$ is a linear combination of bits in $\mathbf{u}_0^{i-1}=(u_0,u_1,\ldots,u_{i-1})$.
This implies that the parity bits are \textit{causally} generated in terms of $\mathbf{u}$,
and at the receiver, their values can be directly determined from the bits estimated earlier in sequential decoding.

The combination of outer encoding and sub-channel allocation
can be represented by a different configuration as depicted in Fig.~\ref{fig:block} (b).
This just rearranges the order of operations to obtain an equivalent model.
Note that it is useful in describing general concatenated polar coding schemes with parity bits \cite{Arikan2019,Coskun2020}.

The input information vector $\mathbf{a}$ is first mapped to a precoder input vector $\mathbf{v}\in\mathbb{F}_2^{N}$ using $\mathcal{A}$ such that $\mathbf{v}_{\mathcal{A}}=\mathbf{a}$ and $\mathbf{v}_{\mathcal{A}^{\mathsf{c}}}=\mathbf{0}$.
This bit allocation procedure is called \textit{rate-profiling}.
Then, $\mathbf{v}$ is transformed into $\mathbf{u}$ by the linear transformation
\begin{equation}\label{eq:vT}
\mathbf{u} = \mathbf{v}\mathbf{T},
\end{equation}
where $\mathbf{T}\in\mathbb{F}_2^{N\times N}$ is the precoding matrix reflecting the effect of the outer code.
Each column of $\mathbf{T}$ is determined as follows:
\begin{itemize}
\item[-] For $i\in\mathcal{A}$, $\mathbf{T}_{i,i}=1$ and $\mathbf{T}_{k,i}=0$ for $k\neq i$.
\item[-] For $i\in\mathcal{F}$, $\mathbf{T}_{\ast,i}=\mathbf{0}$.
\item[-] For $i\in\mathcal{P}$, $\mathbf{T}_{0:i-1,i}$ is given such that $u_i=\mathbf{u}_0^{i-1} \mathbf{T}_{0:i-1,i}$ and $\mathbf{T}_{i:N-1,i}=\mathbf{0}$.
\end{itemize}
It is clear that $\mathbf{T}$ is an upper triangular matrix because of the causal generation of parity bits.
A lower triangular generator matrix $\mathbf{G}$ is then multiplied to generate a polar codeword $\mathbf{x}$.
The end-to-end concatenated polar coding is represented by $\mathbf{x}=\mathbf{v}\mathbf{T}\mathbf{G}$,
so that $\mathbf{TG}$ is regarded as the generator matrix for the entire encoding procedure.

Throughout the paper, we use the following polar code as a toy example.
\begin{exam}\label{ex:toy}
An $(8,3)$ concatenated polar code is constructed with $\mathcal{A}=\{3, 5, 7\}$, $\mathcal{F}=\{0, 1, 2, 4\}$,  $\mathcal{P}=\{6\}$,
where the parity bit $u_6$ is generated by $u_6=u_3 + u_5$.
From the definition of $\mathbf{T}$, we have
\begin{equation}\nonumber
\mathbf{T}=
{
\small
\setlength\arraycolsep{\matsep}\def\arraystretch{\matstretch}
\begin{bmatrix}
0 & 0 & 0 & 0 & 0 & 0 & 0 & 0 \\
0 & 0 & 0 & 0 & 0 & 0 & 0 & 0 \\
0 & 0 & 0 & 0 & 0 & 0 & 0 & 0 \\
0 & 0 & 0 & 1 & 0 & 0 & 1 & 0 \\
0 & 0 & 0 & 0 & 0 & 0 & 0 & 0 \\
0 & 0 & 0 & 0 & 0 & 1 & 1 & 0 \\
0 & 0 & 0 & 0 & 0 & 0 & 0 & 0 \\
0 & 0 & 0 & 0 & 0 & 0 & 0 & 1 \\
\end{bmatrix}
}
~\text{ and }~
\mathbf{TG}=
{
\small
\setlength\arraycolsep{\matsep}\def\arraystretch{\matstretch}
\begin{bmatrix}
0 & 0 & 0 & 0 & 0 & 0 & 0 & 0 \\
0 & 0 & 0 & 0 & 0 & 0 & 0 & 0 \\
0 & 0 & 0 & 0 & 0 & 0 & 0 & 0 \\
0 & 1 & 0 & 1 & 1 & 0 & 1 & 0 \\
0 & 0 & 0 & 0 & 0 & 0 & 0 & 0 \\
0 & 1 & 1 & 0 & 0 & 1 & 1 & 0 \\
0 & 0 & 0 & 0 & 0 & 0 & 0 & 0 \\
1 & 1 & 1 & 1 & 1 & 1 & 1 & 1 \\
\end{bmatrix}
}.
\end{equation}
\end{exam}


\subsection{SC Decoding}

Let $W:\mathcal{X}\to\mathcal{Y}$ be a B-DMC, where $\mathcal{X}$ ($=\mathbb{F}_2$) and $\mathcal{Y}$ are input and output alphabets, respectively.
A polar codeword $\mathbf{x}\in\mathcal{X}^N$ is transmitted through $N$ copies of the B-DMC $W$,
and the vector channel is denoted by $W^N:\mathcal{X}^N\to\mathcal{Y}^N$.
At the receiver, a channel output $\mathbf{y}\in\mathcal{Y}^N$ is observed with transition probability
\begin{equation}
W^N(\mathbf{y}\mmid \mathbf{x}) = \prod_{j=0}^{N-1} W(y_j\mmid x_j).
\end{equation}

The polar encoder combines a bundle of B-DMCs from $\mathbf{u}$ to $\mathbf{y}$ to produce a vector channel $W_N:\mathcal{X}^N\to\mathcal{Y}^N$ in a recursive manner.
The SC decoder splits $W_N$ back into a set of $N$ sub-channels $W_N^{(i)}:\mathcal{X}\to\mathcal{X}^{i-1}\times \mathcal{Y}^N$, $i\in\mathbb{Z}_N$, whose transition probability is defined by
\begin{equation}\label{eq:org_subchannel}
W_N^{(i)}\left(\mathbf{y},\mathbf{u}_{0}^{i-1} \mmid u_i\right) \triangleq
\frac{1}{2^{N-1}}
\sum\nolimits_{\mathbf{u}_{i+1}^{N-1}\in\mathcal{X}^{N-i-1}} W_N \left(\mathbf{y} \mmid \mathbf{u} \right),
\end{equation}
where
$W_N(\mathbf{y}\mmid\mathbf{u}) = W^N(\mathbf{y}\mmid \mathbf{x})$
because of the one-to-one correspondence between $\mathbf{u}$ and $\mathbf{x}=\mathbf{uG}$.
Here,
$\mathbf{u}_0^{i-1}$ is given by earlier operations of the SC decoder,
and the sum over $\mathcal{X}^{N-1-i}$ is interpreted as marginalizing out all the future bits to be decoded later.
Ar{\i}kan demonstrated in \cite{Arikan2009} that the likelihood in \eqref{eq:org_subchannel} can be efficiently calculated using recursive formulas
with a computational complexity of $O(N\log_2 N)$.
Furthermore, the recursive decoding operation makes performance analysis tractable
and plays a key role in proving the achievability of the symmetric capacity.

\section{Problem Formulation} \label{sec:problem}

\subsection{Suboptimality of SC Decoding}

As clearly noted in \cite{Arikan2009},
SC decoding is suboptimal because it treats all the future frozen bits as RVs rather than known bits.
Likewise, the parity bits causally generated in a concatenated coding scheme make SC decoding suboptimal when they are dealt with as RVs.
The suboptimality comes from the relaxation of
the marginalization space $\big\{\mathbf{u}_{i+1}^{N-1}\in\mathcal{X}^{N-i-1}\big\}$ in \eqref{eq:org_subchannel}.
Even if the relaxation is sufficient in proving the channel polarization,
it would be better to refine the marginalization space to achieve better performance for finite-length polar codes.

\begin{figure}[t]
	\centering
	\includegraphics[width=\figwidth]{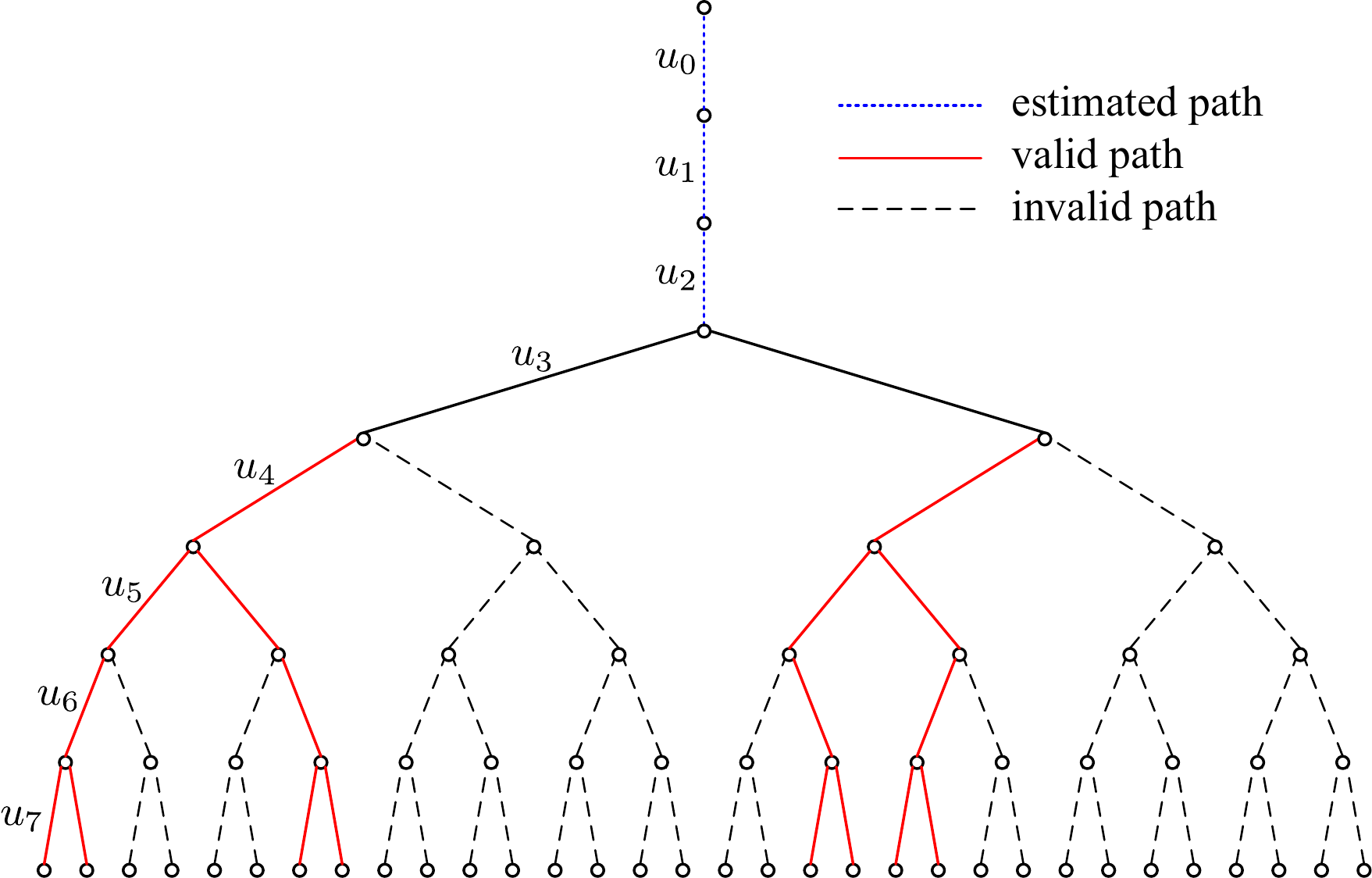}
	\caption{
    A binary decision tree constructed for decoding $u_3$ in Example~\ref{ex:toy}.
    Two FCs are fixed as $u_4=0$ and $u_6=u_3 + u_5$,
    and invalid paths are depicted as gray dashed lines.}
	\label{fig:tree}
\end{figure}

A binary decision tree for decoding the polar code given in Example~\ref{ex:toy} is shown in Fig.~\ref{fig:tree}.
In the estimation of $u_3$,
the conventional SC decoder marginalizes out the future bits in $\mathbf{u}_4^7$ according to \eqref{eq:org_subchannel}
by treating them as RVs.
Accordingly, all the sixteen paths for $\mathbf{u}_4^7\in\mathbb{F}_2^4$ are considered as possible codewords in the decision tree.
However, two future bits $u_4$ and $u_6$ are constrained as a frozen bit and a parity bit, respectively.
Taking these constraints into account, the number of valid paths is only four for each value of $u_3$.

We define future constraints as the future bits whose values are fixed or constrained. 
Specifically, we have the following definition.
\begin{define}[\textit{Future constraint}]
    For two integers $i,k\in\mathbb{Z}_N$ with $k\geq i$,
    bit $u_k$ is called a future constraint (FC) of $u_i$ if $k\in\mathcal{A}^{\mathsf{c}}$.
    The index set of future constraints for $u_i$ is denoted by
    \begin{equation}
    \mathcal{L}_{i} \triangleq \left\{k \geq i\mmid k\in \mathcal{A}^{\mathsf{c}}\right\}.
    \end{equation}
\end{define}
\if\doccolumn1
\noindent\textbf{Remark:} Clearly, $i\in\mathcal{L}_i$ if $u_i$ is either a frozen bit or a parity bit.
\else
\noindent\textit{Remark:} Clearly, $i\in\mathcal{L}_i$ if $u_i$ is either a frozen bit or a parity bit.
\fi

\subsection{Parity-Check Matrix on Encoder Input}

In order to specify the code space for precoding,
we define a parity-check matrix\footnote{For simple and concise presentation throughout the paper, we define the parity-check matrix without taking its transpose, instead of the standard notion of the parity-check matrix in coding theory.} $\mathbf{H}\in\mathbb{F}_2^{N\times N}$ such that
$\mathbf{u}\mathbf{H}=\mathbf{0}$ for any precoded word $\mathbf{u}$.
Based on \eqref{eq:vT}, we have
\begin{equation}
\mathbf{0}=\mathbf{u}\mathbf{H}=\mathbf{v}\mathbf{T}\mathbf{H}
\end{equation}
for any valid precoding input vector $\mathbf{v}$.
Therefore, the parity-check matrix $\mathbf{H}$ is constructed so that $\mathbf{T}\mathbf{H}=\mathbf{0}$ is satisfied.
The $i$-th column of $\mathbf{H}$ is chosen as follows:
\begin{itemize}
\item[-] For $i\in\mathcal{A}$, $\mathbf{H}_{\ast,i}=\mathbf{0}$
\item[-] For $i\in\mathcal{F}$,
$\mathbf{H}_{i,i}=1$ and $\mathbf{H}_{k,i}=0$ for $k \neq i$.
\item[-] For $i\in\mathcal{P}$, $\mathbf{H}_{0:i-1,i}=\mathbf{T}_{0:i-1,i}$, $\mathbf{H}_{i,i}=1$, $\mathbf{H}_{i+1:N-1,i}=\mathbf{0}$.
\end{itemize}
As an example, the parity-check matrix $\mathbf{H}$ for the code in Example~\ref{ex:toy} is given by
\begin{equation}\nonumber
\mathbf{H}=
{
\small
\setlength\arraycolsep{\matsep}\def\arraystretch{\matstretch}
\begin{bmatrix}
1 & 0 & 0 & 0 & 0 & 0 & 0 & 0 \\
0 & 1 & 0 & 0 & 0 & 0 & 0 & 0 \\
0 & 0 & 1 & 0 & 0 & 0 & 0 & 0 \\
0 & 0 & 0 & 0 & 0 & 0 & 1 & 0 \\
0 & 0 & 0 & 0 & 1 & 0 & 0 & 0 \\
0 & 0 & 0 & 0 & 0 & 0 & 1 & 0 \\
0 & 0 & 0 & 0 & 0 & 0 & 1 & 0 \\
0 & 0 & 0 & 0 & 0 & 0 & 0 & 0 \\
\end{bmatrix}.
}
\end{equation}
Note that $\mathbf{H}$ is also an upper triangular matrix, like the precoding matrix $\mathbf{T}$.
Since the columns of $\mathbf{H}$ corresponding to the information bits are zero vectors,
$\mathbf{u}\mathbf{H}=\mathbf{0}$ can be simplified to
\begin{equation}
\mathbf{u} \mathbf{H}_{\ast,\mathcal{A}^{\mathsf{c}}} = \mathbf{0}.
\end{equation}
Letting $\mathbf{H}'\triangleq \mathbf{H}_{\ast,\mathcal{A}^{\mathsf{c}}}$,
we consider the simplified form $\mathbf{u}\mathbf{H}'=\mathbf{0}$ henceforth.

\subsection{Bitwise-MAP-SC Decoding}

While maintaining the principle of SC decoding,
the bitwise-MAP estimation is made by modifying the marginalization space in \eqref{eq:org_subchannel}.
We refer to such decoding as \textit{bitwise-MAP-SC decoding}.
Given $\mathbf{u}_0^{i}$,
we write $\mathbf{u}_{i+1}^{N-1}:\mathbf{u}\mathbf{H}'=\mathbf{0}$ to denote the set of subvectors satisfying the given parity-check equation, that is,
\begin{equation}
\mathbf{u}_{i+1}^{N-1}\in\left\{\mathbf{a}\in\mathbb{F}_2^{N-i-1} \mmid \left(\mathbf{u}_0^{i}, \mathbf{a}\right) \mathbf{H}'=\mathbf{0} \right\},
\end{equation}
where $(\mathbf{u}_0^i,\mathbf{a})\in\mathbb{F}_2^N$ is the serial concatenation of $\mathbf{u}_0^i$ and $\mathbf{a}$.
The $i$-th modified sub-channel after channel splitting is then defined as
\begin{equation}\label{eq:new_subchannel}
\dot{W}_N^{(i)}\left(\mathbf{y},\mathbf{u}_{0}^{i-1} \mmid u_i\right)
\triangleq
\frac{1}{2^{K-1}}\sum\nolimits_{\mathbf{u}_{i+1}^{N-1}:\mathbf{u}\mathbf{H}'=\mathbf{0}} W_N(\mathbf{y} \mmid \mathbf{u}), \\
\end{equation}
where the marginalization space is refined to have valid $\mathbf{u}_{i+1}^{N-1}$ in terms of $\mathbf{u}\mathbf{H}'=\mathbf{0}$.

Using \eqref{eq:new_subchannel}, the bitwise-MAP estimation is achieved in sequential decoding for polar codes.
\begin{thm}\label{thm:map}
Given $\mathbf{y}$ and $\mathbf{u}_0^{i-1}$,
the bitwise-MAP estimate for $u_i$ is given by
\begin{equation}
    \hat{u}_i^{\text{MAP}}\left(\mathbf{y},\mathbf{u}_0^{i-1}\right)
    =
    \argmax_{u_i\in\{0,1\}} \dot{W}_N^{(i)}\big(\mathbf{y},\mathbf{u}_0^{i-1} \mmid  u_i\big).
\end{equation}
\end{thm}
\begin{proof}
Let $p(u_i\mmid \mathbf{y},\mathbf{u}_0^{i-1})$ denote the \textit{a posteriori} probability of $u_i$ given $\mathbf{y}$ and $\mathbf{u}_0^{i-1}$.
The statement is proved by
\if\doccolumn2
\begin{equation}\nonumber
\begin{split}
\hat{u}_i^{\text{MAP}}(\mathbf{y},\mathbf{u}_0^{i-1})
&= \argmax_{u_i\in\{0,1\}} p\big(u_i \mmid \mathbf{y}, \mathbf{u}_0^{i-1}\big) \\
&= \argmax_{u_i\in\{0,1\}} \sum_{\mathbf{u}_{i+1}^{N-1}} p\big(u_i,\mathbf{u}_{i+1}^{N-1} \mmid \mathbf{y}, \mathbf{u}_0^{i-1}\big) \\
&= \argmax_{u_i\in\{0,1\}} \sum_{\mathbf{u}_{i+1}^{N-1}} \frac{p\big(\mathbf{y} \mmid \mathbf{u}\big) p\big(\mathbf{u}\big)} {p\big(\mathbf{y},\mathbf{u}_0^{i-1}\big)} \\
&= \argmax_{u_i\in\{0,1\}} \sum_{\mathbf{u}_{i+1}^{N-1}} W_N\big(\mathbf{y} \mmid \mathbf{u}\big) \cdot \mathbbm{1}(\mathbf{u}\mathbf{H}'=\mathbf{0}) \\
&= \argmax_{u_i\in\{0,1\}} \sum\nolimits_{\mathbf{u}_{i+1}^{N-1}:\mathbf{u}\mathbf{H}'=\mathbf{0}} W_N\big(\mathbf{y} \mmid \mathbf{u}\big),
\end{split}
\end{equation}
\else
\begin{equation}\nonumber
\begin{split}
\hat{u}_i^{\text{MAP}}(\mathbf{y},\mathbf{u}_0^{i-1})
&= \argmax_{u_i\in\{0,1\}} p\big(u_i \mmid \mathbf{y}, \mathbf{u}_0^{i-1}\big)
= \argmax_{u_i\in\{0,1\}} \sum_{\mathbf{u}_{i+1}^{N-1}} p\big(u_i,\mathbf{u}_{i+1}^{N-1} \mmid \mathbf{y}, \mathbf{u}_0^{i-1}\big) \\
&= \argmax_{u_i\in\{0,1\}} \sum_{\mathbf{u}_{i+1}^{N-1}} \frac{p\big(\mathbf{y} \mmid \mathbf{u}\big) p\big(\mathbf{u}\big)} {p\big(\mathbf{y},\mathbf{u}_0^{i-1}\big)}
= \argmax_{u_i\in\{0,1\}} \sum_{\mathbf{u}_{i+1}^{N-1}} W_N\big(\mathbf{y} \mmid \mathbf{u}\big) \cdot \mathbbm{1}(\mathbf{u}\mathbf{H}'=\mathbf{0}) \\
&= \argmax_{u_i\in\{0,1\}} \sum\nolimits_{\mathbf{u}_{i+1}^{N-1}:\mathbf{u}\mathbf{H}'=\mathbf{0}} W_N\big(\mathbf{y} \mmid \mathbf{u}\big),
\end{split}
\end{equation}
\fi
where $\mathbbm{1}(\cdot)$ is the indicator function that returns 1 if the given argument is true, and 0 otherwise.
Note that the fourth equality holds because $p(\mathbf{y},\mathbf{u}_0^{i-1})$ is constant with respect to both $u_i$ and $\mathbf{u}_i^{N-1}$.
\end{proof}

\subsection{Comparison and Remarks}

Given the code in Example~\ref{ex:toy},
the performance of decoding algorithms is evaluated over the binary-input additive white Gaussian noise channel (BI-AWGNC).
A polar codeword $\mathbf{x}$ is mapped to the symbol vector $\mathbf{s}\in\mathbb{R}^N$
such that $s_j=1-2x_j$ for $j\in\mathbb{Z}_N$.
The symbol vector $\mathbf{s}$ is transmitted through $N$ copies of the BI-AWGNC,
and $\mathbf{y}\in\mathbb{R}^N$ is received such that $y_j=s_j+z_j$ for $j\in\mathbb{Z}_N$, where $z_j\sim\mathcal{N}(0,\sigma^2)$
and $\sigma^2$ is the noise variance.
The channel transition probability is given by
\begin{equation}
    W(y_i \mmid x_i) = \frac{1}{\sqrt{2\pi\sigma^2}} \exp\left(-\frac{1}{2\sigma^2}\left(y_i- s_i\right)^2\right).
\end{equation}

The following three decoding algorithms are considered in the evaluation.
\vspace{0.5em}
\subsubsection{SC decoding}
The SC decoder makes an estimate of $u_i$ for $i\in\mathcal{A}$ by
\if\doccolumn2
\begin{equation}\label{eq:sc_est}
\begin{split}
  \hat{u}_i^{\text{SC}}\big(\mathbf{y},\mathbf{u}_0^{i-1}\big)
  &= \argmax_{u_i\in\{0,1\}} W_N^{(i)}\left(\mathbf{y},\hat{\mathbf{u}}_0^{i-1}\mmid u_i\right) \\
  &= \argmax_{u_i\in\{0,1\}} \sum\nolimits_{\mathbf{u}_{i+1}^{N-1}\in\mathcal{X}^{N-i-1}}
  \prod_{j=0}^{N-1} p(y_j\mmid x_j).\\
\end{split}
\end{equation}
\else
\begin{equation}\label{eq:sc_est}
  \hat{u}_i^{\text{SC}}\big(\mathbf{y},\mathbf{u}_0^{i-1}\big)
  = \argmax_{u_i\in\{0,1\}} W_N^{(i)}\left(\mathbf{y},\hat{\mathbf{u}}_0^{i-1}\mmid u_i\right)
  = \argmax_{u_i\in\{0,1\}} \sum\nolimits_{\mathbf{u}_{i+1}^{N-1}\in\mathcal{X}^{N-i-1}}
  \prod_{j=0}^{N-1} W(y_j\mmid x_j).
\end{equation}
\fi
Note that the SC decoder in \eqref{eq:sc_est} yields the same results as those of the decoders in \cite{Leroux2013,BalatsoukasStimming2015}, which are implemented using the log-likelihood ratio (LLR) in a recursive way.
\vspace{0.5em}
\subsubsection{Bitwise-MAP-SC decoding}
Like SC decoding, the bits in $\mathbf{u}$ are sequentially estimated in ascending order of the bit index.
By Theorem~\ref{thm:map}, the bitwise-MAP-SC estimate is made by
\if\doccolumn2
\begin{equation}\label{eq:mapsc_est}
\begin{split}
  \hat{u}_i^{\text{MAP}}\big(\mathbf{y},\mathbf{u}_0^{i-1}\big)
  &= \argmax_{u_i\in\{0,1\}} \dot{W}_N^{(i)}\left(\mathbf{y},\hat{\mathbf{u}}_0^{i-1}\mmid u_i\right) \\
  &= \argmax_{u_i\in\{0,1\}} \sum\nolimits_{\mathbf{u}_{i+1}^{N-1}:\mathbf{u}\mathbf{H}'=\mathbf{0}} \prod_{j=0}^{N-1} W(y_j\mmid x_j).
\end{split}
\end{equation}
\else
\begin{equation}\label{eq:mapsc_est}
  \hat{u}_i^{\text{MAP}}\big(\mathbf{y},\mathbf{u}_0^{i-1}\big)
  = \argmax_{u_i\in\{0,1\}} \dot{W}_N^{(i)}\left(\mathbf{y},\hat{\mathbf{u}}_0^{i-1}\mmid u_i\right)
  = \argmax_{u_i\in\{0,1\}} \sum\nolimits_{\mathbf{u}_{i+1}^{N-1}:\mathbf{u}\mathbf{H}'=\mathbf{0}} \prod_{j=0}^{N-1} W(y_j\mmid x_j).
\end{equation}
\fi
To compute \eqref{eq:mapsc_est}, the valid subvectors $\mathbf{u}_{i+1}^{N-1}$ need to be identified in terms of the given parity-check equation $\mathbf{u}\mathbf{H}'=\mathbf{0}$.
\vspace{0.5em}
\subsubsection{Blockwise-MAP decoding}
The optimal blockwise-MAP estimate is given by
\if\doccolumn2
\begin{equation}\label{eq:map_est}
\begin{split}
  \hat{\mathbf{u}}^{\text{MAP}}(\mathbf{y})
  &= \argmax_{\mathbf{u}} p\left(\mathbf{u}\mmid \mathbf{y}\right) \\
  &= \argmax_{\mathbf{u}} W_N\left(\mathbf{y}\mmid \mathbf{u}\right) p\left(\mathbf{u}\right)  \\
  &= \argmax_{\mathbf{u}} W_N\left(\mathbf{y}\mmid \mathbf{u}\right) \cdot \mathbbm{1}(\mathbf{u}\mathbf{H}'=\mathbf{0})  \\
  &= \argmax_{\mathbf{u}:\mathbf{u}\mathbf{H}'=\mathbf{0}} \prod_{j=0}^{N-1} W(y_j\mmid x_j).
\end{split}
\end{equation}
\else
\begin{equation}\label{eq:map_est}
\begin{split}
  \hat{\mathbf{u}}^{\text{MAP}}(\mathbf{y})
  &= \argmax_{\mathbf{u}} p\left(\mathbf{u}\mmid \mathbf{y}\right)
  = \argmax_{\mathbf{u}} W_N\left(\mathbf{y}\mmid \mathbf{u}\right) p\left(\mathbf{u}\right)  \\
  &= \argmax_{\mathbf{u}} W_N\left(\mathbf{y}\mmid \mathbf{u}\right) \cdot \mathbbm{1}(\mathbf{u}\mathbf{H}'=\mathbf{0})
  = \argmax_{\mathbf{u}:\mathbf{u}\mathbf{H}'=\mathbf{0}} \prod_{j=0}^{N-1} W(y_j\mmid x_j).
\end{split}
\end{equation}
\fi
As described in \cite[Chapter 2]{modern2008},
the estimate of $u_i$ in $\hat{\mathbf{u}}^{\text{MAP}}(\mathbf{y})$ can be written as
\begin{equation}\label{eq:map_est_bit}
\begin{split}
  \left(\hat{\mathbf{u}}^{\text{MAP}}(\mathbf{y})\right)_i
  &= \argmax_{u_i\in\{0,1\}} \max_{\sim u_i: \mathbf{u}\mathbf{H}'=\mathbf{0}} \prod_{j=0}^{N-1} W(y_j\mmid x_j),
\end{split}
\end{equation}
where $\sim{u_i}\triangleq\left\{\mathbf{u}_0^{i-1},\mathbf{u}_{i+1}^{N-1}\right\}$ with an abuse of notation.
\vspace{0.5em}

\begin{figure}[t!]
	\centering
	\includegraphics[width=\figwidth]{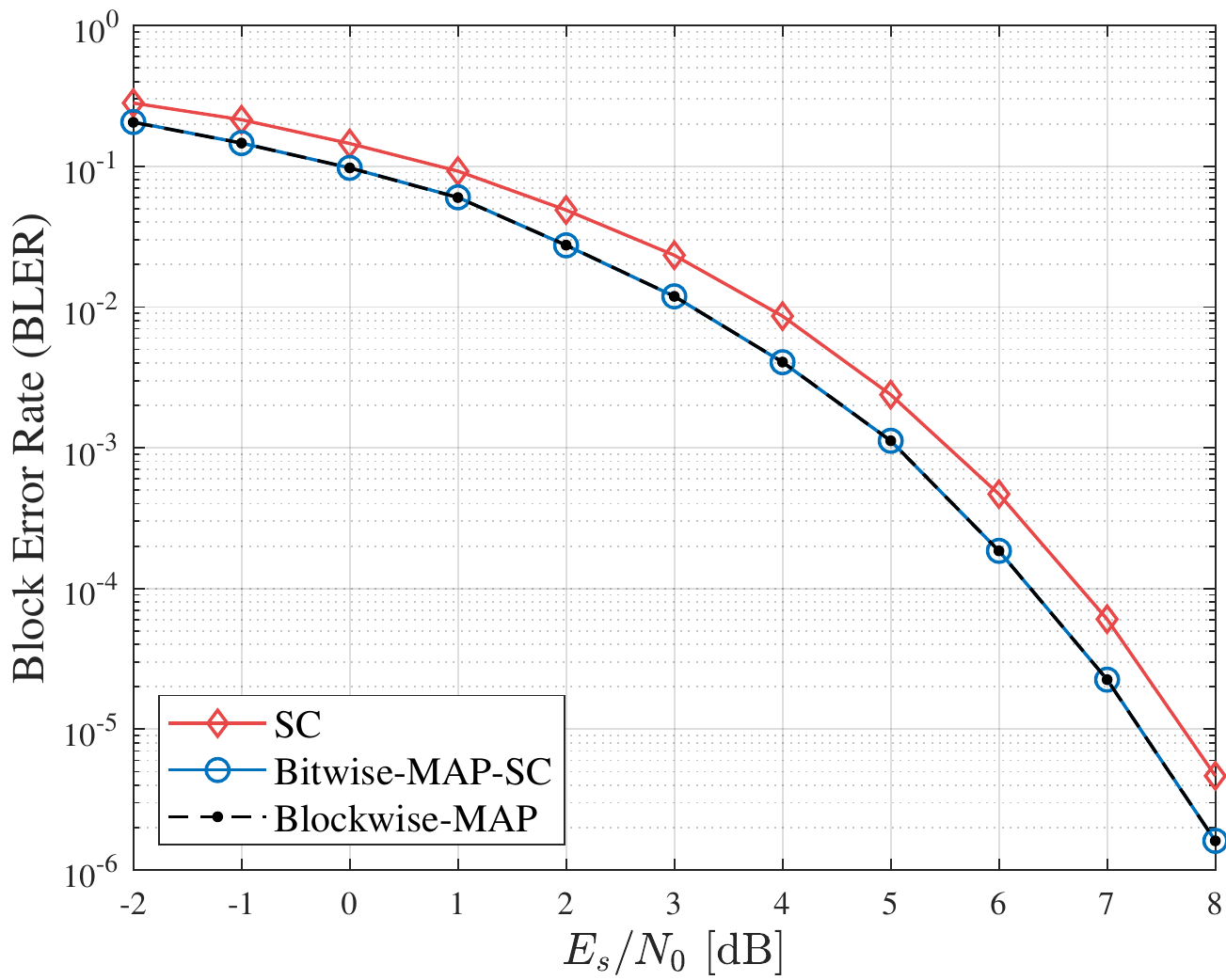}
	\caption{BI-AWGNC BLER performance of decoding schemes for the $(8,3)$ polar code in Example~\ref{ex:toy}.}
	\label{fig:toy_bler}
\end{figure}

Fig.~\ref{fig:toy_bler} shows the block error rate (BLER) curves of the above three decoding schemes for the code in Example~\ref{ex:toy}.
The BLER is evaluated over the BI-AWGNC with signal-to-noise ratio $E_s/N_0$, where $E_s$ is the coded bit energy and $N_0$ is the one-sided noise power spectral density.
Due to the small code dimension,
the estimates in \eqref{eq:mapsc_est} and \eqref{eq:map_est} are feasible,
even though they are NP-hard in general.
The bitwise-MAP-SC decoder performs better than the SC decoder thanks to refining the likelihood as in \eqref{eq:new_subchannel}.
For the given code, the bitwise-MAP-SC decoder behaves almost the same as the blockwise-MAP decoder due to the short block length,
although the estimate in \eqref{eq:mapsc_est} is different from that in \eqref{eq:map_est_bit} in principle.

The bitwise-MAP-SC estimation in \eqref{eq:mapsc_est} is, however, not generally feasible, because it is NP-hard.
Besides, sequentially performing the bitwise-MAP decoder on every information bit is more complicated than the optimal blockwise-MAP decoder.
Roughly analyzing, the blockwise-MAP and bitwise-MAP-SC estimations require approximately $2^K \times N$ and $\sum_{i\in\mathcal{A}} 2^{N-i} \times N$ multiplications, respectively, where the latter is generally larger than the former.
The development of practical decoding schemes that resemble the bitwise-MAP-SC decoder is hence a compelling issue.

From the literature, we see that conventional improved SC-based decoders such as the SCL, SCS, SCF, SC-Fano, and SCI decoders are also designed to tackle the suboptimality.
In particular, we have a remark on the SCL decoding algorithm \cite{Tal2015,BalatsoukasStimming2015},
the most popular practical scheme for polar codes.

\begin{remark}\label{remark:scl}
The SCL decoding algorithm is interpreted as performing post-compensation of the suboptimality brought by not reflecting FCs.
It is worth noting that the SCL decoder does not skip decoding of frozen and parity bits, even if their values are deterministic.
To be specific, the SCL decoder estimates information bits based on the original likelihood in \eqref{eq:org_subchannel}
and computes decoding metrics such as LLRs for frozen and parity bits to update the path metrics (PMs).
These decoding metrics penalize the invalid paths that erroneously survive due in part to the relaxation for the FCs.
Based on the behavior of SCL decoding,
it is possible to show that such processing for frozen and parity bits is not needed if FCs are immediately reflected.
\end{remark}

As shown in Remark~\ref{remark:scl},
the conventional improved SC-based decoders do not consider FCs in each bit decoding in order to keep the operation simple and recursive.
Instead, they compensate for this relaxation by directly processing frozen and parity bits afterward.
In contrast to these conventional methods, our goal is to develop practical techniques that immediately incorporate entire FCs while still maintaining the recursive formulas of SC decoding.
The efficient recursion of SC decoding is possible because all future bits are treated as pure noise, as noted in \cite{Arikan2016}.
Since this treatment is no longer valid, integrating FCs into the recursive decoding operation is not straightforward.

\section{FC-Aided Decoding}

In this section, we propose two elementary FC-aided decoding techniques.
An SCC decoding algorithm is first developed to directly incorporate FCs close to the target information bit.
To further leverage the other FCs that the SCC technique does not cover,
we find their equivalent constraints on the encoder output (i.e., the decoder input), which are available to the recursive decoding operation.
We then devise a BP-SCC decoding algorithm that runs message-passing over the converted equivalent constraints in order to enhance decoding symbols or metrics.
Based on the above two proposed decoding algorithms, a tree search technique is further designed to solve the CSP formulated by the FCs.

\subsection{SCC Decoding}

Consider decoding an information bit $u_i$ (i.e., $i\in\mathcal{A}$).
In conventional SC-based decoders,
a decoding metric for $u_i$ is computed and its estimate $\hat{u}_i$ is obtained accordingly.
On the other hand, in the proposed SCC decoding algorithm,
the estimation of $u_i$ is suspended until the next information bit,
and decoding is performed by capturing all the FCs in between.

Letting $u_i$ be the target information bit to be decoded,
we write $u_{\ell_i}$ to denote its corresponding processing bit,
where $\ell_i$ is given by
\begin{equation}\label{eq:processing_bit}
    \ell_i \triangleq \min\{k\geq i \mmid (k+1\in\mathcal{A}) \vee (k+1=N) \}.
\end{equation}
In other words, the processing bit is the bit right ahead of the next information bit.
If $i+1\in\mathcal{A}$, the processing bit is determined to be the target bit itself (i.e., $\ell_i=i$),
and normal SC decoding is performed.
If $\ell_i>i$, a hypothesis $\mathcal{H}_{i,b}$ is established to check $u_i = b$ for $b\in\mathbb{F}_2$.
Specifically, let $\bar{\mathbf{u}}\langle \mathcal{H}_{i,b} \rangle\in\mathbb{F}_2^{\ell_i+1}$ denote the binary vector tentatively generated for $\mathcal{H}_{i,b}$.
The $j$-th component of $\bar{\mathbf{u}}\langle \mathcal{H}_{i,b} \rangle$ is generated by the applied code construction rule, that is,
\begin{equation}\label{eq:scc_rule}
\bar{\mathbf{u}}\langle \mathcal{H}_{i,b} \rangle_j=
\begin{cases}
\hat{u}_j, &\text{for } j<i,\\
b, &\text{for } j=i,\\
\bar{\mathbf{u}}\langle \mathcal{H}_{i,b} \rangle_0^{j-1} \cdot \mathbf{T}_{0:j-1,j}, &\text{for } i<j\leq \ell_i,\\
\end{cases}
\end{equation}

For the code in Example~\ref{ex:toy}, two information bits $u_3$ and $u_5$ are handled by the SCC approach.
In decoding $u_3$, the processing bit is determined to be $u_4$
and two vectors $\bar{\mathbf{u}}\langle \mathcal{H}_{3,0}\rangle =(0,0,0,0,0)$ and $\bar{\mathbf{u}}\langle \mathcal{H}_{3,1}\rangle=(0,0,0,1,0)$ are made.
For $u_5$, the processing bit is given as $u_6$ because the next information bit is $u_7$.
According to \eqref{eq:scc_rule}, we have $\bar{\mathbf{u}}\langle \mathcal{H}_{5,0} \rangle = (0,0,0,\hat{u}_3,0,0,\hat{u}_3+0)$ and
$\bar{\mathbf{u}}\langle \mathcal{H}_{5,1} \rangle = (0,0,0,\hat{u}_3,0,1,\hat{u}_3+1)$.
As a result, the SCC decoder traverses valid partial paths from $u_i$ to $u_{\ell_i}$ in the binary decision tree shown in Fig.~\ref{fig:tree}.

After establishing the hypotheses,
the bits in $\bar{\mathbf{u}}\langle \mathcal{H}_{i,b} \rangle$ are cancelled out by assuming them as true.
Then, the original SC-based decoding operation employing the recursive formula is exploited to process $u_{\ell_i}$ and
determines the hypothesis that is most likely, based on the given channel observations.
The SCC decoding operation is interpreted as making a decision on $u_i$ by additionally using $\bar{\mathbf{u}}\langle \mathcal{H}_{i,b} \rangle_{i+1}^{\ell_i}$,
leading to better estimation than \eqref{eq:org_subchannel}.
In this way, the FCs with indices in $\{i+1:\ell_i\}$ are directly reflected in the process of estimating $u_i$.

Subsequently, the SC-based decoding, employing the recursive formula, is employed to decode $u_{\ell_i}$ and determine the hypothesis that is most likely based on the observed channel conditions.

\subsection{FC Conversion}

When applying the SCC decoding technique,
the FCs with indices in $\mathcal{A}^{\mathsf{c}}\cap \{i+1:\ell_i\}$ are incorporated into the decoding process of the target bit $u_i$,
while the other FCs are left unused.
In this subsection, we introduce FC conversion rules designed to harness the potential of these untapped FCs,
which are not addressed within the overarching recursive SC decoding formula by the SCC decoding approach.
Specifically, we convert the FCs with indices in $\mathcal{A}^{\mathsf{c}}\cap \{\ell_i+1:N-1\}$ into equivalent ones for $\mathbf{x}$.

We begin by presenting the following proposition to derive a parity-check equation for $\mathbf{x}$.

\begin{prop}\label{prop:Q}
    Assume that an encoder input vector $\mathbf{u}$ is constrained by $\mathbf{u}\mathbf{H}'=\mathbf{0}$.
    For its encoder output vector $\mathbf{x}=\mathbf{u}\mathbf{G}$,
    the equivalent parity-check equation is given by
    \begin{equation}\label{eq:Q}
        \mathbf{x}\mathbf{Q} = \mathbf{0},
    \end{equation}
    where $\mathbf{Q} = \mathbf{G}\mathbf{H}'$.
\end{prop}
\begin{proof}
    Note that $\mathbf{u}=\mathbf{x}\mathbf{G}^{-1}=\mathbf{x}\mathbf{G}$
    because $\mathbf{G}$ is an involutory matrix over $\mathbb{F}_2$.
    Hence, we get an equivalent parity-check equation on $\mathbf{x}$ given by
    \begin{equation}\label{eq:xGH}
    \mathbf{0} = \mathbf{u}\mathbf{H}'=\mathbf{x}\mathbf{G}\mathbf{H}'.
    \end{equation}
\end{proof}

\begin{figure}[t]
	\centering
	\includegraphics[width=\figwidth]{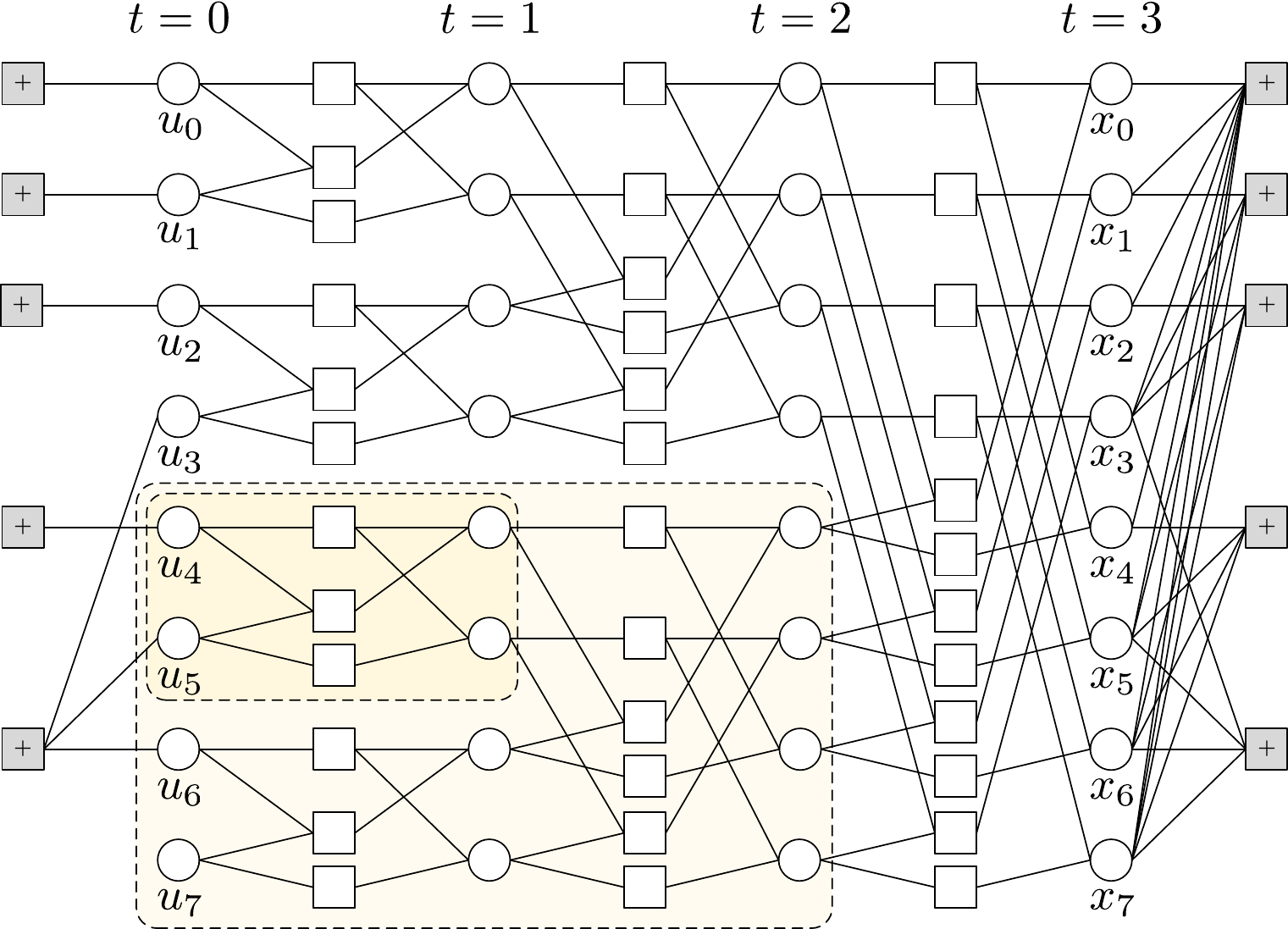}
	\caption{Bipartite graph of the polar code given in Example~\ref{ex:toy} with constraints induced by frozen bits and parity bits.
    The components drawn in the center represent a typical polar code graph,
    in which each $2\times 2$ polarization kernel is depicted in more detail by using three CNs.
    Gray-filled CNs at both ends correspond to constraints (left, $\mathbf{uH}'=\mathbf{0}$) and their conversions (right, $\mathbf{xQ}=\mathbf{0}$).
    Two subgraph structures for the component codes of lengths 2 and 4, which involve $u_5$, are highlighted by using dashed-lined boxes.}
	\label{fig:full_graph}
\end{figure}

For the code in Example~\ref{ex:toy}, we have
\begin{equation}\nonumber
\mathbf{Q}=
{
\small
\setlength\arraycolsep{\matsep}\def\arraystretch{\matstretch}
\begin{bmatrix}
1 & 1 & 1 & 1 & 1 & 1 & 1 & 1 \\
0 & 1 & 0 & 1 & 0 & 1 & 0 & 1 \\
0 & 0 & 1 & 1 & 0 & 0 & 1 & 1 \\
0 & 0 & 0 & 0 & 1 & 1 & 1 & 1 \\
0 & 0 & 0 & 1 & 0 & 1 & 1 & 1 \\
\end{bmatrix}^\mathsf{T}
},
\end{equation}
whose corresponding polar code graph is shown in Fig.~\ref{fig:full_graph}.
This bipartite graph consists of two kinds of vertices:
variable nodes (VNs, circles) and check nodes (CNs, squares).
A VN corresponds to a bit,
while a CN represents a parity-check equation showing that the binary sum of all connected VNs equals 0.
Typical graph components representing $\mathbf{x}=\mathbf{u}\mathbf{G}$ are drawn in the center,
where each $2\times 2$ polarization kernel for $\mathbf{F}=\left[\begin{smallmatrix} 1 & 0 \\ 1 & 1 \end{smallmatrix}\right]$ is depicted using three CNs with more accuracy and detail than the conventional polar code graph.
The graph is derived from the recursive construction of polar codes
and consists of $n+1$ stages indexed from $0$ (leftmost) to $n$ (rightmost).
The VNs at stages $0$ and $n$ correspond to the encoder input $\mathbf{u}$ and the encoder output $\mathbf{x}$, respectively.

New components representing the constraints on $\mathbf{u}$ and $\mathbf{x}$ are added to the graph by using gray-filled CNs.
The original constraints given by $\mathbf{u}\mathbf{H}'=\mathbf{0}$ are shown on the left side of the graph,
while the equivalent parity-check equation $\mathbf{x}\mathbf{Q}=\mathbf{0}$ is depicted on the right side.
Each constraint is presented by a corresponding CN.
By Proposition \ref{prop:Q}, the constraint $u_3+u_5+u_6=0$ on the left side is converted into the equivalent one $x_3+x_5+x_6+x_7=0$ on the right side, and so on.

Looking at Proposition~\ref{prop:Q},
the FCs that cannot be utilized in conventional SC decoding now appear on the encoder output.
It turns out that they are available to the decoder input at the receiver.
To be specific,
$\mathbf{xQ}=\mathbf{0}$ is modified into use as sequential decoding proceeds,
and the following theorem presents how to obtain an instant parity-check equation for each bit.
\begin{thm}\label{thm:full_Q}
    Assume that we are in a position to decode $u_i$ based on past estimates $\hat{\mathbf{u}}_0^{i-1}$.
    Then, the parity-check equation on $\mathbf{x}$ induced by FCs is given by
    \begin{equation}\label{eq:thm:full_Q}
    \hat{\mathbf{u}}_0^{i-1}\mathbf{H}_{0:i-1,\mathcal{L}_i} + \mathbf{x} \mathbf{Q}^{(i)} = \mathbf{0},
    \end{equation}
    where $\mathbf{Q}^{(i)} = \mathbf{G}_{\ast,i:N-1} \mathbf{H}_{i:N-1,\mathcal{L}_i}$.
\end{thm}
\begin{proof}
    Given $\hat{\mathbf{u}}_0^{i-1}$, the parity-check equation $\mathbf{u}\mathbf{H}=\mathbf{0}$ is broken down into
    \begin{equation}
        \hat{\mathbf{u}}_0^{i-1}\mathbf{H}_{0:i-1,\ast}
        +
        \mathbf{u}_{i}^{N-1} \mathbf{H}_{i:N-1,\ast}
        = \mathbf{0}.
    \end{equation}
    Since $\mathbf{u}=\mathbf{x}\mathbf{G}$ and $\mathbf{u}_{i}^{N-1}=\mathbf{x}\mathbf{G}_{\ast,i:N-1}$,
    we have
    \begin{equation}\label{eq:new_constraint}
        \hat{\mathbf{u}}_0^{i-1}\mathbf{H}_{0:i-1,\ast} + \mathbf{x}\mathbf{G}_{\ast,i:N-1} \mathbf{H}_{i:N-1,\ast} = \mathbf{0}.
    \end{equation}
    By removing the columns corresponding to past estimates and information bits for which parity-check equations do not need to be identified,
    we make \eqref{eq:new_constraint} into \eqref{eq:thm:full_Q} in a more compact form.
\end{proof}

The parity-check equation in \eqref{eq:thm:full_Q} is tractable in decoding $u_i$,
where the first term corresponds to the past estimates $\hat{\mathbf{u}}_0^{i-1}$ while the second one is determined by $\mathbf{x}$.
The number of converted constraints on $\mathbf{x}$ is exactly the same as that of the FCs on $\mathbf{u}$ (i.e. $|\mathcal{L}_i|$).

\begin{cor}\label{cor:full_Q}
    Applying the SCC decoding algorithm together,
    we have
    \begin{equation}\label{eq:scc_compact_constraint}
        \bar{\mathbf{u}}\langle \mathcal{H}_{i,b} \rangle \mathbf{H}_{0:\ell_i,\mathcal{L}_{\ell_i+1}} + \mathbf{x} \mathbf{Q}^{(\ell_i+1)} = \mathbf{0}
    \end{equation}
    for the hypothesis $\mathcal{H}_{i,b}$.
\end{cor}
\begin{proof}
    It is simply derived from \eqref{eq:thm:full_Q} by considering the processing bit index $\ell_i$ instead of the target bit index $i$ and replacing $\hat{\mathbf{u}}_0^{i-1}$ with $\bar{\mathbf{u}}\langle \mathcal{H}_{i, b} \rangle$ to model the SCC decoding behavior.
\end{proof}

\subsection{Subgraph-Based FC Conversion}\label{sec:subgraph}

In \eqref{eq:thm:full_Q} and \eqref{eq:scc_compact_constraint},
each row of $\mathbf{Q}^{(i)}$ and $\mathbf{Q}^{(\ell_i+1)}$ generally exhibits a high degree, indicating a substantial number of ones.
Consequently, the number of variables in $\mathbf{x}$ associated with each component parity-check equation tends to be notably large.
This characteristic may give rise to underdetermined or ill-posed decoding.
Based on the fact that a polar code graph is recursively constructed by polar codes of smaller lengths,
we find a compact conversion rule in this subsection.

Suppose that $u_i$ is a target bit and $u_j$ ($j\in\mathcal{L}_i$) is its FC.
If both $u_i$ and $u_j$ belong to the same component polar code of a smaller length,
the converted constraint brought by $u_j$ can be identified in this code, rather than the mother code.
In Fig.~\ref{fig:full_graph}, two subgraph structures including $u_5$ are depicted,
where $\mathbf{u}_4^5$ and $\mathbf{u}_4^7$ are to be encoder input vectors for the polar codes of length $2$ and $4$, respectively.
Clearly,
the additional constraint induced by the parity bit $u_6$ can be found in the polar code of length 4.

For $i\in \mathbb{Z}_N$ and $t\in\{1:n\}$, we write $\mathcal{T}{(i,t)}\triangleq \left\{ s(i,t)+k \mmid k\in\mathbb{Z}_{2^t} \right\}$ to indicate the index set of the bits in $\mathbf{u}$ that together belong to the subgraph of size $2^t$, where $s(i,t) \triangleq 2^t \times \lfloor  i/2^t\rfloor$.
Clearly, $\mathcal{T}(i,t)\supset\mathcal{T}(i,t-1)$.
For example, we have $\mathcal{T}{(5,1)}=\{4:5\}$ and $\mathcal{T}{(5,2)}=\{4:7\}$ as shown in Fig.~\ref{fig:full_graph}.
Let $\mathbf{x}_{i}^{(t)}\in\mathbb{F}_2^{2^t}$ be the intermediate codeword generated by encoding $\mathbf{u}_{\mathcal{T}(i,t)}\in\mathbb{F}_2^{2^t}$,
that is,
\begin{equation}\label{eq:subcode0}
\mathbf{x}_i^{(t)} = \mathbf{u}_{\mathcal{T}(i,t)}\mathbf{F}^{\otimes t},
\end{equation}
where $\mathbf{F}^{\otimes t}\in\mathbb{F}_2^{2^t\times 2^t}$ is the generator matrix of the component polar code of length $2^t$.
Note that $\mathbf{x}_i^{(t)}$ appears at stage $t$ in the polar code graph as shown in Fig.~\ref{fig:full_graph}.
In addition,
let $\mathcal{T}'(i,t)\triangleq \mathcal{T}(i,t) \cap \{i:N-1\}$ denote the index set of the bits in $\mathbf{u}_{\mathcal{T}(i,t)}$ behind $u_{i-1}$,
and let $\mathcal{L}_{i,t} \triangleq \mathcal{L}_i \cap \left(\mathcal{T}(i,t)\backslash \mathcal{T}(i,t-1)\right) $ be the index set of the FCs only belonging to the subgraph of length $2^t$, but not to its nested subgraph of length $2^{t-1}$.

\begin{thm}\label{thm:sub_Q}
Assume that we are in a position to decode $u_i$ based on past estimates $\hat{\mathbf{u}}_0^{i-1}$.
For $t\in\{1:n\}$, the parity-check equation on $\mathbf{x}_i^{(t)}$ induced by the FCs in the subgraph of length $2^t$  is given by
\begin{equation}\label{eq:thm:sub_Q}
\hat{\mathbf{u}}_{0}^{i-1}\mathbf{H}_{0:i-1,\mathcal{L}_{i,t}}
+\mathbf{x}_i^{(t)}\mathbf{Q}^{(i,t)} = \mathbf{0},
\end{equation}
where $\mathbf{Q}^{(i,t)} = \left(\mathbf{F}^{\otimes t}\right)_{\ast,-s(i,t)+\mathcal{T}'(i,t)} \mathbf{H}_{\mathcal{T}'(i,t),\mathcal{L}_{i,t}}$.
\end{thm}
\begin{proof}
Since $\mathbf{F}^{\otimes t}$ is an involutory matrix, we clearly have
\begin{equation}\label{eq:subcode1}
\mathbf{u}_{\mathcal{T}(i,t)} = \mathbf{x}_i^{(t)}\mathbf{F}^{\otimes t}.
\end{equation}
The generation of $\mathbf{u}_{\mathcal{T}'(i,t)}$ is then written by
\begin{equation}\label{eq:subcode2}
\mathbf{u}_{\mathcal{T}'(i,t)} = \mathbf{x}_i^{(t)}\left(\mathbf{F}^{\otimes t}\right)_{\ast,-s(i,t)+\mathcal{T}'(i,t)},
\end{equation}
where $-s(i,t)+\mathcal{T}'(i,t)=\{-s(i,t)+a\mmid a\in\mathcal{T}'(i,t) \}$.
Focusing only on the subgraph of interest, we have
\begin{equation}\label{eq:subcode3}
\hat{\mathbf{u}}_{0}^{i-1}\mathbf{H}_{0:i-1,\mathcal{L}_{i,t}}
+\mathbf{u}_{\mathcal{T}'(i,t)}\mathbf{H}_{\mathcal{T}'(i,t),\mathcal{L}_{i,t}} = \mathbf{0}
\end{equation}
in the same way as in Theorem~\ref{thm:full_Q},
where the bits of $\mathbf{u}$ behind the subgraph do not contribute to \eqref{eq:subcode3} because of the upper-triangular structure of $\mathbf{H}$.
By putting \eqref{eq:subcode2} into the second term of \eqref{eq:subcode3}, we have \eqref{eq:thm:sub_Q}.
\end{proof}

By the subgraph-based conversion for decoding $u_i$, the set $\mathcal{L}_{i}$ is divided into disjoint subsets $\mathcal{L}_{i,t}$ for $t\in\{1:n\}$, that is,
\begin{equation}
    \bigcup_{t=1}^{n} \mathcal{L}_{i,t} = \mathcal{L}_{i}.
\end{equation}
Therefore, the total number of constraints is preserved even under the subgraph-based conversion.

Considering the processing bit index $\ell_i$ for \eqref{eq:thm:sub_Q} instead of the target bit index $i$,
we have the following result for SCC decoding.
\begin{cor}\label{cor:sub_Q}
Under SCC decoding for $u_i$ with hypothesis $\mathcal{H}_{i,b}$,
the subgraph-based FC conversion gives a parity-check equation 
\begin{equation}\label{eq:cor:sub_Q}
    \bar{\mathbf{u}}\langle \mathcal{H}_{i,b} \rangle \mathbf{H}_{0:\ell_i,\mathcal{L}_{\ell_i+1,t}} + \mathbf{x}_{\ell_i+1}^{(t)} \mathbf{Q}^{(\ell_i+1,t)} = \mathbf{0}.
\end{equation}
\end{cor}

\subsection{Graph Construction and Belief Propagation}\label{sec:bp}

A given polar code graph is modified by the FC conversion.
In sequential decoding of each bit,
a part of the full graph (e.g., Fig.~\ref{fig:full_graph}) is selectively active while the previous estimates are cancelled.
The instantly constructed graph for decoding each bit looks like a tree in which there are $2^t$ VNs at stage $t$.
For $t\in\{0:n-1\}$ and $k\in\mathbb{Z}_{2^t}$,
a VN corresponding to $x_k^{(t)}$ has two child VNs for $x_k^{(t+1)}$ and $x_{k+2^t}^{(t+1)}$,
where their connections are determined by the current processing bit index $\ell_i$.
Let $\langle \ell_i \rangle_t \in\mathbb{F}_2$ denote the $t$-th element of the binary representation of $\ell_i$ such that $\ell_i = \sum_{i=0}^{n-1} \langle \ell_i \rangle_t 2^t$.
\if\doccolumn2
If $\langle \ell_i \rangle_t =0$, a single parity-check (SPC) node representing
\begin{equation}
x_k^{(t)} = x_k^{(t+1)}+x_{k+2^t}^{(t+1)}
\end{equation}
is constructed.
On the other hand, a repetition node of
\begin{equation}
x_k^{(t)}=x_k^{(t+1)}+\beta_k^{(t)}=x_{k+2^t}^{(t+1)}
\end{equation}
appears in the graph if $\langle \ell_i \rangle_t = 1$, where
$\beta_k^{(t)}$ corresponds to the estimate for $x_k^{(t)}$ and is obtained by back-propagating $\bar{\mathbf{u}}\langle \mathcal{H}_{i,b} \rangle_{0:\ell_i-1}$ to stage $t$ in the sequential cancellation operation.
\else
If $\langle \ell_i \rangle_t =0$, a single parity-check (SPC) node representing
$x_k^{(t)} = x_k^{(t+1)}+x_{k+2^t}^{(t+1)}$
is constructed.
On the other hand, a repetition node of
$x_k^{(t)}=x_k^{(t+1)}+\beta_k^{(t)}=x_{k+2^t}^{(t+1)}$
appears in the graph if $\langle \ell_i \rangle_t = 1$, where $\beta_k^{(t)}$ is the estimate of $x_k^{(t+1)}+x_{k+2^t}^{(t+1)}$ obtained by the previous SC operation for the corresponding SPC node.
\fi
The construction rule of an instant graph and the back-propagation operation are well documented in the literature (\textit{cf.} \cite{Leroux2013,BalatsoukasStimming2015}).

\begin{figure}[t]
	\centering
\if\doccolumn1
	\includegraphics[width=\figwidthA]{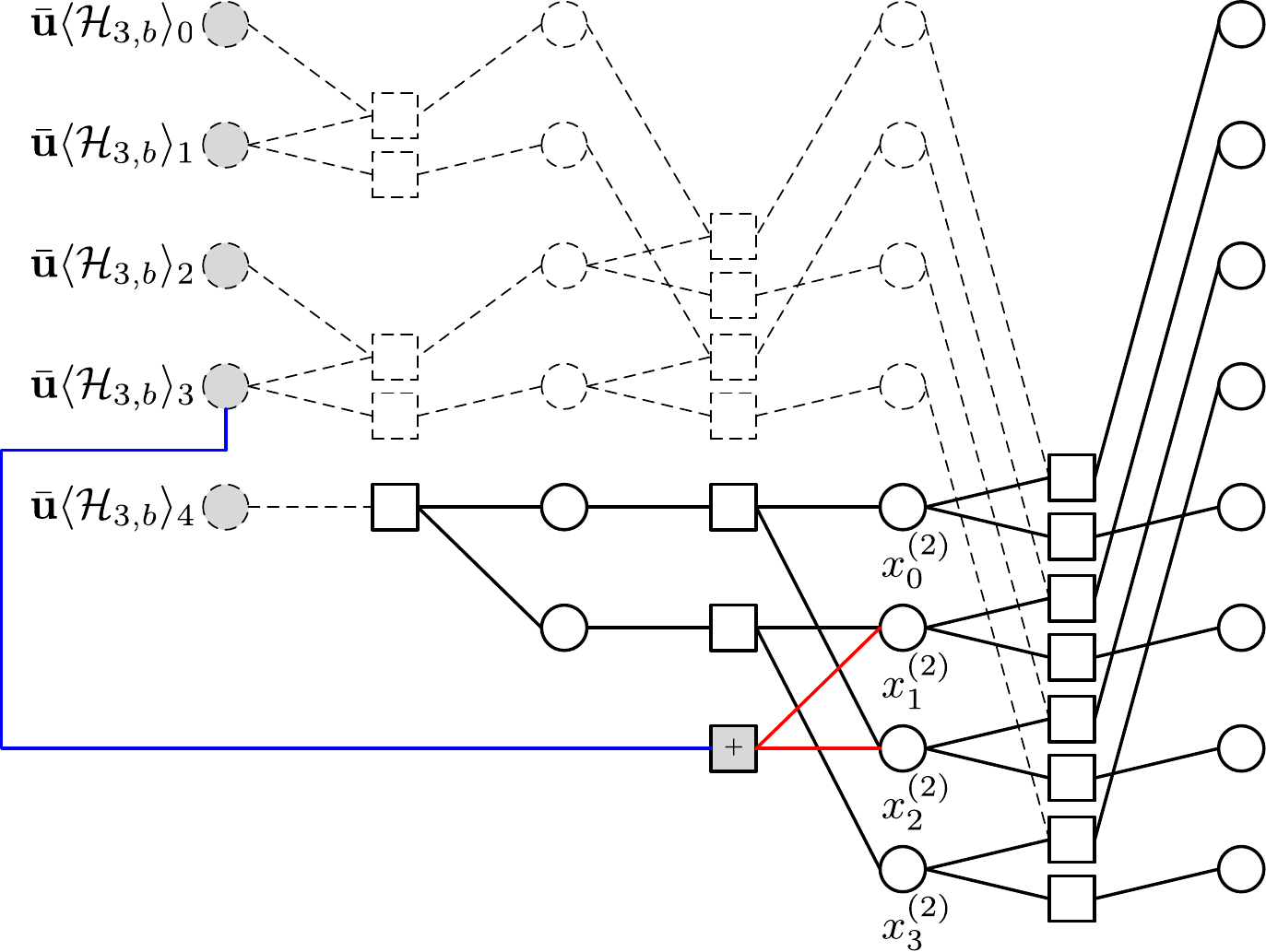}
\else
    \includegraphics[width=\figwidth]{Fig_FC_Graph.pdf}
\fi
	\caption{Instant graph construction with subgraph-based FC conversion for decoding $u_3$ in Example~\ref{ex:toy}.
            White squares represent normal CNs in the polar code graph, while a gray-filled square corresponds to an FCCN representing $x_1^{(2)}+x_2^{(2)}+\bar{\mathbf{u}}\langle \mathcal{H}_{3,b}\rangle_3=0$, converted from the FC $u_6=u_3 + u_5$.}
	\label{fig:conversion}
\end{figure}

An instant graph can be constructed by focusing on the combination of SCC decoding and the subgraph-based FC conversion given in Corollary~\ref{cor:sub_Q}.
According to \eqref{eq:cor:sub_Q},
$|\mathcal{L}_{\ell_i+1,t}|$ new CNs are added at stage $t$,
and are associated with two sets of VNs corresponding to $\bar{\mathbf{u}}\langle \mathcal{H}_{i,b} \rangle$ and $\mathbf{x}_{\ell_i+1}^{(t)}$, respectively.
These CNs are referred to as \textit{FC check nodes (FCCNs)} at stage $t$.
We rewrite \eqref{eq:cor:sub_Q} into
\begin{equation}\label{eq:new_sub_Q}
\mathbf{x}_{\ell_i+1}^{(t)} \mathbf{Q}^{(\ell_i+1,t)}
=
\bar{\mathbf{u}}\langle \mathcal{H}_{i,b} \rangle \mathbf{H}_{0:\ell_i,\mathcal{L}_{\ell_i+1,t}},
\end{equation}
where the left-hand side (LHS) is represented by the connections between the FCCNs and the VNs of $\mathbf{x}_{\ell_i+1}^{(t)}$,
while the right-hand side (RHS) is simply given as a linear combination of the bits whose values are already determined in $\bar{\mathbf{u}}\langle \mathcal{H}_{i,b}\rangle$.
Clearly, the RHS serves as non-zero constraint values to check the given parity-check equation.
Let $\mathcal{Q}_j^{(\ell_i+1,t)}$ be the index set of VNs in $\mathbf{x}_{\ell_i}^{(t)}$, connected to the $j$-th FCCN.
Then, we have
\begin{equation}
\mathcal{Q}_{j}^{(\ell_i+1,t)} \triangleq \left\{ q \in\mathbb{Z}_{2^{t}} \mmid \mathbf{Q}^{(\ell_i+1,t)}_{q,j}=1 \right\}.
\end{equation}

Fig.~\ref{fig:conversion} shows the instant graph connection when decoding $u_3$ in the code of Example~\ref{ex:toy}.
The processing bit index $\ell_3$ is determined to be $4$ by \eqref{eq:processing_bit},
and a basic tree-like graph for $u_{\ell_3}=u_4$ is first formed.
From \eqref{eq:new_sub_Q},
$\mathbf{x}_5^{(2)} \mdot [0110]^\mathsf{T} = \bar{\mathbf{u}} \langle \mathcal{H}_{3,b} \rangle \mdot [00010]^\mathsf{T}$ is given,
and an FCCN representing $x_1^{(2)}+x_2^{(2)} = \bar{\mathbf{u}}\langle \mathcal{H}_{3,b} \rangle_3 $ is supplemented to the graph.

The instant polar code graph is modified by supplementing FCCNs and their corresponding edges,
so it is worth introducing a new decoding scheme to take advantage of these new components.
We apply the iterative BP method \cite{Gallager1962, modern2008} to polar decoding over the modified graph.
Two major changes are made in our scheme.
First, message-passing operations are performed with an appropriate number of iterations in order to utilize the FCCNs.
Second, decoding metrics are updated in both directions, whereas messages are delivered only in one direction (descending order of stage index $t$) in the conventional SC-based decoding.
We formalize a combination of BP and SCC methods as BP-SCC decoding.

\if\doccolumn1
\begin{figure}[t!]
    \centering
    \subfloat[FCCN processing]{
        \includegraphics[width=\halfsubfigwidthA]{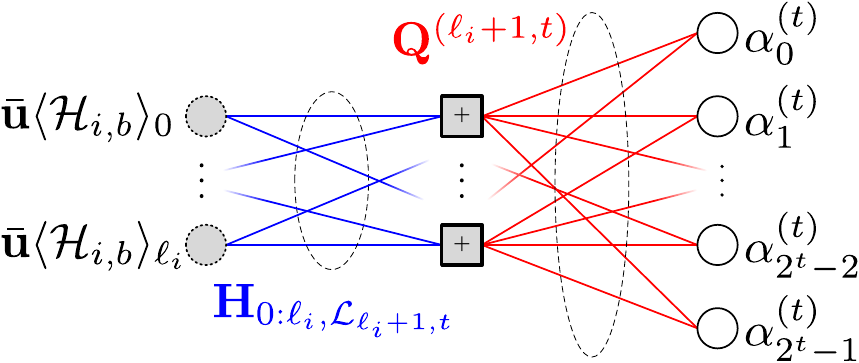}} $~$
    \subfloat[CN processing (SPC node)]{
        \includegraphics[width=\halfsubfigwidth]{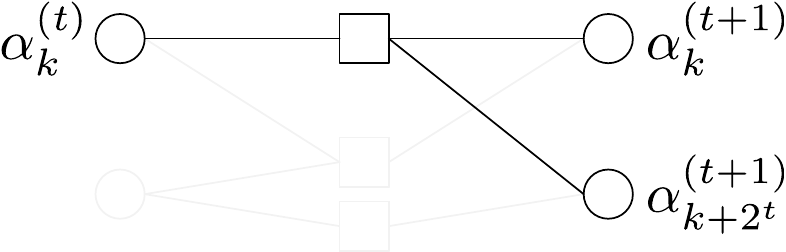}} $~$
    \subfloat[CN processing (repetition node)]{
        \includegraphics[width=\halfsubfigwidth]{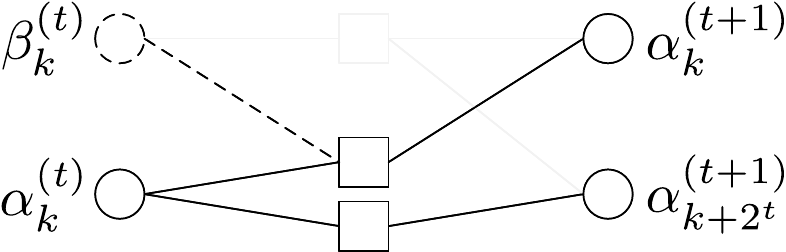}}\\
  \caption{Three component operations at stage $t$ in BP-SCC decoding.}
  \label{fig:component}
\end{figure}
\else
\begin{figure}[t!]
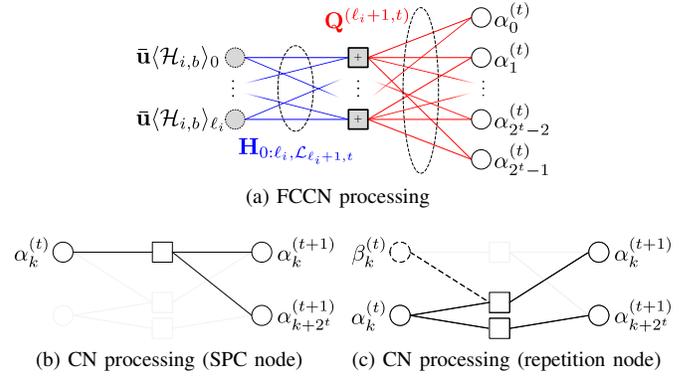

    \centering
    \subfloat[FCCN processing]{
        \includegraphics[width=0.62\linewidth]{Fig_Component_PC1.pdf}} \\
    \subfloat[CN processing (SPC node)]{
        \includegraphics[width=0.48\linewidth]{Fig_Component_Degrade.pdf}} \hfill
    \subfloat[CN processing (repetition node)]{
        \includegraphics[width=0.48\linewidth]{Fig_Component_Upgrade.pdf}}\\
  \caption{Three component operations at stage $t$ in BP-SCC decoding.}
  \label{fig:component}
\end{figure}
\fi

From now on, we present a brief sketch of BP-SCC decoding operations over the modified graph with FCCNs.
In order to represent decoding metrics,
the conventional notation of the tree-like data structure
    $\big\{ \alpha_k^{(t)} \mmid t\in\{0:n\}, k\in\mathbb{Z}_{2^t } \big\}$
is used, where $\alpha_k^{(t)}$ is a decoding metric for the $k$-th variable node at stage $t$.
The decoding metrics at stage $n$ (i.e., $\alpha_k^{(n)}$ for $k\in\mathbb{Z}_N$) are initialized using the given channel observations $\mathbf{y}$.
Then, the BP-SCC decoding operations for $u_i$ ($i\in\mathcal{A}$) are conducted as follows:
\begin{enumerate}
  \item
  The processing bit index $\ell_i$ is determined by \eqref{eq:processing_bit} and $\bar{\mathbf{u}}\langle \mathcal{H}_{i,b}\rangle$ is generated for $b\in\{0,1\}$.
  A tree-like graph is constructed for processing $u_{\ell_i}$, and FCCNs are supplemented to each stage by \eqref{eq:new_sub_Q}.
  \item
  For $t=n-1,\ldots,0$ in descending order, the following component operations are sequentially performed:
  \begin{enumerate}
    \item \textit{FCCN operation:}
        The decoding metrics at stage $t+1$ (i.e., $\alpha_k^{(t+1)}$ for $k\in\mathbb{Z}_{2^{t+1}}$) are first updated via the FCCNs as depicted in Fig.~\ref{fig:component} (a).
        Each FCCN supplemented at stage $t+1$ takes variable-to-check (V2C) messages from its neighbor VNs,
        computes check-to-variable (C2V) messages,
        and passes them back to its neighbors.
        Then, VNs at stage $t+1$ update their decoding metrics based on the delivered C2V messages.
        In the message calculation, the bits in $\bar{\mathbf{u}}\langle \mathcal{H}_{i,b}\rangle$ may cause non-zero constraint values,
        and thus, the message-passing results vary with the hypothesis value $b$.
    \item \textit{CN operation:}
        For each $k\in\mathbb{Z}_{2^t}$,
        three decoding metrics $\alpha_k^{(t)}$, $\alpha_k^{(t+1)}$, $\alpha_{k+2^t}^{(t)}$ are mutually updated by message-passing.
        The message-passing operation type at stage $t$ is determined by $\langle \ell_i \rangle_t$.
        As shown in Fig.~\ref{fig:component} (b) and (c),
        the operation for an SPC node is carried out if $\langle \ell_i \rangle_t=0$,
        while that for a repetition node is done using the past estimate $\beta_k^{(t)}$ if $\langle \ell_i \rangle_t=1$.
        The message-passing operation is performed back and forth,
        that is, not only $\alpha_k^{(t)}$ but also $\alpha_k^{(t+1)}$ and $\alpha_{k+2^t}^{(t)}$ are updated.
    \item \textit{Processing bit operation:}
        After finishing the CN operation at stage $t=0$, the final metric $\alpha_0^{(0)}$ is obtained.
        It is further updated by considering the given processing bit $\bar{\mathbf{u}}\langle \mathcal{H}_{i,b} \rangle_{\ell_i}$.
  \end{enumerate}
  \item
  The operations at Step 2) are repeated at most $I_{\max}$ times, where $I_{\max}$ is a predetermined maximum number of iterations.
  At each iteration, a validation check for the tentative estimates (including all the intermediate nodes in the graph) is performed,
  and the iteration can be early terminated depending on the check result.
  \item
  After performing the above operations for each $b\in\mathbb{F}_2$,
  a decision on the hypothesis $\mathcal{H}_{i,b}$ is made to determine the traversal over the binary decision tree shown in Fig.~\ref{fig:tree}.
  Let $r(\mathcal{H}_{i,b})\in\{0,1\}$ denote the report for $\mathcal{H}_{i,b}$,
  where $r(\mathcal{H}_{i,b})=0$ if a conflict between the channel observations $\mathbf{y}$ and the hypothesis $\bar{\mathbf{u}}\langle \mathcal{H}_{i,b} \rangle$ is detected,
  and $r(\mathcal{H}_{i,b})=1$ otherwise.
  The decision result $r(\mathcal{H}_{i,b})$ is made by the validation check performed at Step 3),
  and soft metrics such as LLRs and PMs can be further applied to elaborate the decision result.
\end{enumerate}

According to the operations described above, the decoding metrics are improved by additional message-passing over the FCCNs, compared with conventional SC-based decoding methods.
In fact, the BP-SCC decoding algorithm aided by FCs enhances the estimation of $\bar{\mathbf{u}}\langle \mathcal{H}_{i,b} \rangle$.
Generally speaking, the specific definition of message-passing operations depends on the channel type.
As an example, that for the BEC is introduced in the next section.

\subsection{Tree Search with Stack-based Backjumping (SBJ)}\label{sec:csp}

Sequential decoding for polar codes can be seen as solving the search problem for a binary decision tree given in Fig.~\ref{fig:tree}.
The original SC decoding algorithm employs a DFS heuristic,
while the SCL decoder exploits a BFS method with limited branching based on soft metrics (i.e., PMs).
In the proposed decoding algorithms,
FCs serve as hard validity conditions that a given partial solution must satisfy.
Therefore, the FC-aided decoding algorithms are interpreted as solving a sort of dynamic CSP \cite{Meseguer1989,Mittal1990} formulated by FCs,
which adaptively change with the decisions made in sequential estimation.
The dynamic CSP that we have consists of a set of variables $\mathbf{u}_{\mathcal{A}}$,
each of them associated with the alphabet $\mathbb{F}_2$,
and a set of constraints $\mathbf{uH'}=\mathbf{0}$, equivalent to $\mathbf{xQ}=\mathbf{0}$ by Proposition~\ref{prop:Q}.
It is decomposed into sub-problems sequentially making a decision on $u_i$ under the dynamically changed parity-check equation as given in Theorems \ref{thm:full_Q} and \ref{thm:sub_Q}, and Corollaries \ref{cor:full_Q} and \ref{cor:sub_Q}.

The SC-based sequential processing with FCs is modeled by tree search to solve the dynamic CSP,
and this framework is well-established, especially in the artificial intelligence (AI) research area \cite{Kumar1998,Brailsford1999}.
Based on the previous works in the AI field, a pertinent tree search policy is designed using SBJ.
In the tree search,
some techniques for returning to predecessors (such as backtracking and backjumping) are generally used to avoid invalid solutions to CSPs and efficiently traverse other valid candidates.
Backtracking methods triggered by soft metrics were also exploited and evaluated for the SC with backtracking (SC-BT) decoder \cite{Vajha2019} and Fano decoders \cite{Jeong2019,Arikan2019,Rowshan2021}.
In these schemes, soft metrics are commonly compared with dynamic thresholds to determine whether to move forward or backward.
On the other hand, in the proposed method of this paper,
we have hard validity conditions induced by FCs.
If the decoder does not observe any conflict for a certain branch under these conditions,
it can safely jump back to the corresponding node later when the decoder runs into a dead-end.
For a later visit, we operate a stack $\mathcal{S}$ to store branches of which bit assignments are determined to be valid in terms of FCs.

To minimize the number of node visits, we propose an efficient tree traversal algorithm as follows:
\begin{itemize}
  \item When decoding $u_i$ with $i\in\mathcal{A}$, we only generate and verify a hypothesis $\mathcal{H}_{i,0}$ for bit value 0 using BP-SCC decoding.
  \begin{itemize}
    \item[-] If no conflict is detected (i.e., $r(\mathcal{H}_{i,0})=1$), the decoder proceeds to the branch associated with $\mathcal{H}_{i,0}$ and pushes the branch corresponding to $\mathcal{H}_{i,1}$ onto a stack $\mathcal{S}$ for potential later visits.
    \item[-] If a conflict is detected (i.e., $r(\mathcal{H}_{i,0})=0$), the second hypothesis $\mathcal{H}_{i,1}$ is generated and verified by BP-SCC decoding.
    \begin{itemize}
        \item If no conflict is detected (i.e., $r(\mathcal{H}_{i,1})=1$),
        the decoder proceeds to the branch associated with $\mathcal{H}_{i,1}$.
        \item if a conflict is detected (i.e., $r(\mathcal{H}_{i,1})=0$),
        the decoder jumps back to the last unvisited branch retrieved from the stack $\mathcal{S}$.
        If the stack $\mathcal{S}$ is empty, the decoder declares a decoding failure and terminates the decoding process.
    \end{itemize}
  \end{itemize}
\end{itemize}
This opportunistic approach to selectively visiting branches corresponding to bit value 1 significantly reduces the number of node visits.
If no error occurs over the channel, the expected number of node visits is $3N/2$.

\if\doccolumn1
\begin{figure}[t]
	\centering
	\includegraphics[width=9.5cm]{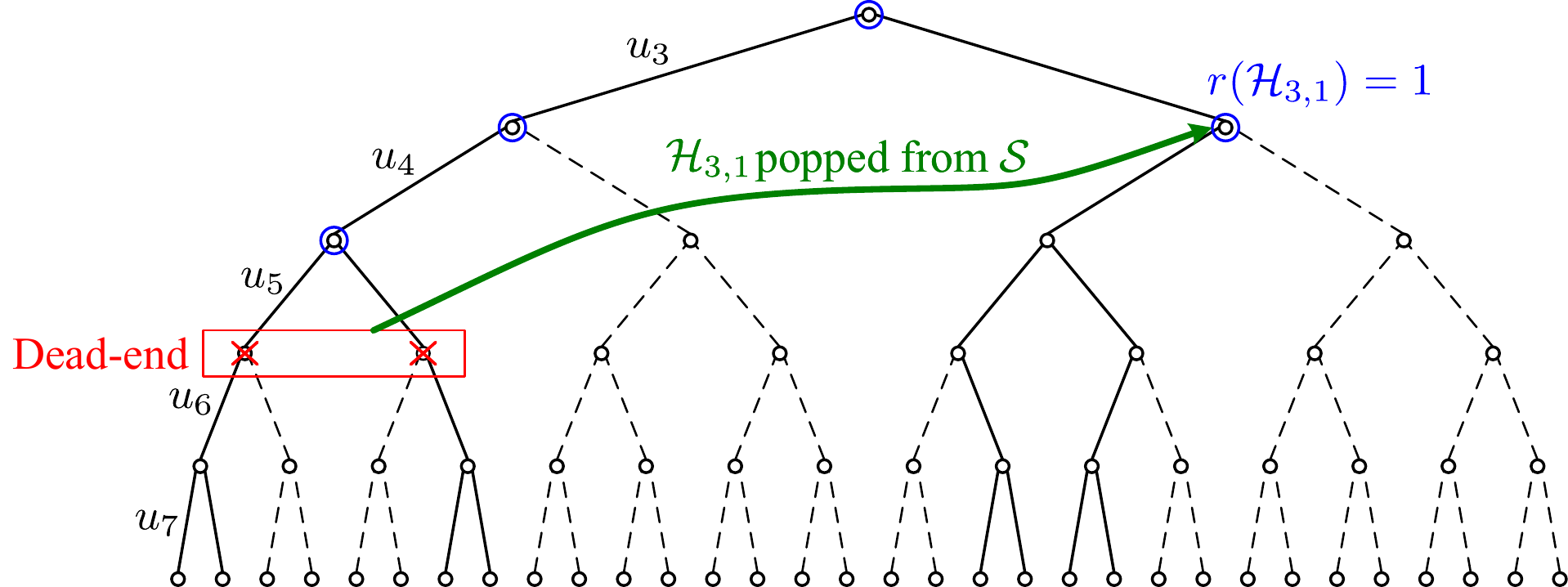}
	\caption{Stack-based backjumping (SBJ) with a stack $\mathcal{S}$ using the constraint satisfaction check results.}
	\label{fig:backjumping}
\end{figure}
\else
\begin{figure}[t]
	\centering
	\includegraphics[width=\figwidth]{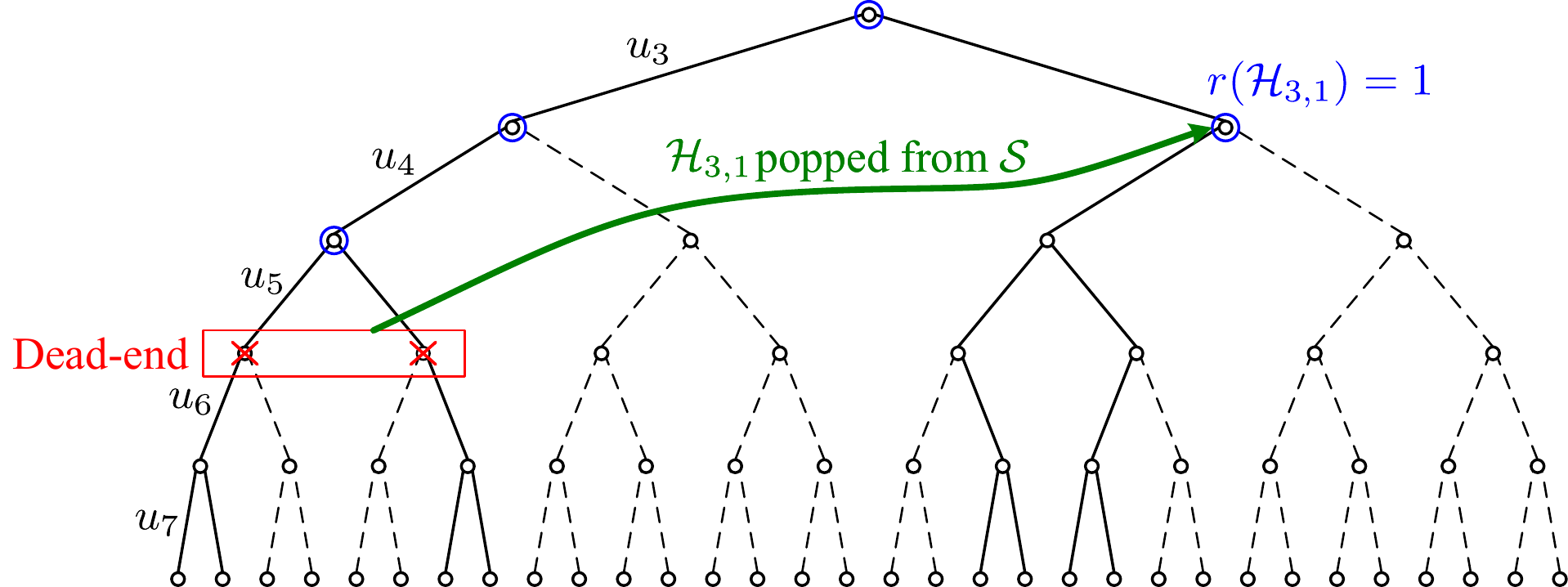}
	\caption{Example of stack-based backjumping (SBJ) using the constraint satisfaction check results for the code given in Example~\ref{ex:toy}.}
	\label{fig:backjumping}
\end{figure}
\fi

Fig.~\ref{fig:backjumping} shows a simple example of the tree search operation with SBJ for the code given in Example~\ref{ex:toy}.
First, suppose that the proposed decoder identifies that $\mathcal{H}_{3,0}$ does not violate the given constraints with respect to FCs.
According to the behavior described above, the decoder proceeds to the branch of $\mathcal{H}_{3,0}$ and pushes the branch of $\mathcal{H}_{3,1}$ to $\mathcal{S}$.
After that, we suppose that neither $\mathcal{H}_{5,0}$ nor $\mathcal{H}_{5,1}$ passes the CSPs given by FCs.
Due to the dead-end identified by the given CSPs, the decoder abandons the progress so far and jumps back to the branch of $\mathcal{H}_{3,1}$, which is popped from $\mathcal{S}$.
In this way, the CSP-based tree traversal finds a consistent bit assignment in sequential decoding.

The proposed SBJ technique is fundamentally distinct from SCS decoding \cite{Niu2012a},
although the former is similar to the latter in the sense that they both use a stack as a data structure.
One distinct point is whether FCs are incorporated or not.
The former is triggered by FCs, while the latter does not exploit FCs.
Another point is how the stack is used.
The stack in the proposed SBJ technique is used to push the unvisited branches that are identified not to violate the CSPs formulated by FCs.
On the other hand, the stack in SCS decoding is used to manage the limited set of the most promising nodes in tree search.

\section{Decoding Over the BEC}\label{sec:alg}

This section presents a specific BP-SCC decoding algorithm aided by FCs for the BEC.
Because no bit error is assumed, the decoding operation for the BEC is simple and understood clearly without any ambiguity.
We expect that the study on the proposed decoding over the BEC provides a vital intuition to understand that over general B-DMCs.

\subsection{Algorithm Description}

The transition probability of the BEC is given by
\begin{equation}
W(y_k \mmid x_k) = \begin{cases}
1-p_{\epsilon}, &\text{for } y_k = x_k,\\
p_{\epsilon},   &\text{for } y_k = \epsilon,\\
\end{cases}
\end{equation}
where $p_{\epsilon}$ is the channel erasure probability and $\epsilon$ is the symbol representing an erasure of the transmitted bit.
Unlike conventional decoding schemes for the BEC,
we add another symbol $\eta$ to denote a conflict possibly detected during decoding.
Summarizing them, we write $\mathbb{E}\triangleq \{0,1,\epsilon,\eta\}$ to denote the alphabet of decoding symbols $\alpha_k^{(t)}$ for $t\in\{0:n\}$ and $k\in\mathbb{Z}_{2^t}$.

In order to describe the decoding operations, we first define two elementary operators $\boxplus,\boxdot:\mathbb{E}\times\mathbb{E}\to\mathbb{E}$,
where $\boxplus$ is used to identify the sum of two inputs (i.e., SPC node) while $\boxdot$ is employed to check that two inputs are the same (i.e., repetition node).
Specifically, the results of these two operators are given by
\begin{equation}
    a \boxplus b =\begin{cases}
    \eta,       &\text{if } a=\eta \text{ or } b=\eta,\\
    \epsilon,   &\text{else if } a=\epsilon \text{ or } b=\epsilon,\\
    a + b,  &\text{otherwise},
    \end{cases}
\end{equation}
and
\if\doccolumn1
\begin{equation}
    a \boxdot b =
    \begin{cases}
    \eta,       &\text{if } (a \boxplus b = \eta) \text{ or } (a \boxplus b \neq \eta \text{ and } a \neq b),\\
    \epsilon,   &\text{else if } a=\epsilon \text{ and } b=\epsilon,\\
    b,          &\text{else if } a=\epsilon,\\
    a,          &\text{otherwise.}
    \end{cases}
\end{equation}
\else
\begin{equation}
    a \boxdot b =
    \begin{cases}
    \eta,       &\text{if } (a \boxplus b = \eta) \text{ or } (a \boxplus b \neq \eta \text{ and } a \neq b),\\
    \epsilon,   &\text{else if } a=\epsilon \text{ and } b=\epsilon,\\
    b,          &\text{else if } a=\epsilon,\\
    a,          &\text{otherwise.}
    \end{cases}
\end{equation}
\fi
It is clear that both $\boxplus$ and $\boxdot$ are commutative.

Using the two component operators,
the BP-SCC decoding algorithm for the BEC is specified.
Note that we write $a \leftarrow b$ to denote that $b$ is assigned to $a$ or $a$ is updated as $b$.
Given the channel observations $\mathbf{y}=(y_0,y_1,\ldots,y_{N-1})$,
the initial value of the decoding symbol $\alpha_k^{(t)}$ for $k\in\mathbb{Z}_{2^t}$ is determined by
\begin{equation}
\alpha_k^{(t)} \leftarrow \begin{cases}
y_k, &\text{if } t=n,\\
\epsilon, &\text{otherwise.}
\end{cases}
\end{equation}
In the process of decoding $u_i$ ($i\in\mathcal{A}$),
the values of the FCs preceding the next information bit, i.e., $\bar{\mathbf{u}}\langle \mathcal{H}_{i,b}\rangle_{i+1:\ell_i}$
are determined based on \eqref{eq:scc_rule} for the hypothesis $\mathcal{H}_{i,b}$.
Subsequently, the decoding operations for processing $u_{\ell_i}$ follow the descriptions outlined in Subsection~\ref{sec:bp}.
The following component operations are performed in descending order of the stage index $t\in\{0:n-1\}$.
\begin{enumerate}
    \item \textit{FCCN operation:}
        For stage $t+1$, the decoding symbols are updated by using the given FCCNs.
        We write $c_j$ to denote the $j$-th FCCN at this stage
        and $\phi_j\triangleq (\bar{\mathbf{u}}\langle \mathcal{H}_{i,b} \rangle \mathbf{H}_{0:\ell_i,\mathcal{L}_{\ell_i+1,t+1}})_j$ to indicate its corresponding nonzero constraint value given in the RHS of \eqref{eq:new_sub_Q}.
        Let $\mathcal{C}_k$ be the index set of FCCNs connected to $x_k^{(t+1)}$,
        and let $\mathcal{V}_j$ denote the index set of VNs neighbored by $c_j$.
        The C2V message from $c_j$ to $x_k^{(t+1)}$ is computed as
        \begin{equation}\label{eq:c2v}
            \gamma_{j\to k} \leftarrow \left(\bigboxplus\nolimits_{l \in \mathcal{V}_j \backslash \{k\} } \alpha_l^{(t+1)}\right) \boxplus \phi_j,
        \end{equation}
        where the big operator indicates serially repeated operations for variables whose indices are designated by the subscript.
        Based on these C2V messages, the decoding symbol for $x_k^{(t+1)}$ is then updated as
        \begin{equation}\label{eq:ap_v}
            \alpha_k^{(t+1)} \leftarrow \left(\bigboxdot\nolimits_{l\in\mathcal{C}_k} \gamma_{l \to k} \right) \boxdot \alpha_k^{(t+1)}.
        \end{equation}
    \item \textit{SPC node operation:}
        If $\langle \ell_i \rangle_t=0$ for the given processing bit index $\ell_i$,
        then three variables $\alpha_k^{(t)},\alpha_k^{(t+1)},\alpha_{k+2^t}^{(t+1)}$ for each $k\in\mathbb{Z}_{2^t}$ are concurrently updated as
        \begin{equation}
        \begin{split}
        &\alpha_k^{(t)} \leftarrow \alpha_k^{(t)} \boxdot \big( \alpha_k^{(t+1)} \boxplus \alpha_{k+2^t}^{(t+1)} \big),\\
        &\alpha_k^{(t+1)} \leftarrow \alpha_k^{(t+1)} \boxdot \big( \alpha_k^{(t)} \boxplus \alpha_{k+2^t}^{(t+1)} \big),\\
        &\alpha_{k+2^t}^{(t+1)} \leftarrow \alpha_{k+2^t}^{(t+1)} \boxdot \big( \alpha_k^{(t)} \boxplus \alpha_k^{(t+1)} \big).
        \end{split}
        \end{equation}
    \item \textit{Repetition node operation:}
        If $\langle \ell_i \rangle_t=1$,
        four input variables $\alpha_k^{(t)},\alpha_k^{(t+1)},\alpha_{k+2^t}^{(t+1)},\beta_k^{(t)}$ are fed into this operation,
        where $\beta_k^{(t)}$ is an estimate of $x_k^{(t+1)}+x_{k+2^t}^{(t+1)}$ attained by the previous SC operation as shown in Fig.~\ref{fig:component} (c).
        The first three decoding symbols are simultaneously refined as
        \begin{equation}
        \begin{split}
        &\alpha_k^{(t)} \leftarrow \alpha_k^{(t)} \boxdot \big(\big( \alpha_k^{(t+1)} \boxplus \beta_k^{(t)} \big) \boxdot \alpha_{k+2^t}^{(t+1)} \big),\\
        &\alpha_k^{(t+1)} \leftarrow \big(\alpha_k^{(t)} \boxplus \beta_k^{(t)} \big) \boxdot \alpha_k^{(t+1)},\\
        &\alpha_{k+2^t}^{(t+1)} \leftarrow \alpha_k^{(t)} \boxdot \alpha_{k+2^t}^{(t+1)}.
        \end{split}
        \end{equation}
    \item \textit{Processing bit operation:}
        The decoding symbol $\alpha_0^{(0)}$ is obtained after the operation for $t=0$.
        Depending on the value of $\alpha_0^{(0)}$, the decoder determines the next action as follows:
    \begin{itemize}
      \item If $\alpha_0^{(0)}=\bar{\mathbf{u}}\langle \mathcal{H}_{i,b}\rangle_{\ell_i}$,
        the decoding operation is immediately terminated and $r(\mathcal{H}_{i,b})=1$ is reported.
      \item If $\alpha_0^{(0)}\neq\epsilon$ and  $\alpha_0^{(0)}\neq\bar{\mathbf{u}}\langle \mathcal{H}_{i,b}\rangle_{\ell_i}$,
        the decoder stops and reports $r(\mathcal{H}_{i,b})=0$.
      \item If $\alpha_0^{(0)}=\epsilon$, the decoder repeats the operation at Step 1) again until the number of running iterations reaches $I_{\max}$.
    \end{itemize}
\end{enumerate}

During the operations at Steps 1) to 3),
the decoder early terminates the iterative operations and yields $r(\mathcal{H}_{i,b})=0$ as soon as any of the decoding symbols is updated to be a conflict $\eta$.

\subsection{Numerical Analysis via Density Evolution}

The performance of SCC and BP-SCC decoding schemes is numerically analyzed by density evolution (DE).
As usual, we assume that the all-zero codeword is transmitted without loss of generality.
Then, for each $i\in\mathcal{A}$, the proposed DE method tracks the probability that the decoder does not detect a conflict $\eta$ for the hypothesis $\mathcal{H}_{i,1}$.
The partial bit sequence $\bar{\mathbf{u}}\langle \mathcal{H}_{i,1} \rangle$ is generated by assuming ${u}_i=1$,
and the estimates appearing at the intermediate stages of the instant graph are accordingly determined by the successive cancellation procedure.
Hence, some estimates used in the repetition node (i.e. $\beta_k^{(t)}$ in Fig.~\ref{fig:component} (c))
are determined to be 1, violating the all-zero codeword assumption.
Nevertheless, if a conflict $\eta$ is not detected by the decoder,
it leads to a decoding error with probability $1/2$ due to the random selection for $r(\mathcal{H}_{i,0})=r(\mathcal{H}_{i,1})=1$.

Since the new symbol $\eta$ is added to the alphabet of decoding symbols,
the DE procedure is modified to take the new status into account.
Considering the operators $\boxplus$ and $\boxdot$,
we define two elementary functions handling probability mass functions (PMFs) in order to track the probability of the conflict $\eta$.
For two input PMFs $p_1, p_2:\mathbb{E}\to[0,1]$,
the DE function for SPC is denoted by $p_{\boxplus}=\psi_{\boxplus}(p_1,p_2)$,
where the output PMF $p_{\boxplus}:\mathbb{E}\to[0,1]$ is given by
\if\doccolumn1
\begin{equation}\nonumber
\begin{split}
&p_{\boxplus}[0]= p_1[0]\mdot p_2[0] + p_1[1]\mdot p_2[1];
~~~p_{\boxplus}[1] = p_1[0]\mdot p_2[1] + p_1[1]\mdot p_2[0]; \\
&p_{\boxplus}[\epsilon] =  p_1[\epsilon]\mdot (p_2[0]+p_2[1]+p_2[\epsilon]) + p_2[\epsilon]\mdot(p_1[0]+p_1[1]);
~~~p_{\boxplus}[\eta] = p_1[\eta] + p_2[\eta] - p_1[\eta]\mdot p_2[\eta].
\end{split}
\end{equation}
\else
\begin{equation}\nonumber
\begin{split}
&p_{\boxplus}[0]= p_1[0]\mdot p_2[0] + p_1[1]\mdot p_2[1];\\
&p_{\boxplus}[1] = p_1[0]\mdot p_2[1] + p_1[1]\mdot p_2[0];\\
&p_{\boxplus}[\epsilon] = p_1[\epsilon]\mdot (p_2[0]+p_2[1]+p_2[\epsilon]) + p_2[\epsilon]\mdot(p_1[0]+p_1[1]); \\
&p_{\boxplus}[\eta] = p_1[\eta] + p_2[\eta] - p_1[\eta]\mdot p_2[\eta].
\end{split}
\end{equation}
\fi
For a set of three or more PMFs $\{p_1,p_2,p_3,\ldots\}$,
we have
\begin{equation}\nonumber
\psi_{\boxplus}(\{p_1,p_2,p_3,\ldots\}) = \psi_{\boxplus}(p_1,\psi_{\boxplus}(p_2,\psi_{\boxplus}(p_3,\ldots)))
\end{equation}
because $\boxplus$ is commutative.
Similarly, for two input PMFs $p_1,p_2\in\mathbb{E}\to[0,1]$ and a binary value $b\in\mathbb{F}_2$,
the DE function for repetition is denoted by $p_{\boxdot}=\psi_{\boxdot}(p_1, p_2, b)$ in which the output PMF $p_{\boxdot}:\mathbb{E}\to[0,1]$ is computed by
\if\doccolumn1
\begin{equation}\nonumber
\begin{split}
&p_{\boxdot}[0] = p_1[b]\mdot (p_2[0] + p_2[\epsilon]) + p_1[\epsilon] \mdot p_2[0];
~~~p_{\boxdot}[1] = p_1[\check{b}] \mdot (p_2[1] + p_2[\epsilon]) + p_1[\epsilon] \mdot p_2[1]; \\
&p_{\boxdot}[\epsilon] = p_1[\epsilon] \mdot p_2[\epsilon];
~~~p_{\boxdot}[\eta] = p_1[b] \mdot p_2[1] + p_1[\check{b}] \mdot p_2[0] + p_1[\eta] (1-p_2[\eta]) + p_2[\eta].
\end{split}
\end{equation}
\else
\begin{equation}\nonumber
\begin{split}
&p_{\boxdot}[0] = p_1[b]\mdot (p_2[0] + p_2[\epsilon]) + p_1[\epsilon] \mdot p_2[0];\\
&p_{\boxdot}[1] = p_1[\check{b}] \mdot (p_2[1] + p_2[\epsilon]) + p_1[\epsilon] \mdot p_2[1];\\
&p_{\boxdot}[\epsilon] = p_1[\epsilon] \mdot p_2[\epsilon];\\
&p_{\boxdot}[\eta] = p_1[b] \mdot p_2[1] + p_1[\check{b}] \mdot p_2[0] + p_1[\eta] (1-p_2[\eta]) + p_2[\eta].\\
\end{split}
\end{equation}
\fi

Note that the DE operations are established under the cycle-free assumption,
so the results are not accurate for BP-SCC decoding over the graph with cycles especially when $I_{\max}>1$.
For this reason, we only consider $I_{\max}=1$ for BP-SCC decoding here.
Since no iteration is performed,
we only need to consider the forward update, that is, only $\alpha_k^{(t)}$ is updated at SPC and repetition nodes in Fig.~\ref{fig:component} (b) and (c).

Let $p_k^{(t)}$ be the PMF of the $k$-th VN at stage $t$.
We first initialize the PMFs at stage $n$ by setting $p_k^{(n)}[0]=1-p_{\epsilon}$, $p_k^{(n)}[1]=0$, $p_k^{(n)}[\epsilon]=p_{\epsilon}$, and $p_k^{(n)}[\eta]=0$ for all $k\in\mathbb{Z}_N$.
Then, by modeling the decoding operations, the PMFs at stages $t\in\{0:n-1\}$ are sequentially computed in descending order of $t$.

As the first step of the stage-$t$ operation, the PMFs at stage $t+1$ are updated through FCCNs in the BP-SCC decoder.
This procedure by FCCNs is just skipped for SCC decoding.
Let $q_{j\to k}$ denote the PMF of the C2V message from CN $c_j$ to VN $x_k^{(t+1)}$.
Based on \eqref{eq:c2v}, it is computed as
\begin{equation}
    q_{j\to k} \leftarrow  \psi_{\boxplus} \left( p'_{j}, \psi_{\boxplus}\left(\left\{p_l^{(k+1)} \mmid l\in\mathcal{V}_j \backslash \{k\}\right\} \right) \right),
\end{equation}
where $p'_j:\mathbb{E}\to[0,1]$ is the PMF for the deterministic constraint value $\phi_j$, i.e., $p'_{j}[\delta]=1$ if $\delta=\phi_j$ and $p'_{j}[\delta]=0$ otherwise.
Since there may be cycles between FCCNs and VNs,
the PMF update rule is modified to take only the most influential C2V message in order to mitigate the cycle effect.
We find CN $c_{j'}$ whose C2V message has the highest probability of the conflict $\eta$, that is,
its index $j'$ is given by
\begin{equation}
    j' = \argmax\nolimits_{j\in\{0:|\mathcal{L}_{\ell_i+1,t+1}|-1\}} q_{j\to k}[\eta].
\end{equation}
The PMF of the decoding symbol $\alpha_k^{(t+1)}$ is then updated by $p_k^{(t+1)} \leftarrow \psi_{\boxdot}\big(p_k^{(t+1)}, q_{j' \to k}  \big)$.

Based on the decoding symbols at stage $t+1$,
either an SPC or a repetition node operation is applied depending on the processing bit index $\ell_i$ to calculate $p_k^{(t)}$ as
\begin{equation}
p_k^{(t)} \leftarrow \begin{cases}
\psi_{\boxplus}\left(p_k^{(t+1)},p_{k+2^t}^{(t+1)}\right), &\text{if } \langle \ell_i \rangle_t = 0,\\
\psi_{\boxdot}\left(p_k^{(t+1)},p_{k+2^t}^{(t+1)},\beta_k^{(t)}\right), &\text{if } \langle \ell_i \rangle_t = 1,\\
\end{cases}
\end{equation}
In this way, the conflict probability is gradually computed by reflecting all the bits in $\bar{\mathbf{u}}\langle \mathcal{H}_{i,1}\rangle$, and the corresponding final result is given in $p_0^{(0)}$.

It is definite that $\mathcal{H}_{i,0}$ always passes the conflict check by BP-SCC decoding under the assumption of the all-zero codeword transmission.
If no conflict is also observed for $\mathcal{H}_{i,1}$,
$\hat{u}_i$ is determined to be $1$ with probability $1/2$ due to the random selection.
According to the path selection described in Subsection~\ref{sec:csp}, the bit error probability of $u_i$ is given by
\begin{equation}
P_{b}(i) = \frac{1}{2} \left(p_0^{(0)}\left[\bar{\mathbf{u}}\langle \mathcal{H}_{i,1} \rangle_{\ell_i}\right] + p_0^{(0)}[\epsilon] \right),
\end{equation}
where $p_0^{(0)}\left[\bar{\mathbf{u}}\langle \mathcal{H}_{i,1} \rangle_{\ell_i}\right] + p_0^{(0)}[\epsilon]$ is the probability that $r(\mathcal{H}_{i,1})=1$. Finally, the BLER is estimated by
\begin{equation}
P_{B} \approx 1 - \prod_{i\in\mathcal{A}} \left(1-P_{b}(i)\right).
\end{equation}

\if\doccolumn2
\begin{figure*}[t]
    \centering
    \centering
    \subfloat[$N=64,~K=32$]{
        \includegraphics[width=\subsubfigwidth]{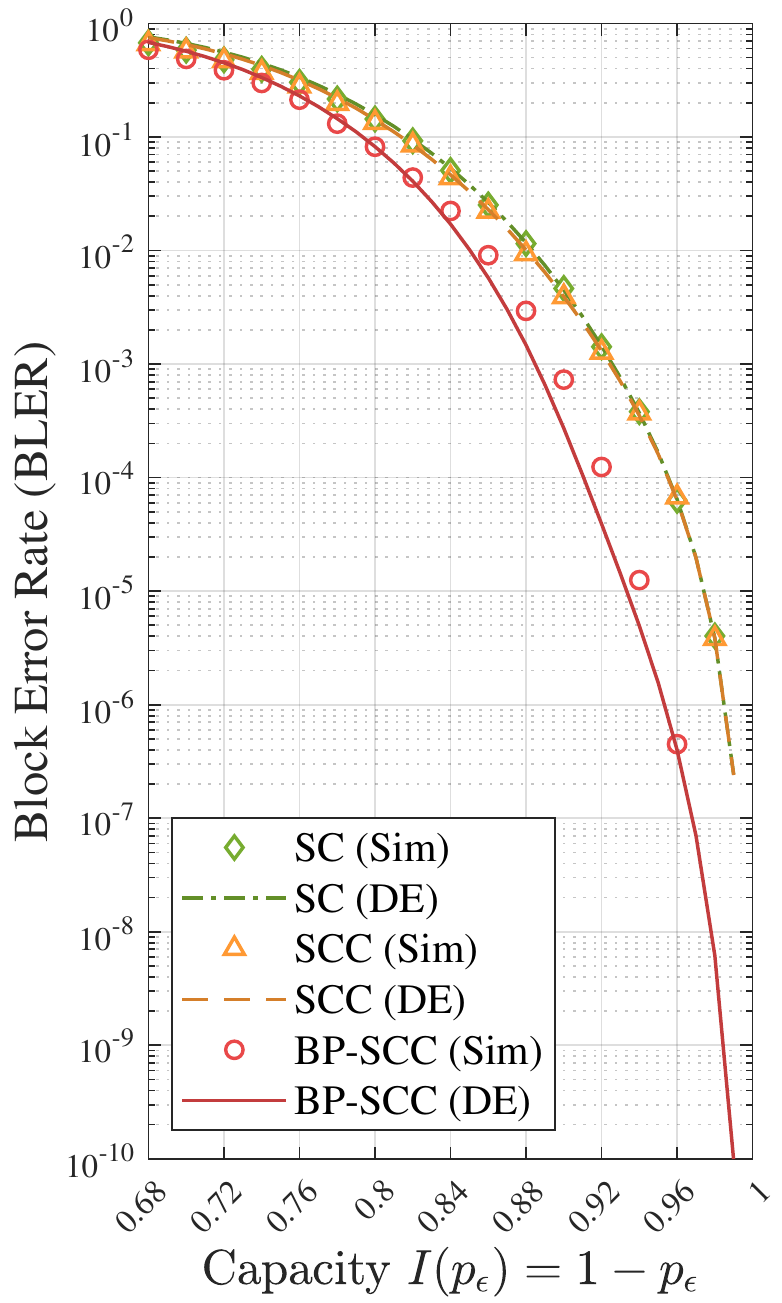}}
    \subfloat[$N=128,~K=64$]{
        \includegraphics[width=\subsubfigwidth]{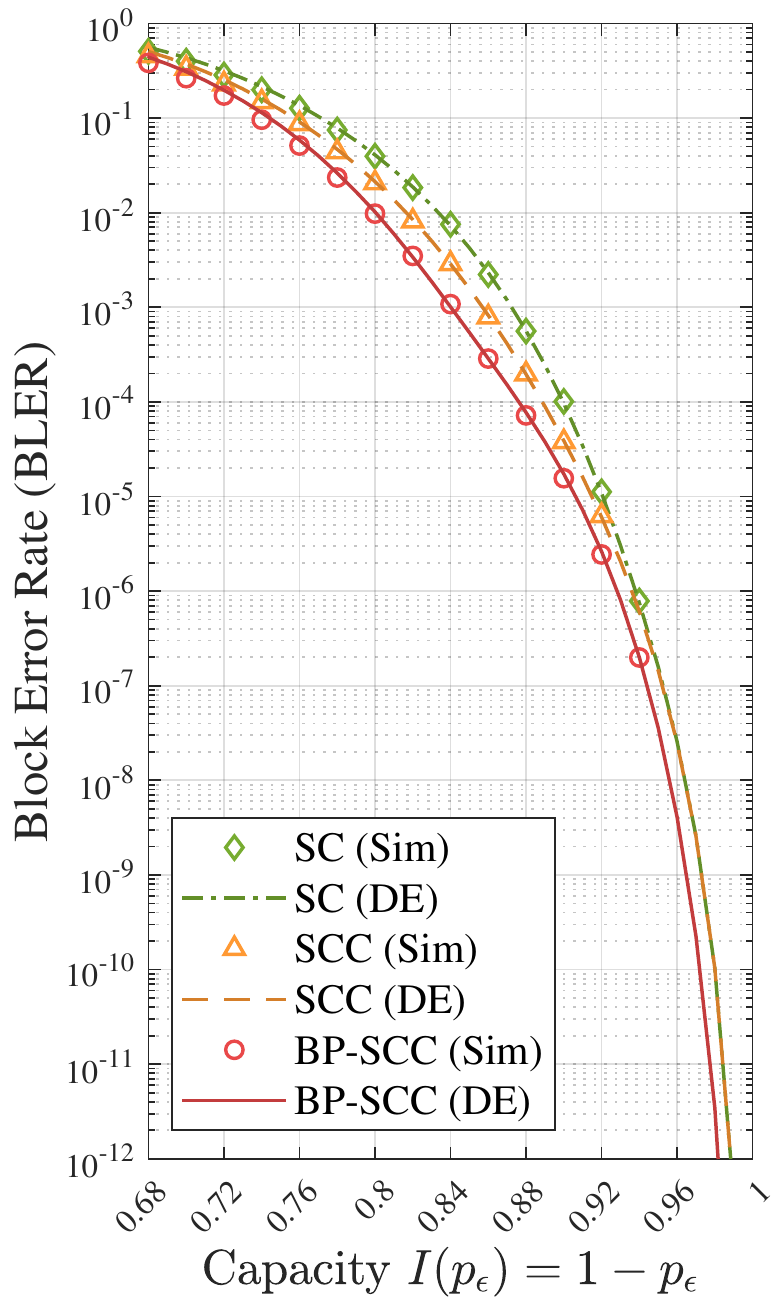}}
    \subfloat[$N=256,~K=128$]{
        \includegraphics[width=\subsubfigwidth]{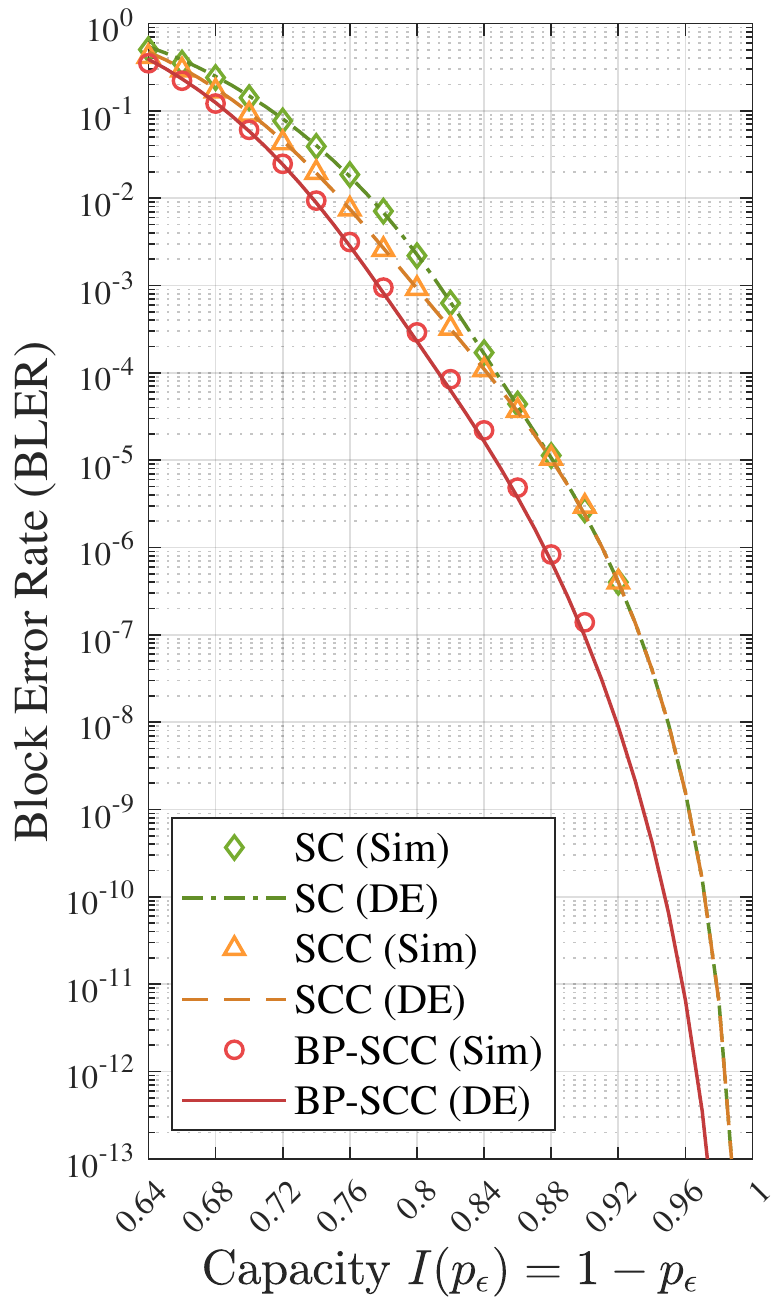}}
    \subfloat[$N=512,~K=256$]{
        \includegraphics[width=\subsubfigwidth]{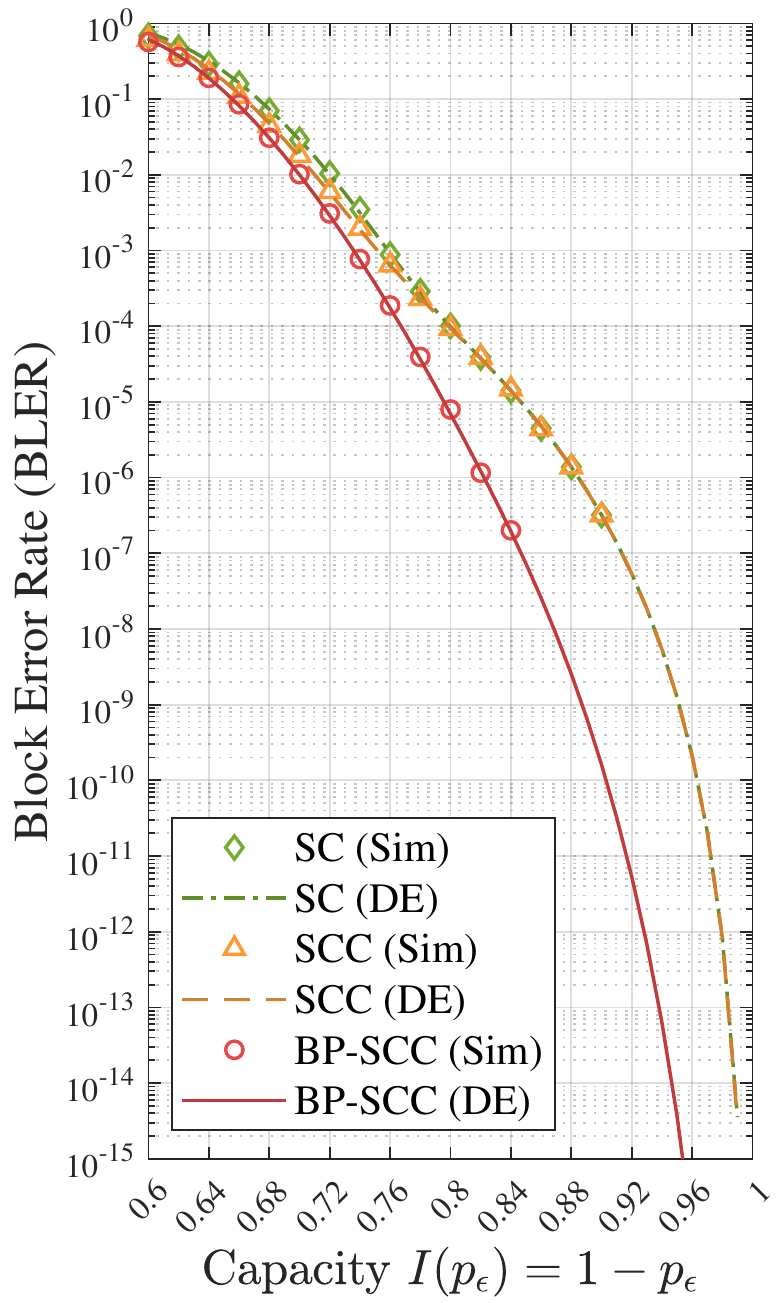}} \\
  \caption{Numerical analysis and simulation results for the half-rate 3GPP NR polar codes having code lengths 64, 128, 256, and 512.
    The number of maximum iterations $I_{\max}$ is set to one for BP-SCC decoding.}
    \label{fig:de_result}
\end{figure*}
\else
\begin{figure}[t!]
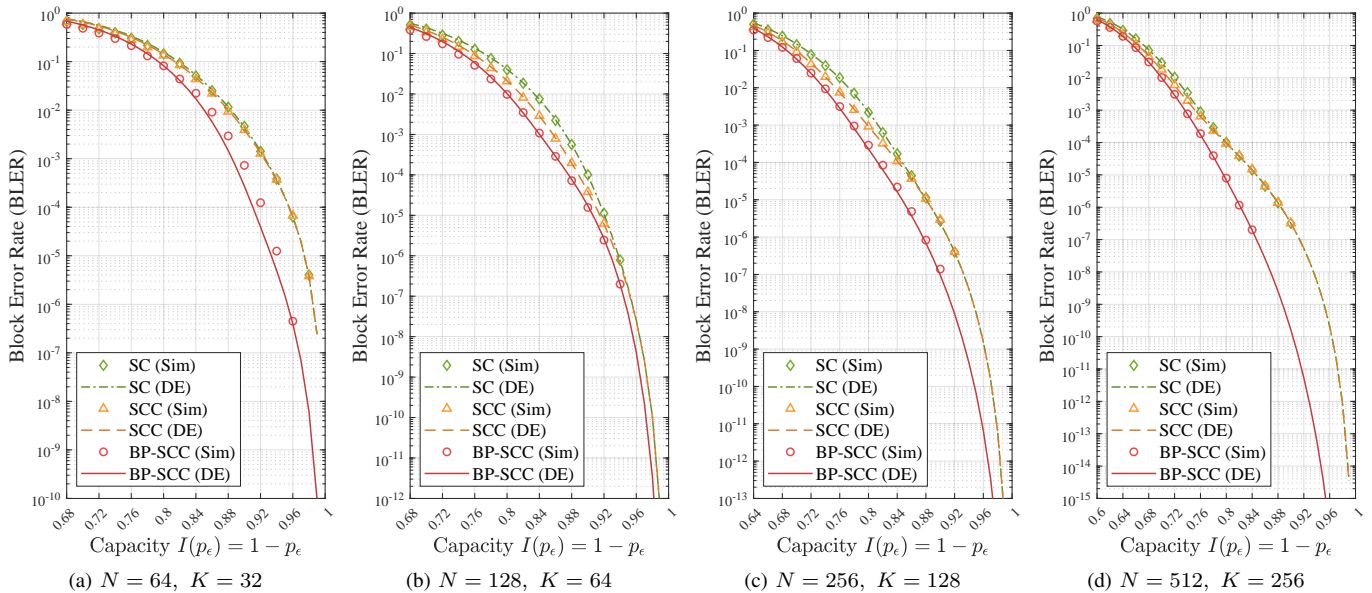

    \centering
    \subfloat[$N=64,~K=32$]{
        \includegraphics[width=\subsubfigwidth]{Fig_DE_N64.pdf}}
    \subfloat[$N=128,~K=64$]{
        \includegraphics[width=\subsubfigwidth]{Fig_DE_N128.pdf}}
    \subfloat[$N=256,~K=128$]{
        \includegraphics[width=\subsubfigwidth]{Fig_DE_N256.pdf}}
    \subfloat[$N=512,~K=256$]{
        \includegraphics[width=\subsubfigwidth]{Fig_DE_N512.pdf}} \\
  \caption{Numerical analysis and simulation results for the half-rate 3GPP NR polar codes having code lengths 64, 128, 256, and 512.
    The number of maximum iterations $I_{\max}$ is set to one for BP-SCC decoding.}
  \label{fig:de_result}
\end{figure}
\fi

In Fig.~\ref{fig:de_result}, the numerical analysis results derived by the proposed DE technique are shown and compared with
practical simulation results.
We consider four half-rate polar codes of length $N\in\{64, 128, 256, 512\}$.
The 5G NR polar code construction method \cite{TS38212} is applied to generate these codes,
in which the 11-bit CRC code with generator polynomial $g(D)=D^{11}+D^{10}+D^{9}+D^{5}+1$ is concatenated as an outer coding scheme.
The eleven parity bits are generated by the CRC encoder and appended to the input vector $\mathbf{a}$ as shown in Fig.~\ref{fig:block} (a).
Three decoding schemes, SC, SCC, and BP-SCC are analyzed by DE and numerically evaluated,
where $I_{\max}$ is set to one for BP-SCC decoding.
As depicted in each subfigure,
the proposed numerical analysis method well estimates the practical BLER performance,
which shows that the behavior of the proposed decoding schemes is analytically tractable.

\subsection{Numerical Results}

We evaluate and compare four decoding algorithms by numerical experiments in this subsection.
Their characteristics are given as follows:
\begin{itemize}
  \item[-] \textit{SC decoding:}
    The original SC decoding algorithm \cite{Arikan2009} is considered, where no FC is entailed.
    The SC decoder estimates each bit based on the original likelihood \eqref{eq:org_subchannel},
    where the DFS is thus used as a tree search method.
  \item[-] \textit{SCC decoding:}
    The FCs close to the target bit are directly used by establishing hypotheses,
    while those behind the next information bit are not taken into account.
    In the case that the processing bit index $\ell_i$ is identical to the target bit index $i$,
    standard SC decoding is performed rather than SCC decoding.
  \item[-] \textit{BP-SCC decoding:}
    All FCs are exploited to improve the decoding of the target information bit.
    The FCs before the next information bit are incorporated by SCC decoding approach,
    while the others are employed by the subgraph-based conversion rule introduced in Subsection~\ref{sec:subgraph}.
    Unlike SCC decoding, simple SC decoding cannot be performed even in the case that $\ell_i=i$,
    because the FCs beyond the next information bit need to be checked.
    Here, we set $I_{\max}=5$ to achieve better performance.
    The decoder immediately stops and reports a decoding failure when it encounters a dead-end in the tree search.
  \item[-] \textit{BP-SCC-SBJ decoding:}
    In addition to the above BP-SCC decoding algorithm,
    the SBJ method is applied in order to efficiently avoid inconsistent partial solutions.
    The details about tree search with SBJ are introduced in Subsection~\ref{sec:csp}.
\end{itemize}

\if\doccolumn1
\begin{figure*}[t!]
    \centering
    \subfloat[$N=64,~K=32,~w_{\text{min}}(\mathbf{G}')=8,~w_{\text{min}}(\mathbf{TG})=20$]{
        \includegraphics[width=\subfigwidth]{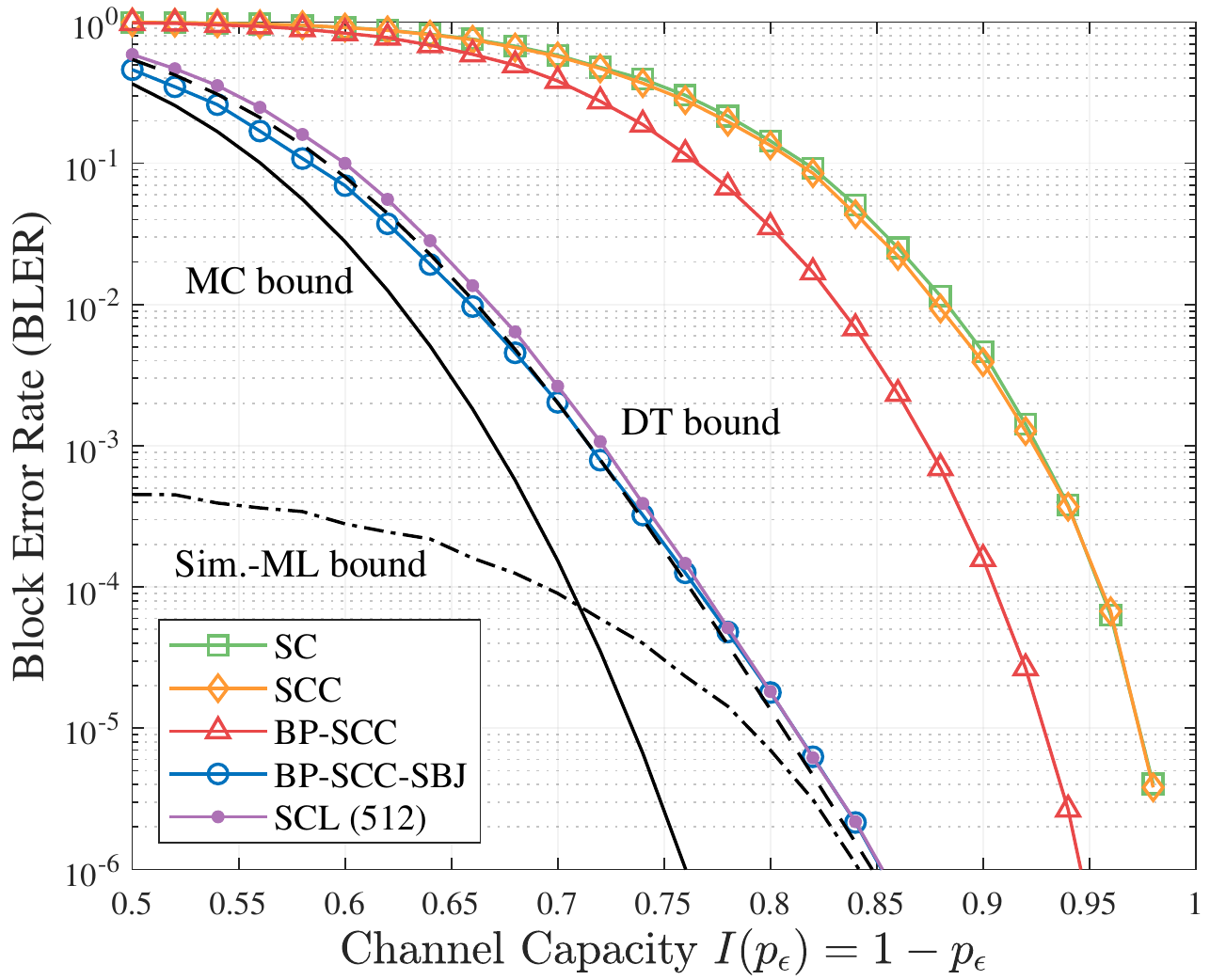}}\hfill
    \subfloat[$N=128,~K=64,~w_{\text{min}}(\mathbf{G}')=8,~w_{\text{min}}(\mathbf{TG})=40$]{
        \includegraphics[width=\subfigwidth]{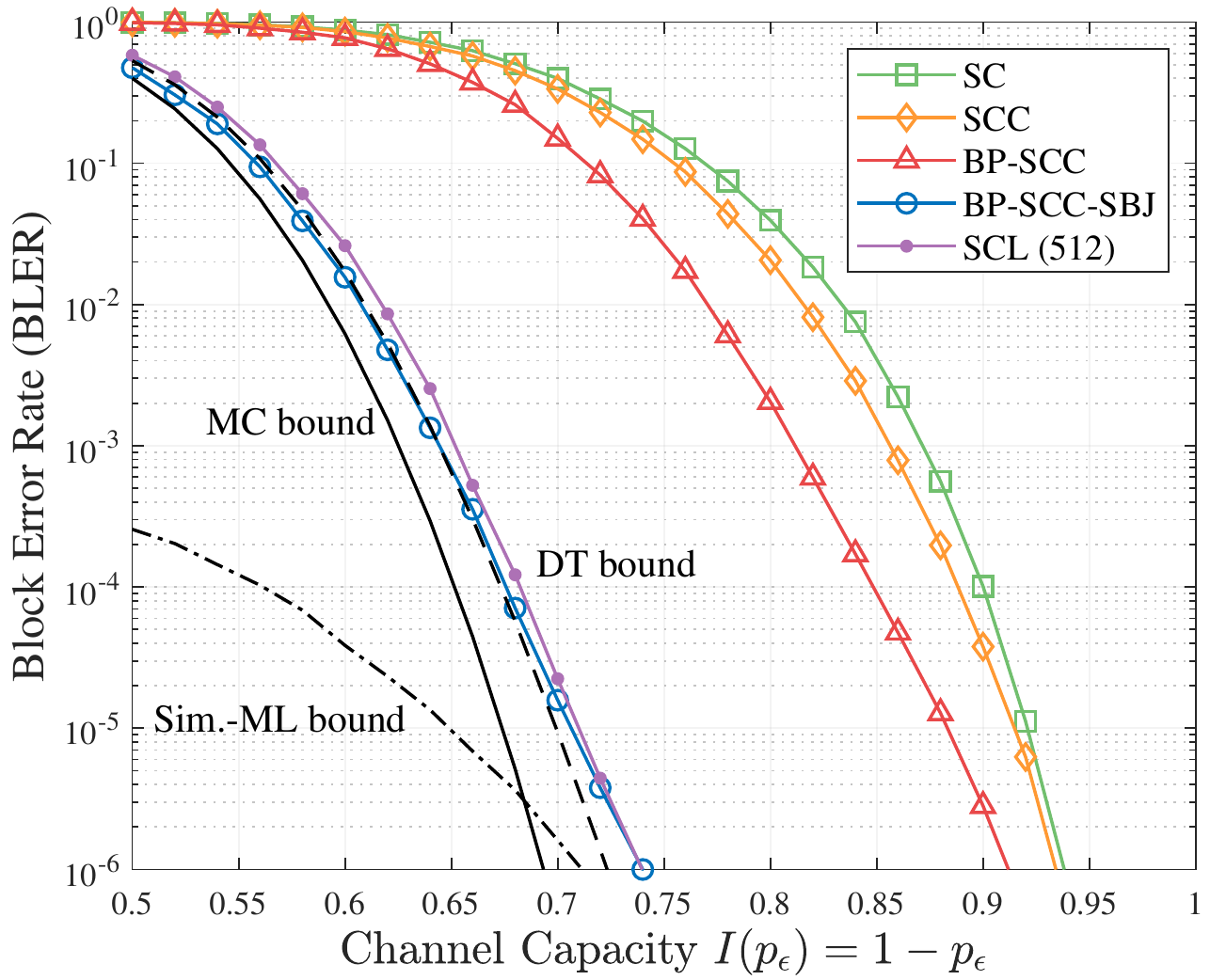}} \\
    \subfloat[$N=256,~K=128,~w_{\text{min}}(\mathbf{G}')=8,~w_{\text{min}}(\mathbf{TG})=72$]{
        \includegraphics[width=\subfigwidth]{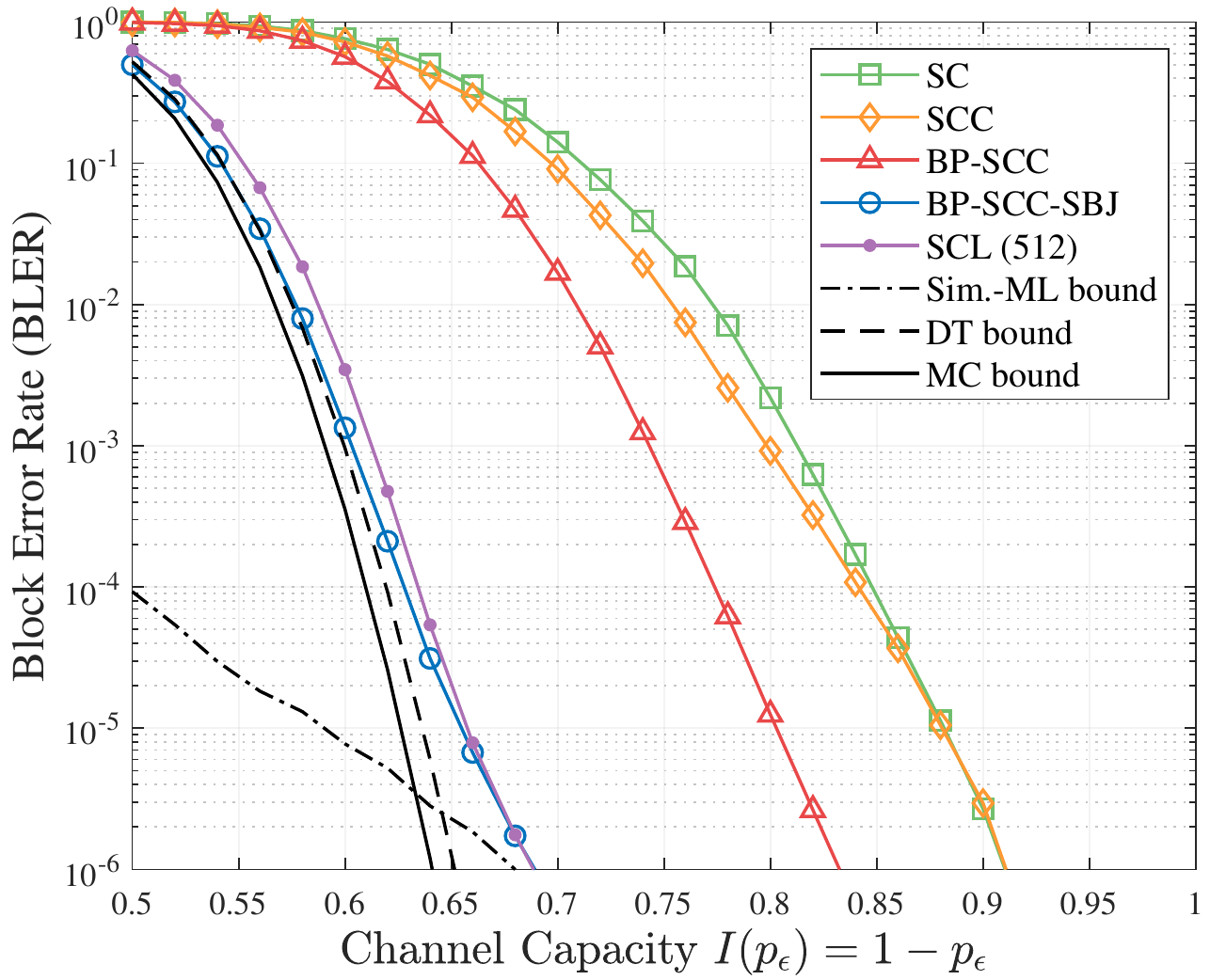}}\hfill
    \subfloat[$N=512,~K=256,~w_{\text{min}}(\mathbf{G}')=8,~w_{\text{min}}(\mathbf{TG})=128$]{
        \includegraphics[width=\subfigwidth]{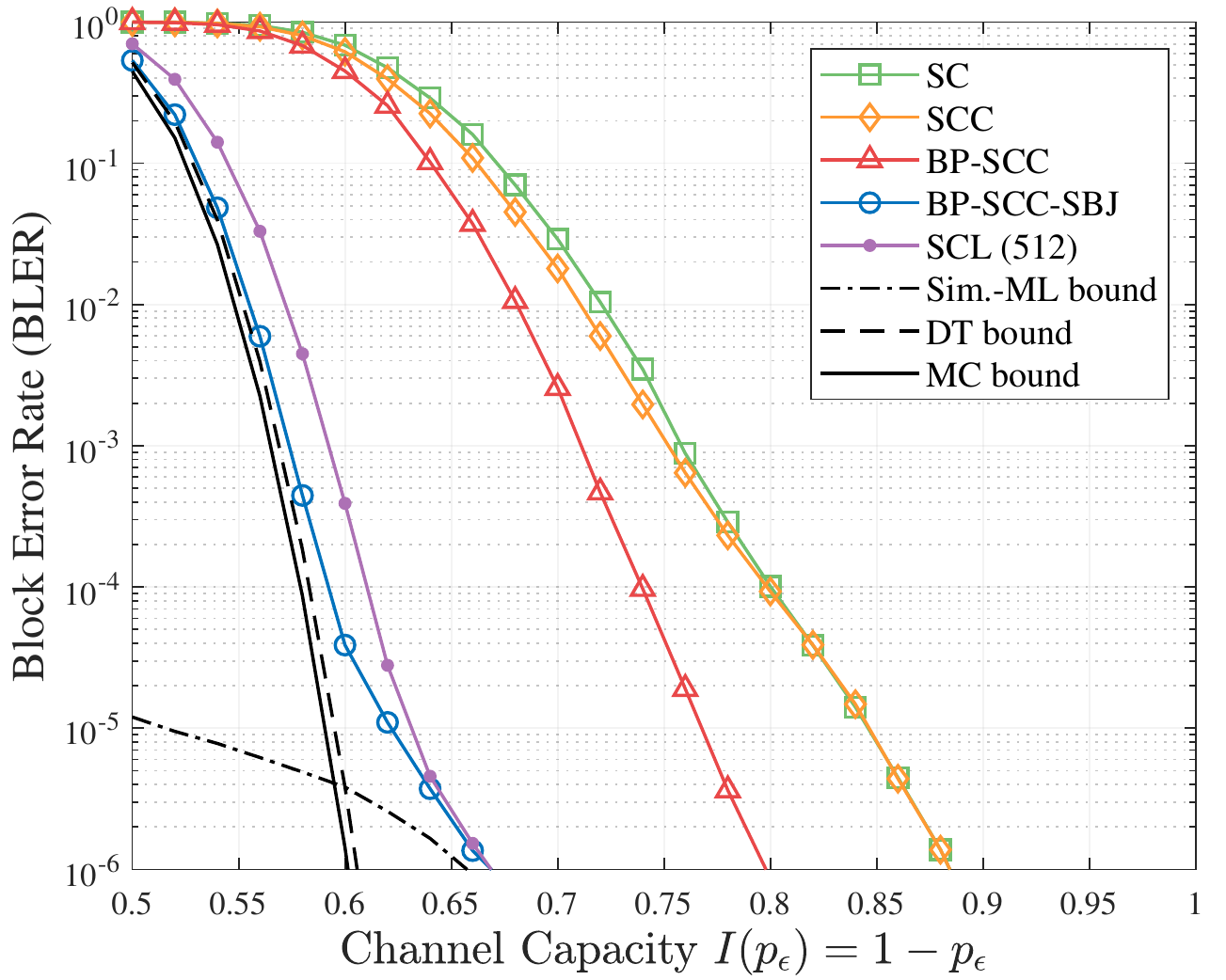}} \\
  \caption{Erasure recovery performance of five decoders over the BEC.
      The 3GPP NR uplink polar code construction scheme specified in \cite{TS38212} is used,
      where the 11-bit CRC code is concatenated.
      In each subfigure, the simulation-based ML bound, the DT bound, and the MC bound are also presented as a dash-dotted line, a dashed line, and a solid line without markers, respectively.}
  \label{fig:bler}
\end{figure*}
\else
\begin{figure*}[t!]
    \centering
    \subfloat[$N=64,~K=32,~w_{\text{min}}(\mathbf{G}')=8,~w_{\text{min}}(\mathbf{TG})=20$]{
        \includegraphics[width=\subfigwidth]{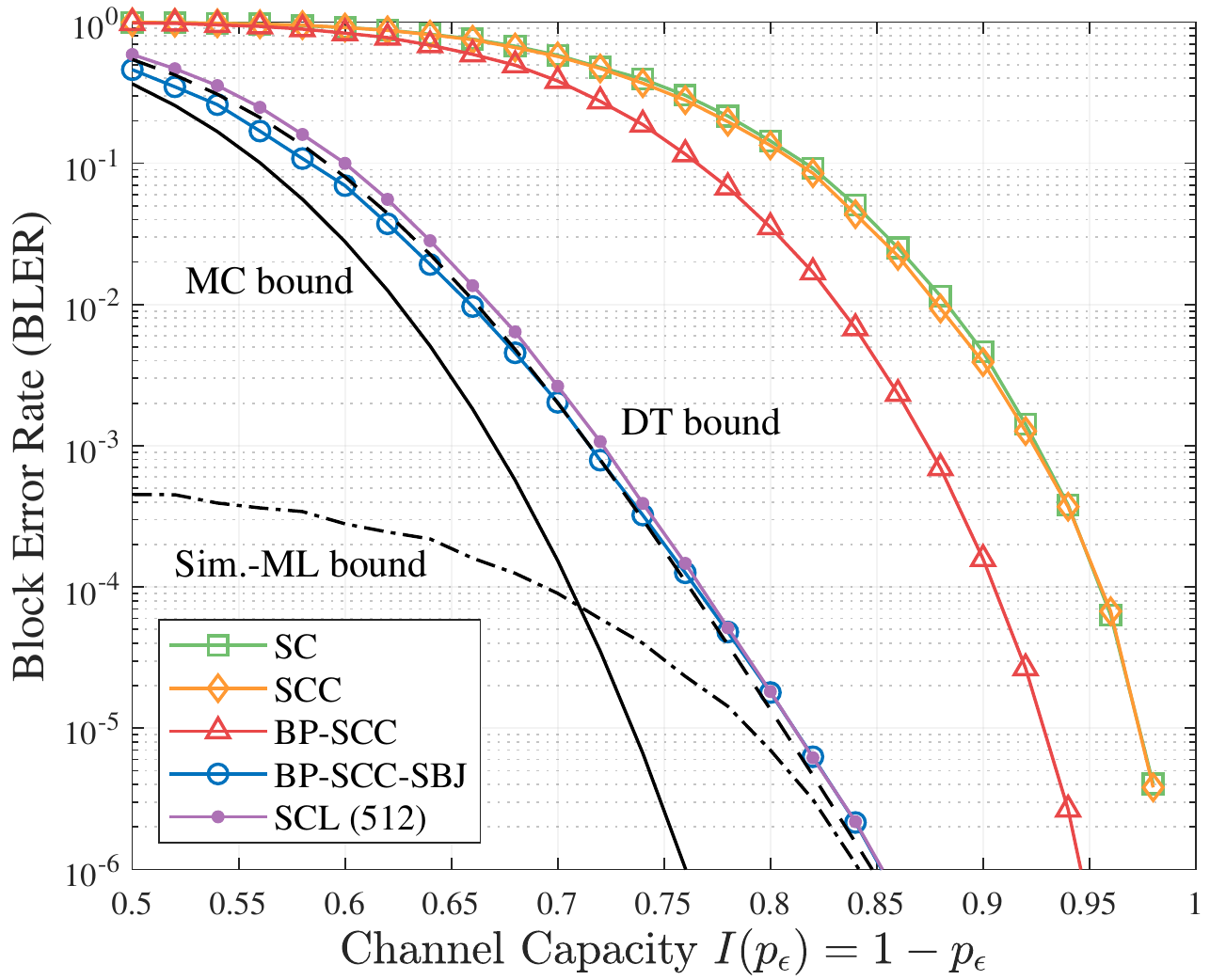}}\hfill
    \subfloat[$N=128,~K=64,~w_{\text{min}}(\mathbf{G}')=8,~w_{\text{min}}(\mathbf{TG})=40$]{
        \includegraphics[width=\subfigwidth]{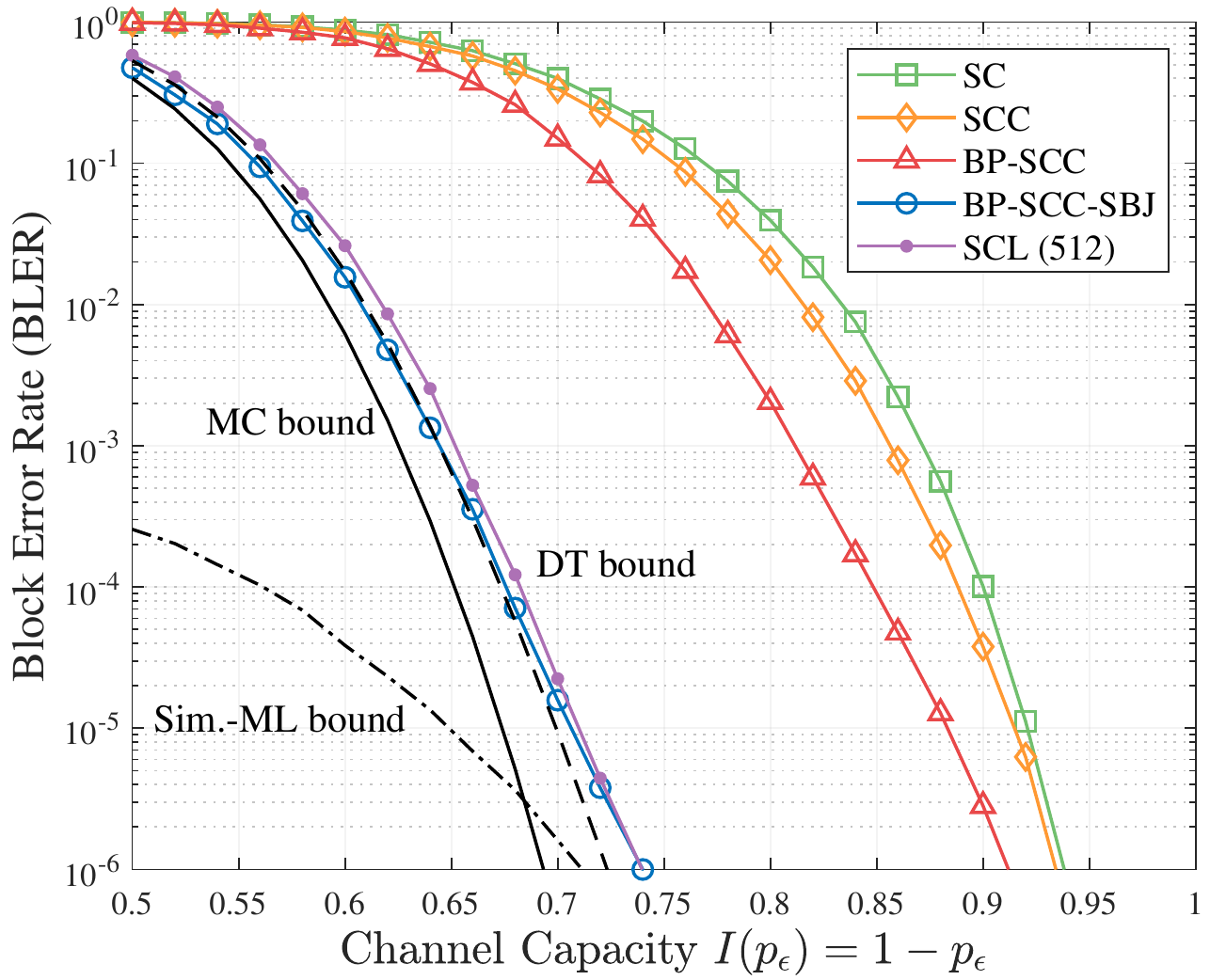}} \\
    \subfloat[$N=256,~K=128,~w_{\text{min}}(\mathbf{G}')=8,~w_{\text{min}}(\mathbf{TG})=72$]{
        \includegraphics[width=\subfigwidth]{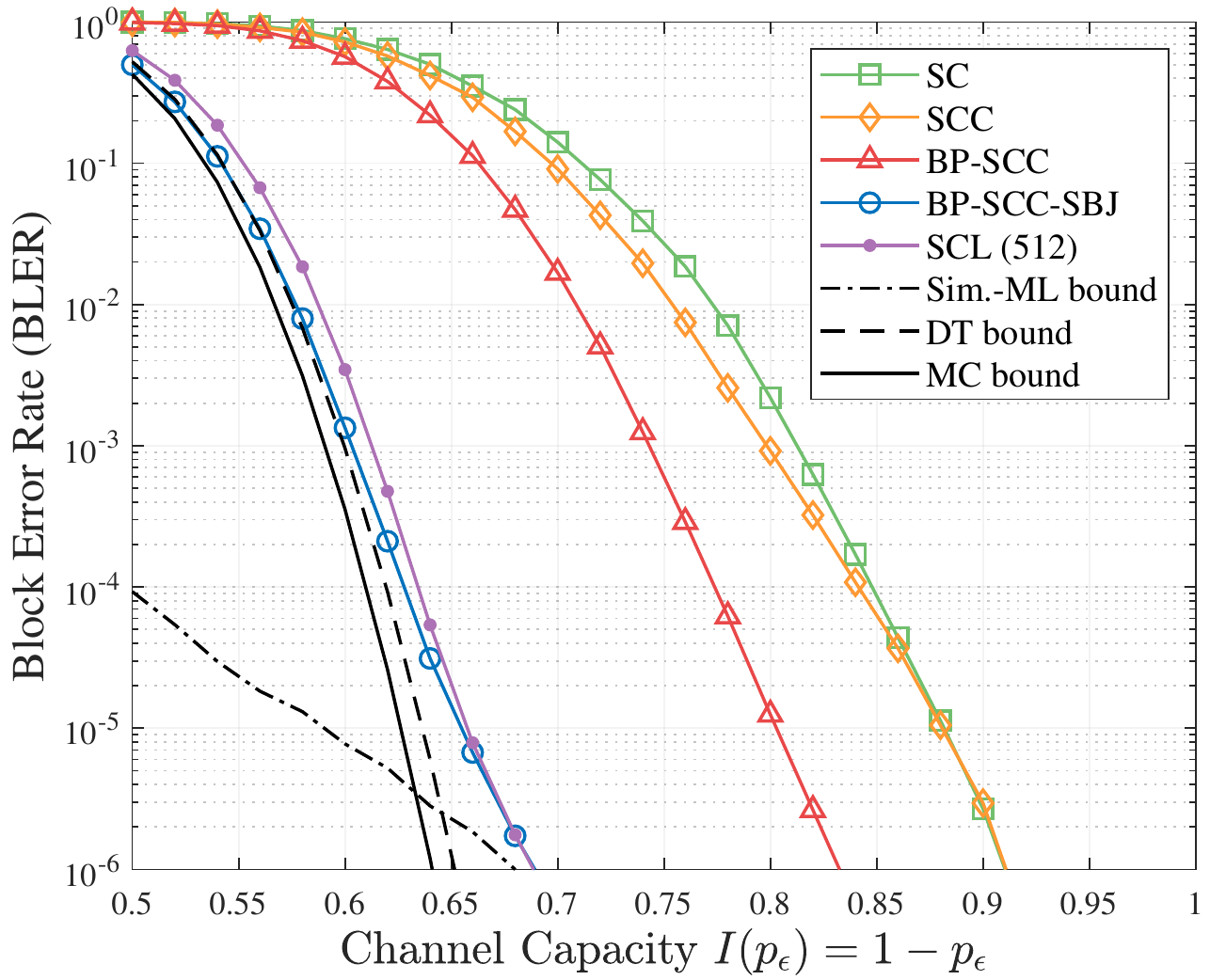}}\hfill
    \subfloat[$N=512,~K=256,~w_{\text{min}}(\mathbf{G}')=8,~w_{\text{min}}(\mathbf{TG})=128$]{
        \includegraphics[width=\subfigwidth]{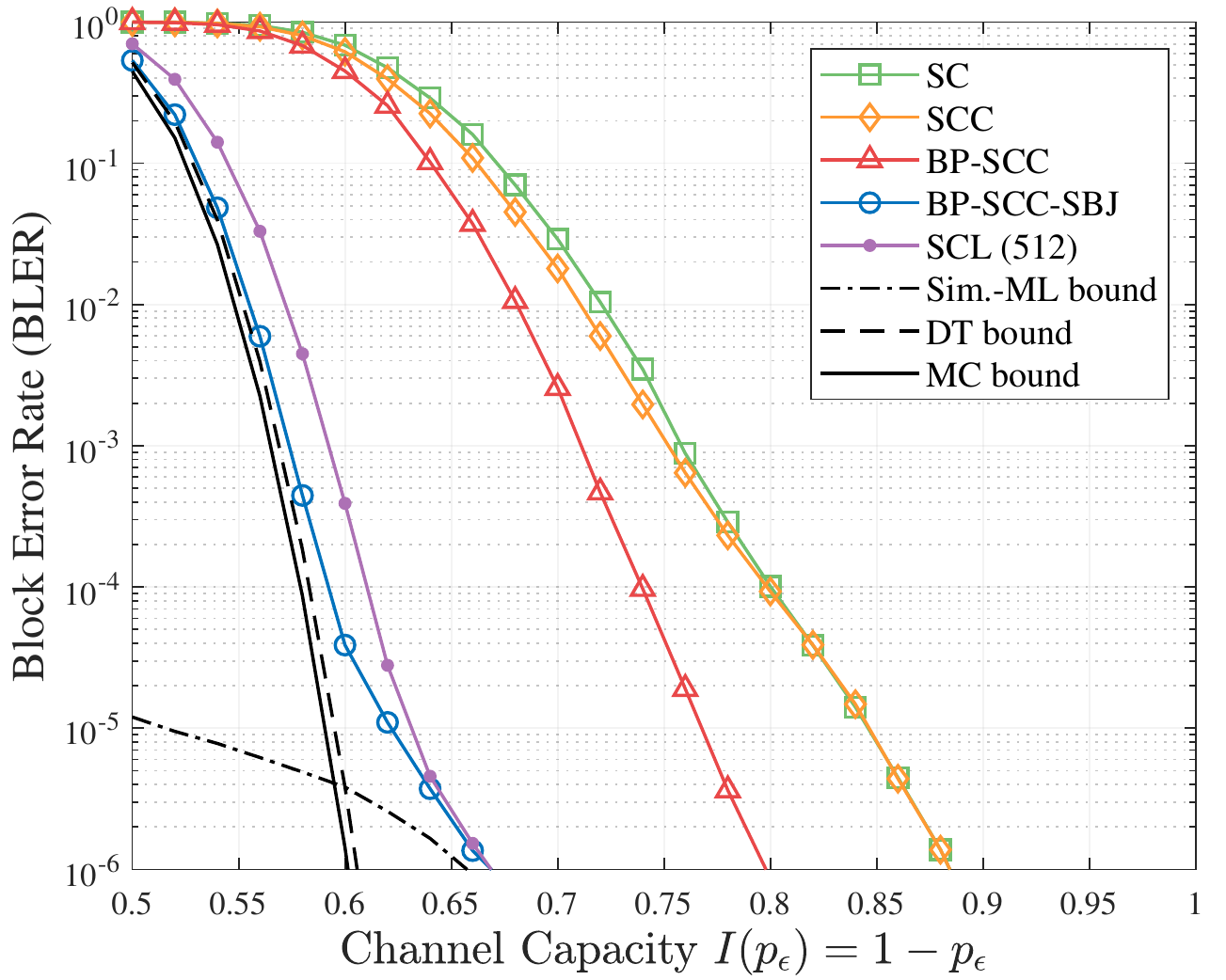}} \\
  \caption{Erasure recovery performance of five decoders over the BEC.
      The 3GPP NR uplink polar code construction scheme specified in \cite{TS38212} is used,
      where the 11-bit CRC code is concatenated.
      In each subfigure, the simulation-based ML bound, the DT bound, and the MC bound are also presented as a dash-dotted line, a dashed line, and a solid line without markers, respectively.}
    \label{fig:bler}
\end{figure*}
\fi

\if\doccolumn1
\begin{figure*}[t!]
    \centering
    \subfloat[$N=64,~K=32,~w_{\text{min}}(\mathbf{G}')=8,~w_{\text{min}}(\mathbf{TG})=20$]{
        \includegraphics[width=\subfigwidth]{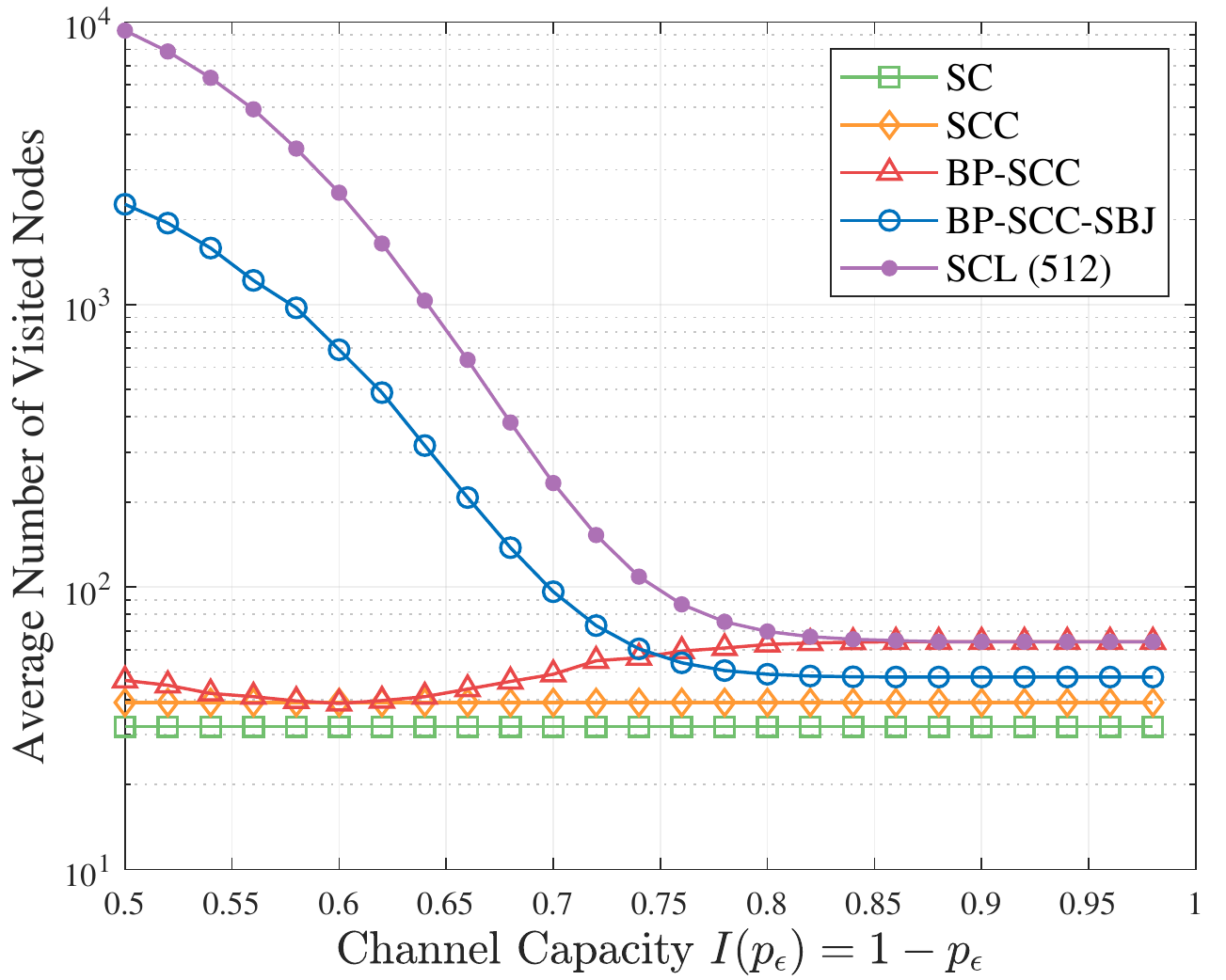}}\hfill
    \subfloat[$N=128,~K=64,~w_{\text{min}}(\mathbf{G}')=8,~w_{\text{min}}(\mathbf{TG})=40$]{
        \includegraphics[width=\subfigwidth]{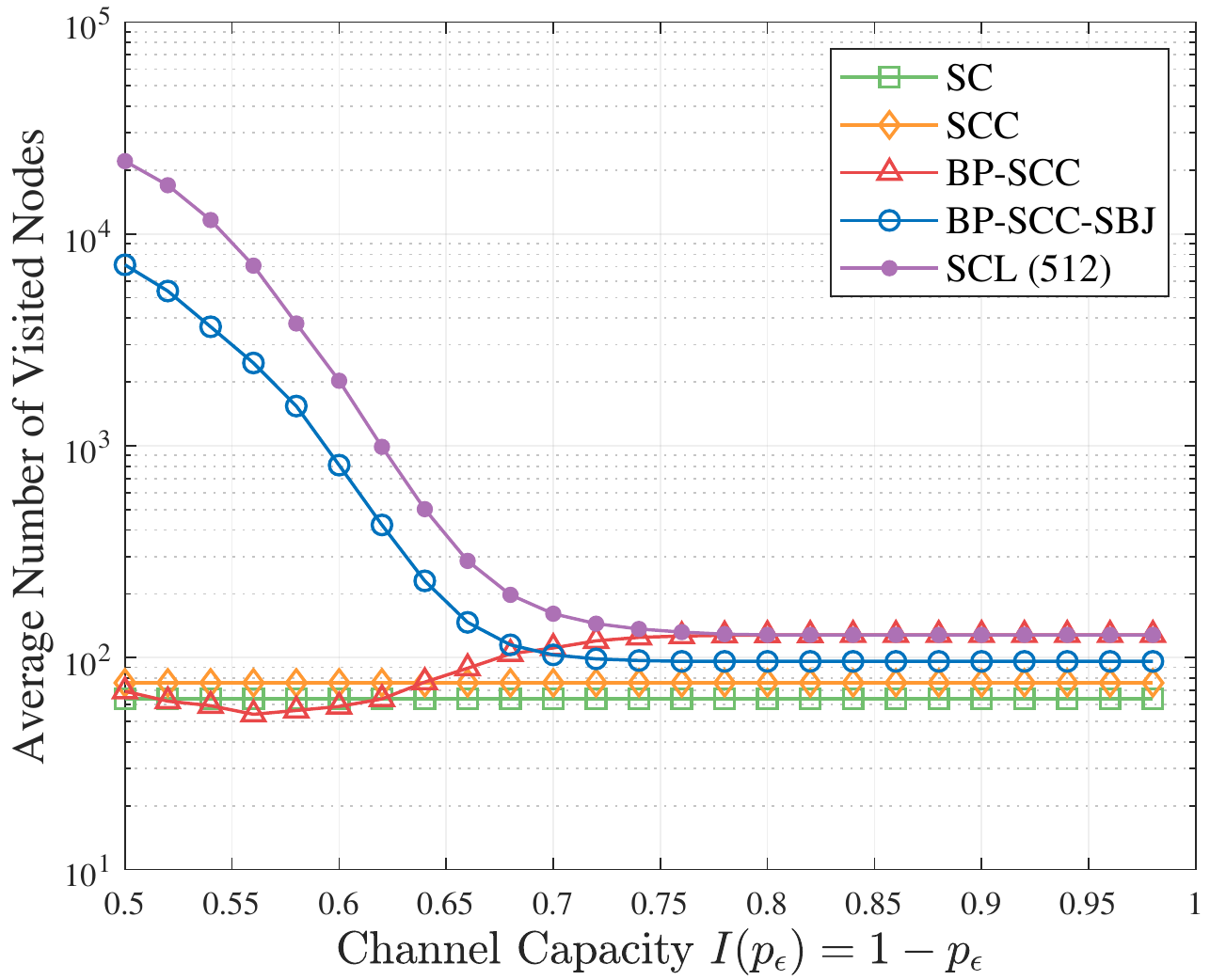}} \\
    \subfloat[$N=256,~K=128,~w_{\text{min}}(\mathbf{G}')=8,~w_{\text{min}}(\mathbf{TG})=72$]{
        \includegraphics[width=\subfigwidth]{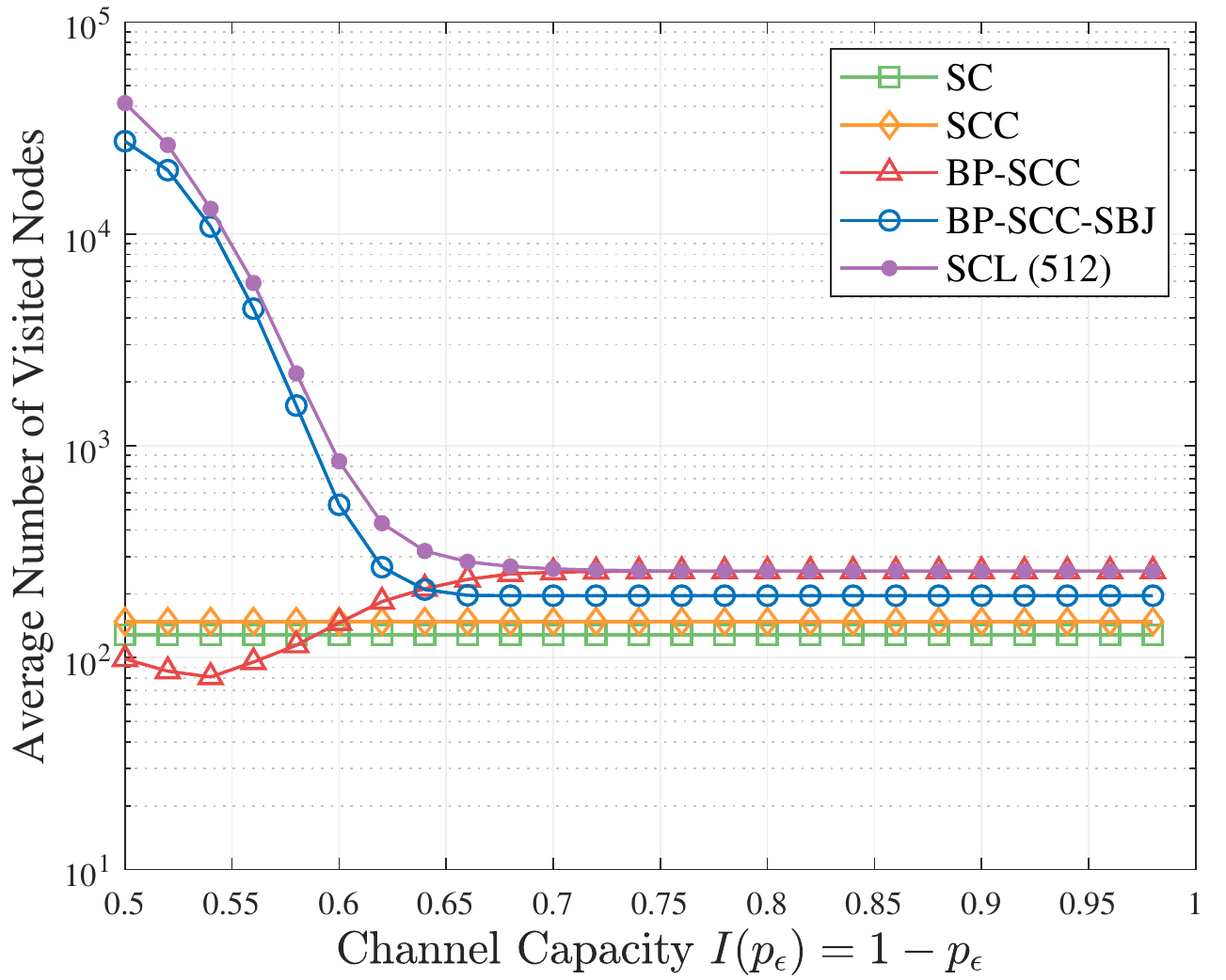}}\hfill
    \subfloat[$N=512,~K=256,~w_{\text{min}}(\mathbf{G}')=8,~w_{\text{min}}(\mathbf{TG})=128$]{
        \includegraphics[width=\subfigwidth]{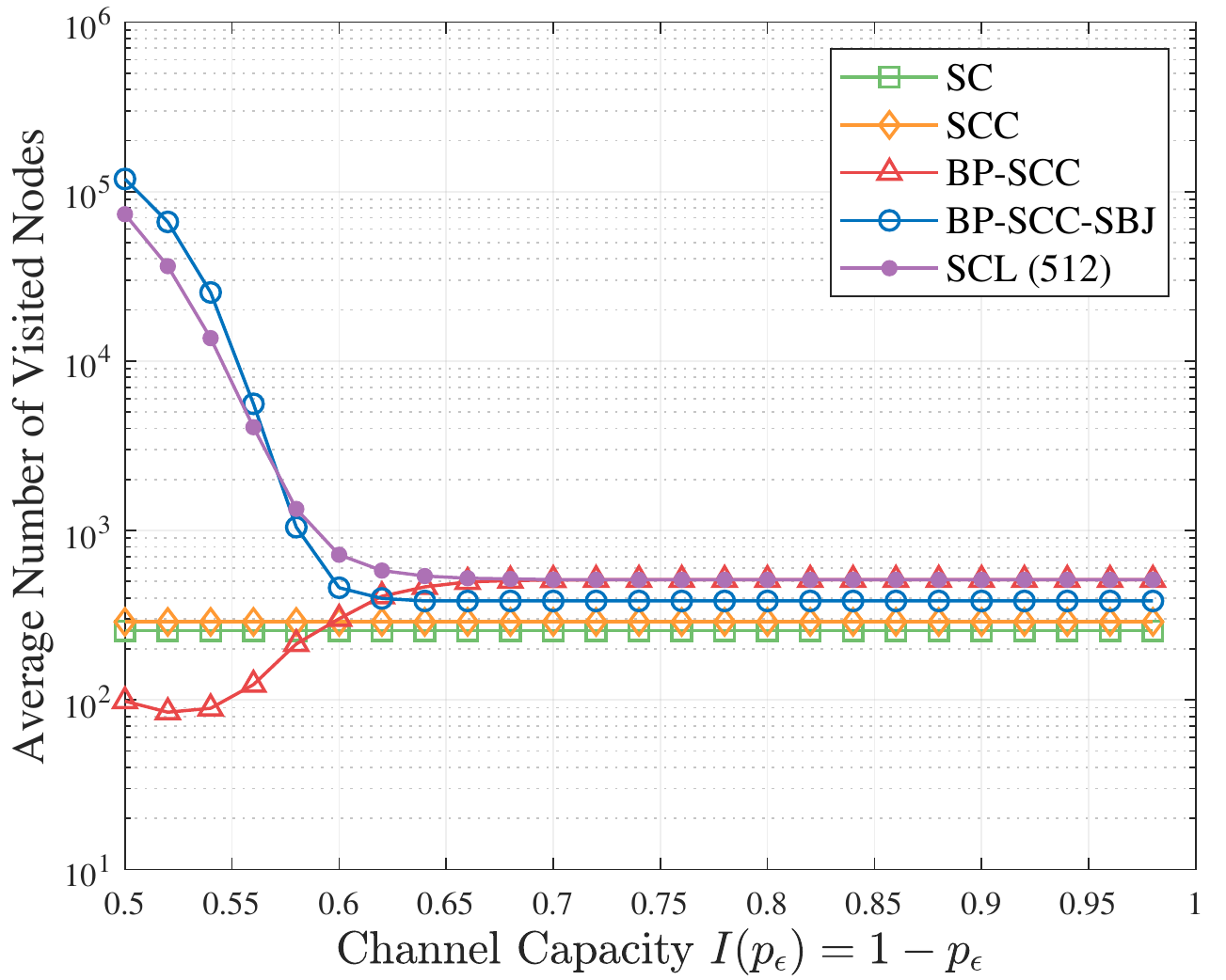}} \\
  \caption{Average number of visited nodes in the binary tree for five decoders over the BEC.
  The 3GPP NR uplink polar code construction scheme specified in \cite{TS38212} is used, where the 11-bit CRC code is concatenated.}
  \label{fig:anv}
\end{figure*}
\else

\begin{figure*}[t!]
    \centering
    \subfloat[$N=64,~K=32,~w_{\text{min}}(\mathbf{G}')=8,~w_{\text{min}}(\mathbf{TG})=20$]{
        \includegraphics[width=\subfigwidth]{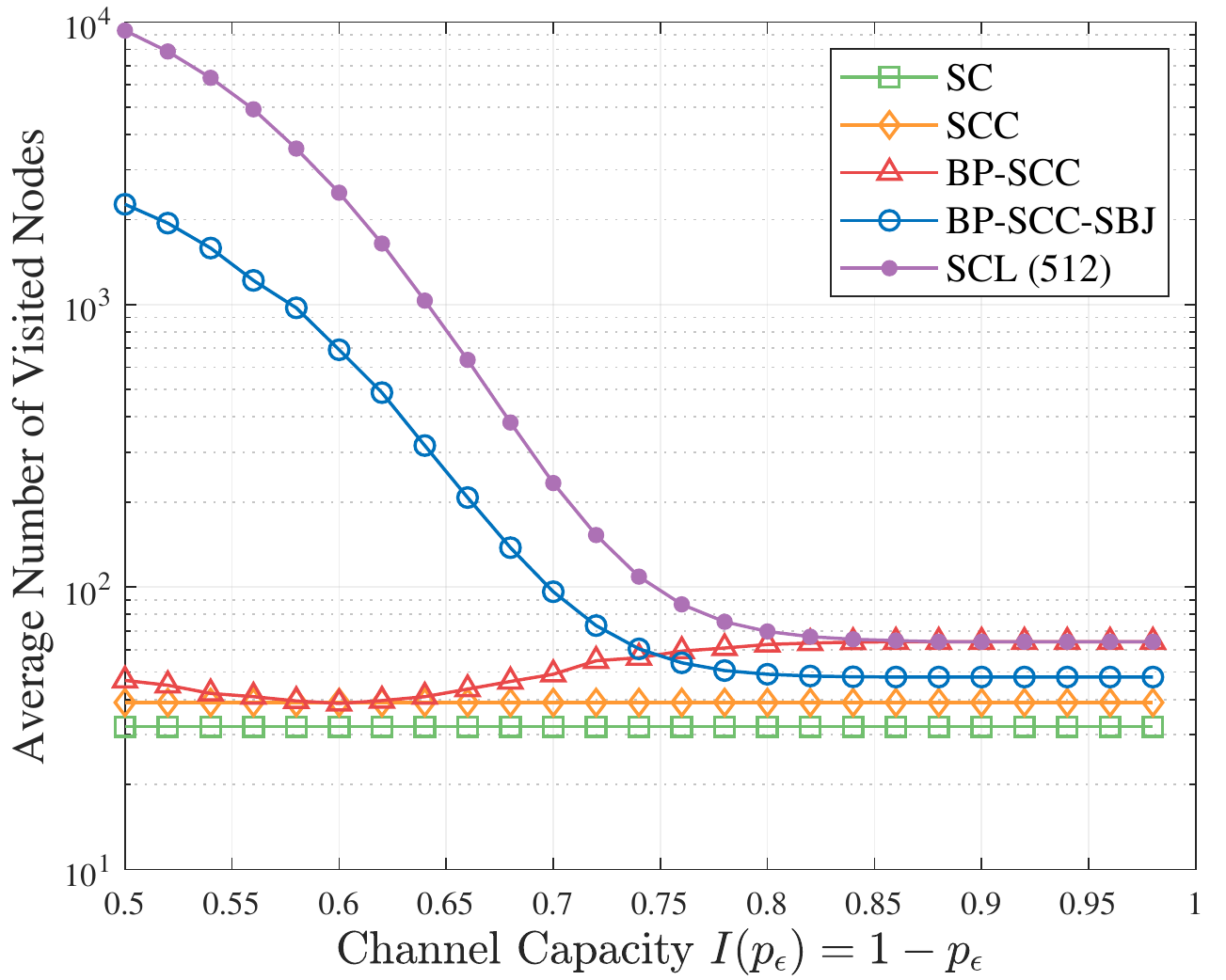}}\hfill
    \subfloat[$N=128,~K=64,~w_{\text{min}}(\mathbf{G}')=8,~w_{\text{min}}(\mathbf{TG})=40$]{
        \includegraphics[width=\subfigwidth]{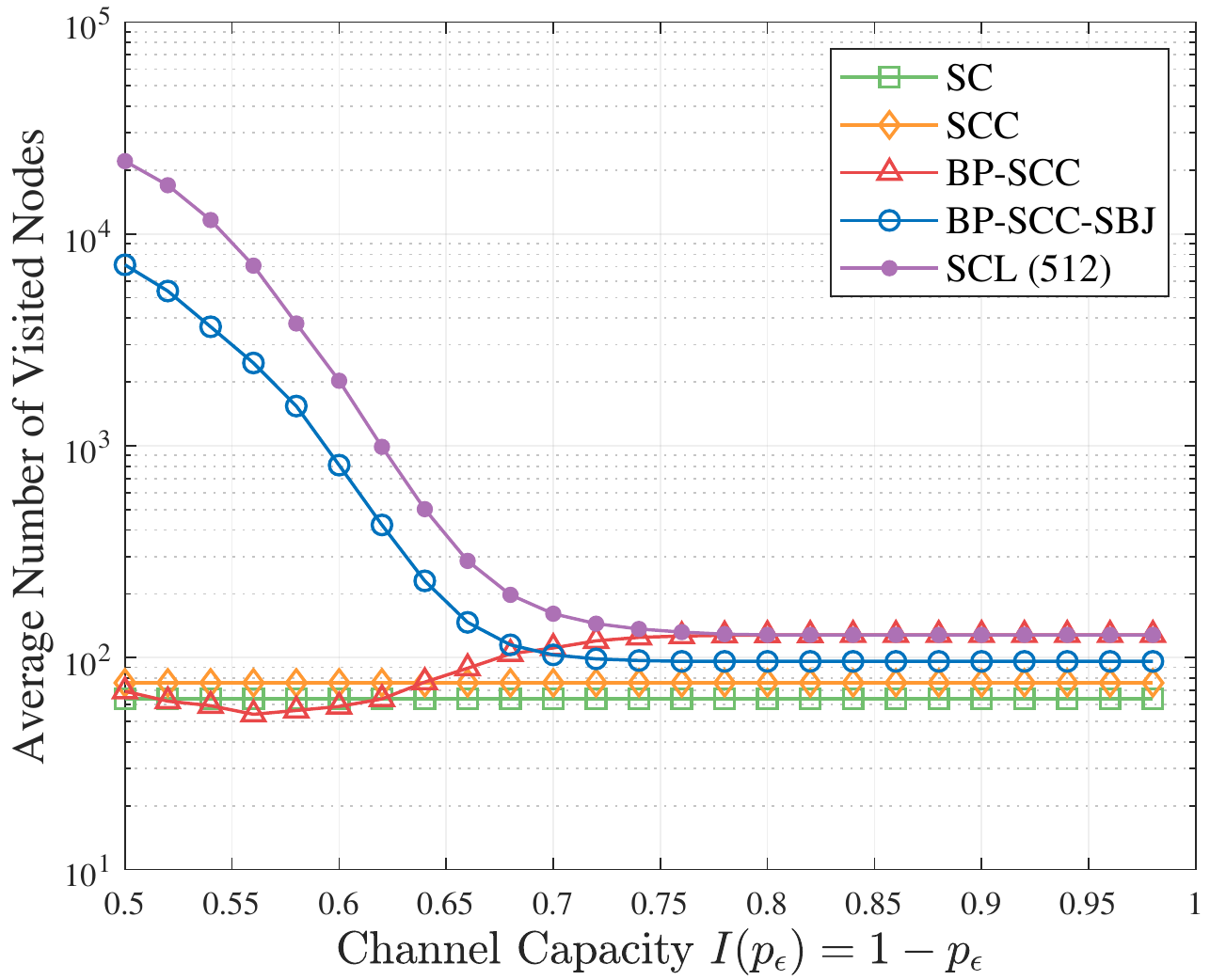}} \\
    \subfloat[$N=256,~K=128,~w_{\text{min}}(\mathbf{G}')=8,~w_{\text{min}}(\mathbf{TG})=72$]{
        \includegraphics[width=\subfigwidth]{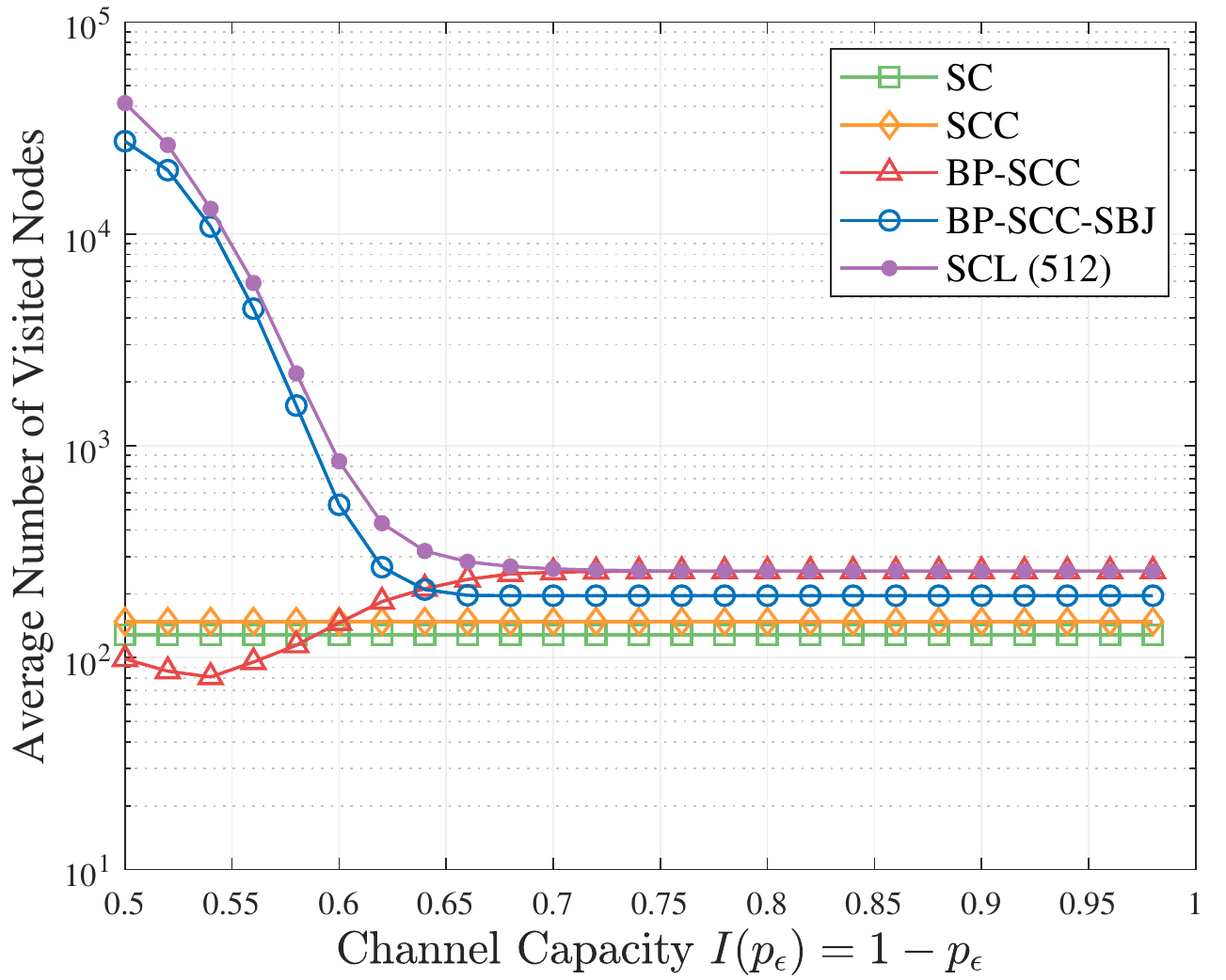}}\hfill
    \subfloat[$N=512,~K=256,~w_{\text{min}}(\mathbf{G}')=8,~w_{\text{min}}(\mathbf{TG})=128$]{
        \includegraphics[width=\subfigwidth]{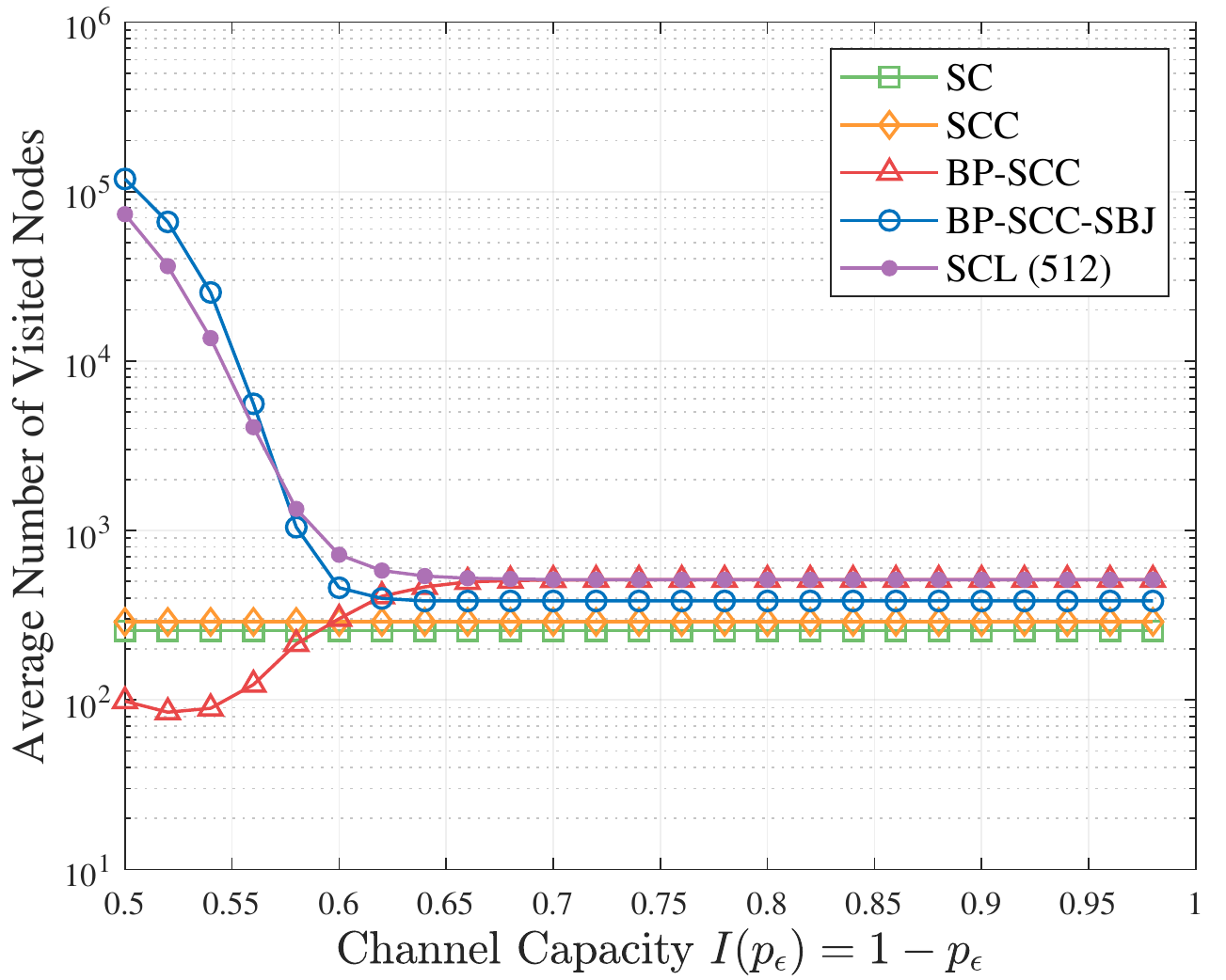}} \\
  \caption{Average number of visited nodes in the binary tree for five decoders over the BEC.
  The 3GPP NR uplink polar code construction scheme specified in \cite{TS38212} is used, where the 11-bit CRC code is concatenated.}
    \label{fig:anv}
\end{figure*}
\fi

The 3GPP NR polar codes introduced in the previous subsection are exploited in these simulations.
In SCC, BP-SCC, and BP-SCC-SBJ algorithms,
the 11 CRC parity bits are exploited as FCs along with frozen bits, thereby improving the likelihood of subchannels by \eqref{eq:new_subchannel}.
For each code, we find the minimum non-zero row weights of the effective generator matrix $\mathbf{G}'\triangleq\mathbf{G}_{\mathcal{A}\cup\mathcal{P},\ast}$
and the corresponding concatenated generator matrix $\mathbf{TG}$,
which are presented in the caption of each subfigure as $w_{\min}(\mathbf{G}')$ and $w_{\min}(\mathbf{TG})$, respectively.

As performance references, three bounds are drawn together.
One is the simulation-based ML bound emulated by using the methodology proposed in \cite{Tal2015}.
Whenever the SC decoder yields a wrong estimate,
we check whether $\sum_{i=0}^{N-1} \mathbbm{1}\left(y_i=\hat{x}_i\right) = \sum_{i=0}^{N-1} \mathbbm{1}\left(y_i=x_i\right)$, that is,
$W_N(\mathbf{y}\mmid \hat{\mathbf{x}}) = W_N(\mathbf{y}\mmid \mathbf{x})$.
The ML bound is given by measuring the ratio of such events.
Another is the DT bound \cite[Thm. 37]{Polyanskiy2010}, which is a strong achievability bound on the average BLER over all the codes of length $N$ and dimension $K$.
This bound is regarded as an approximation of the best achievable performance by any code of length $N$ and dimension $K$.
The last one is the meta-converse (MC) bound \cite[Thm. 38]{Polyanskiy2010}, which gives a lower bound on the BLER of any code of length $N$ and dimension $K$.

Additionally, an SCL decoder for the BEC is implemented by referring to \cite{Coskun2020,Mondelli2014} and its performance results are provided as a benchmark.
The SCL decoding operation for the BEC in the literature is not directly comparable to the proposed algorithms,
because it declares an error as soon as the number of current candidate paths exceeds a given list size.
In such cases, the SCL algorithm is slightly modified to select candidates randomly for fair comparison, instead of giving up decoding.
The list size of SCL decoding is chosen to be 512, resulting in a similar complexity to the BP-SCC-SBJ algorithm.

The decoding performance and complexity measures of the five decoding algorithms under consideration are shown in Figs. \ref{fig:bler} and \ref{fig:anv}.
As a complexity measure, we use the average number of visited nodes in the binary decision tree, which is a widely accepted metric in the literature \cite{Yao2020,Moradi2020}.
Since the proposed SCC, BP-SCC, and BP-SCC-SBJ algorithms make use of FCs in the estimation of each bit,
they outperform SC decoding, as shown in Fig.~\ref{fig:bler}.
The more techniques applied, the better the performance is.
The most advanced BP-SCC-SBJ algorithm outperforms SCL decoding in terms of error correction performance,
and approaches the DT bound at high channel erasure probability.
This near-bound performance is achieved by existing commercial codes, without the need for designing specially structured codes.
Further refinement of code design can also be explored while considering the utilization of the proposed decoding algorithms.
At low channel erasure probability, the performance is bounded by the ML bound,
which is mainly determined by the minimum distance characteristic of the employed code.

\section{Conclusion and Future Works}

We settled the suboptimal problem of the conventional SC-based decoding algorithms,
which was raised by Ar{\i}kan when polar codes were first proposed.
This paper presented initial works on FC-aided decoding methods.
There are several important open research items.
First, a decoding algorithm using FCs needs to be developed for more general B-DMCs.
The design of hard-decision decoders for the BEC is simple, because a conflict is clearly and explicitly detected.
However, the soft-input decoders for general B-DMCs should be more carefully designed.
For this case, the proposed techniques have the potential to complement existing well-established decoders, including SCL and SC-Fano decoders, since the incorporation of FCs is irrelevant to the underlying decoder structure.
In addition, it is interesting to investigate the structure and characteristics of polar codes suitable to be decoded by FC-aided schemes.
Deep learning techniques are also expected to be utilized for tree search methods solving CSPs with low complexity.



\begin{thebibliography}{30}

\bibitem{Arikan2009}
E.~Ar{\i}kan,
``Channel polarization: a method for constructing capacity-achieving codes for symmetric binary-input memoryless channels,''
\textit{IEEE Trans. Inf. Theory}, vol.~55, no.~7, pp. 3051--3073, 2009.

\bibitem{Tal2015}
I.~Tal and A.~Vardy,
``List decoding of polar codes,''
\textit{IEEE Trans. Inf. Theory}, vol.~61, no.~5, pp. 2213--2226, 2015.

\bibitem{Niu2012a}
K. Niu and K. Chen,
``Stack decoding of polar codes,''
\textit{Electron. Lett.}, vol.~48, no.~12, pp.~695--696, June 2012.

\bibitem{Afisiadis2014}
O. Afisiadis, A. Balatsoukas-Stimming, and A. Burg,
``A low-complexity improved successive cancellation decoder for polar codes,''
in \textit{Proc. Asilomar Conf. Signals, Systems, and Computers},
Pacific Grove, CA, USA, Nov. 2014, pp. 2116--2120.

\bibitem{Jeong2019}
M.-O. Jeong and S.-N. Hong,
``SC-Fano decoding of polar codes,''
\textit{IEEE Access}, vol.~7, pp. 81682--81690, 2019.

\bibitem{Coskun2020}
M. C. Co\c{s}kun, J. Neu, and H. D. Phister,
``Successive Cancellation Inactivation Decoding for Modified Reed-Muller and eBCH Codes,''
in \textit{Proc. IEEE Int. Symp. Inf. Theory (ISIT)}, Los Angeles, CA, USA, June 2020, pp.~437--442.

\bibitem{TS38212}
\textit{{NR} multiplexing and channel coding (Release 16)},
document TSG RAN TS38.212 v17.1.0, 3GPP, Mar. 2022.

\bibitem{Arikan2016}
E.~Ar{\i}kan, ``On the origin of polar coding,''
\textit{IEEE J. Sel. Areas Commun.}, vol.~34, no.~2, pp. 209--223, Feb. 2016.

\bibitem{Trifonov2016}
P.~Trifonov and V. Miloslavskaya,
``Polar subcodes,''
\textit{IEEE J. Sel. Areas Commun.}, vol.~34, no.~2, pp.~254--266, Feb. 2016.

\bibitem{Niu2012}
K.~Niu and K.~Chen,
``{CRC}-aided decoding of polar codes,''
\textit{IEEE Commun. Lett.}, vol.~16, no.~10, pp. 1668--1671, Oct. 2012.

\bibitem{Wang2016}
T.~Wang, D.~Qu, and T.~Jiang,
``Parity-check-concatenated polar codes,''
\textit{IEEE Commun. Lett.}, vol.~20, no.~12, pp. 2342--2345, Dec. 2016.

\bibitem{Miloslavskaya2021}
V.~Miloslavskaya, B.~Vucetic, Y. Li, G. Park, and O.-S.~Park,
``Recursive design of precoded polar codes for SCL decoding,''
\textit{IEEE Trans. Commun.}, vol.~69, no.~12, pp.~7945--7959, Dec. 2021.

\bibitem{Arikan2019}
E.~Ar{\i}kan,
``From sequential decoding to channel polarization and back again,''
2019. [Online]. Available: arXiv:1908.09594


\bibitem{Yuan2014}
B.~Yuan and K. K. Parhi,
``Algorithm and architecture for hybrid decoding of polar codes,''
in \textit{Proc. IEEE Int. Symp. Circuits and Systems (ISCAS)}, Melbourne, Australia, 2014, pp. 209--212.

\bibitem{Zhou2018}
X. Zhou, Y. Shin, X. Tan, X. You, Z. Zhang, and C. Zhang,
``An adjustable hybrid SC-BP polar decoder,''
in \textit{Proc. IEEE Asia Pacific Conf. Circuits and Systems (APCCAS)}, Chengdu, China, 2018, pp. 211--214.

\bibitem{Vajha2019}
M.~Vajha, V.~S.~C. Mukka, and P.~V. Kumar,
``Backtracking and look-ahead decoding algorithms for improved successive cancellation decoding performance of polar codes,''
in \textit{Proc. IEEE Int. Symp. Inf. Theory (ISIT)}, Paris, France, Jul. 2019, pp. 31--35.

\bibitem{Schurch2017}
C.~Sch\"{u}rch,
``On successive cancellation decoding of polar codes and related codes,''
Ph.D Dissertation, ETH Z\"{u}rich, 2017.

\bibitem{Sun2020}
H.~Sun, R.~Liu, and C.~Gao,
``A simplified decoding method of polar codes based on hypothesis testing,''
\textit{IEEE Commun. Lett.}, vol.~24, no.~3, pp. 530--533, Mar. 2020.

\newpage
\bibitem{Ardakani2019}
M.~H.~Ardakani, M.~Hanif, M.~Ardakani, and C.~Tellambura,
``Fast Successive-Cancellation-Based Decoders of Polar Codes,''
\textit{IEEE Trans. Commun.}, vol.~67, no.~7, pp.~4562--4574, July 2019.

\bibitem{Meseguer1989}
P.~Meseguer,
``Constraint satisfaction problems: An overview,''
\textit{AI Commun.}, vol.~2, no.~1, pp. 3--17, Jan. 1989.

\bibitem{Mittal1990}
S.~Mittal and B.~Falkenhainer,
``Dynamic constraint satisfaction problems,''
in \textit{Proc. 8th Nat. Conf. Artif. Intell.}, 1990, pp. 25-32.

\bibitem{Prosser1993}
P.~Prosser,
``Hybrid algorithms for the constraint satisfaction problem,''
\textit{Comput. Intell.}, vol.~9, no.~3, pp. 268--299, Aug. 1993.

\bibitem{Polyanskiy2010}
Y.~Polyanskiy, H.~V. Poor, and S.~Verd\'{u},
``Channel coding rate in the finite blocklength regime,''
\textit{IEEE Trans. Inf. Theory}, vol.~56, no.~5, pp. 2307--2359, May 2010.

\bibitem{Leroux2013}
C.~Leroux, A.~J. Raymond, G.~Sarkis, and W.~J. Gross,
``A semi-parallel successive-cancellation decoder for polar codes,''
\textit{IEEE Trans. Signal Process.}, vol.~61, no.~2, pp. 289--299, Jan. 2013.

\bibitem{BalatsoukasStimming2015}
A.~Balatsoukas-Stimming, M.~B. Parizi, and A.~Burg,
``LLR-based successive cancellation list decoding of polar codes,''
\textit{IEEE Trans. Signal Process.}, vol.~63, no.~19, pp. 5165--5179, Oct. 2015.

\bibitem{Gallager1962}
R.~Gallager,
``Low-density parity-check codes,''
\textit{IRE Trans. Inf. Theory}, vol.~8, no.~1, pp. 21--28, Jan. 1962.

\bibitem{modern2008}
T.~Richardson and R.~Urbanke, \textit{Modern Coding Theory}.
Cambridge, UK: Cambridge University Press, 2008.

\bibitem{Kumar1998}
V.~Kumar,
``Algorithms for constraint satisfaction problems: A survey,''
\textit{A.I. Mag}, vol.~13, no.~1, pp. 32--44, 1992.

\bibitem{Brailsford1999}
S.~Brailsford, C.~Potts, and B.~Smith,
``Constraint satisfaction problems: Algorithms and applications,''
\textit{Eur. J. Oper. Res.}, vol. 119, pp. 557--581, 1999.

\bibitem{Rowshan2021}
M.~Rowshan, A.~Burg, and E.~Viterbo,
``Polarization-adjusted convolutional (PAC) codes: Sequential decoding vs list decoding,''
\textit{IEEE Trans. Veh. Technol.}, vol.~70, no.~2, pp. 1434--1447, Feb. 2021.


\bibitem{Mondelli2014}
M.~Mondelli, S.~H.~Hassani, R. L. Urbanke,
``From polar to Reed-Muller codes: A technique to improve the finite-length performance,''
\textit{IEEE Trans. Commun.}, vol.~62, no.~9, pp.~3084--3091, Sep. 2014.

\bibitem{Yao2020}
H. Yao, A. Fazeli, A. Vardy,
``List decoding of Ar{\i}kan's PAC codes,''
in \textit{Proc. IEEE Int. Symp. Inf. Theory (ISIT)},
Los Angeles, CA, USA, June 2020, pp.~443--448.

\bibitem{Moradi2020}
M. Moradi, A. Mozammel, K. Qin, and E. Ar{\i}kan,
``Performance and complexity of sequential decoding of PAC codes,''
2020. [Online]. Available: arXiv:2012.04990v2.

\bibitem{Jang2023access}
M. Jang, J.-H. Kim, S. Myung, and K. Yang,
``Successive cancellation decoding with future constraints for polar codes over the binary erasure channel,''
\textit{IEEE Access}, vol. 11, pp. 97699--97715, 2023, doi: 10.1109/ACCESS.2023.3312577.

\end{thebibliography}
\end{document}